\documentclass[aps, prd, amsmath, floats, floatfix, twocolumn, nofootinbib, superscriptaddress]{revtex4-1}
\usepackage[pdftex]{graphicx}
\usepackage{color}
\usepackage{latexsym}
\usepackage[colorlinks]{hyperref}

\newcommand{\be}{\begin{eqnarray}}
\newcommand{\ee}{\end{eqnarray}}

\begin{document}

\title{Testing general relativity with supermassive black holes\\using X-ray reflection spectroscopy}

% Review article prepared for a Special Issue of Universe and based on some talks given at the meeting "Recent Progress in Relativistic Astrophysics" (6-8 May 2019, Shanghai, China).

\author{Askar~B.~Abdikamalov}
\affiliation{Center for Field Theory and Particle Physics and Department of Physics, Fudan University, 200438 Shanghai, China}

\author{Dimitry~Ayzenberg}
\affiliation{Center for Field Theory and Particle Physics and Department of Physics, Fudan University, 200438 Shanghai, China}

\author{Cosimo~Bambi}
\email{bambi@fudan.edu.cn}
\affiliation{Center for Field Theory and Particle Physics and Department of Physics, Fudan University, 200438 Shanghai, China}

%\author{Kishalay~Choudhury}
%\affiliation{Center for Field Theory and Particle Physics and Department of Physics, Fudan University, 200438 Shanghai, China}

%\author{Thomas~Dauser}
%\affiliation{Remeis Observatory \& ECAP, Universit\"{a}t Erlangen-N\"{u}rnberg, 96049 Bamberg, Germany}

%\author{Javier~A.~Garc{\'\i}a}
%\affiliation{Cahill Center for Astronomy and Astrophysics, California Institute of Technology, Pasadena, CA 91125, USA}
%\affiliation{Remeis Observatory \& ECAP, Universit\"{a}t Erlangen-N\"{u}rnberg, 96049 Bamberg, Germany}

\author{Sourabh~Nampalliwar}
\affiliation{Theoretical Astrophysics, Eberhard-Karls Universit\"at T\"ubingen, 72076 T\"ubingen, Germany}

\author{Ashutosh~Tripathi}
%\email{ashutosh\_tripathi@fudan.edu.cn}
\affiliation{Center for Field Theory and Particle Physics and Department of Physics, Fudan University, 200438 Shanghai, China}

\author{Jelen~Wong}
\affiliation{Center for Field Theory and Particle Physics and Department of Physics, Fudan University, 200438 Shanghai, China}

\author{Yerong~Xu}
\affiliation{Center for Field Theory and Particle Physics and Department of Physics, Fudan University, 200438 Shanghai, China}

\author{Jinli~Yan}
\affiliation{Center for Field Theory and Particle Physics and Department of Physics, Fudan University, 200438 Shanghai, China}

\author{Yunfeng~Yan}
\affiliation{Center for Field Theory and Particle Physics and Department of Physics, Fudan University, 200438 Shanghai, China}

\author{Yuchan~Yang}
\affiliation{Center for Field Theory and Particle Physics and Department of Physics, Fudan University, 200438 Shanghai, China}

\begin{abstract}
In this paper, we review our current efforts to test General Relativity in the strong field regime by studying the reflection spectrum of supermassive black holes. So far we have analyzed 11~sources with observations of \textsl{NuSTAR}, \textsl{Suzaku}, \textsl{Swift}, and \textsl{XMM-Newton}. Our results are consistent with general relativity, according to which the spacetime metric around astrophysical black holes should be well approximated by the Kerr solution. We discuss the systematic uncertainties in our model and we present a preliminary study on the impact of some of them on the measurement of the spacetime metric.
\end{abstract}

\maketitle

%%%%%%%%%%%%%%%%%%%%%%%%%%%%%%%

\section{Introduction}

The Theory of General Relativity was proposed by Einstein at the end of 1915~\cite{einstein}, and still represents the standard framework for the description of gravitational fields and of the chrono-geometrical structure of spacetime. While the first experimental test can be dated back to the observation of light bending by the Sun by Eddington and collaborators in 1919~\cite{eddington}, systematic tests of General Relativity started much later, since the 1960s with experiments in the Solar System and since the 1970s with observations of radio pulsars. For a review, see, for instance, Ref.~\cite{will}. Note that all these tests are in the so-called weak field regime, where corrections to Newtonian gravity are small and can be treated perturbatively. Today the interest is shifting to test General Relativity in more extreme conditions, in particular on very large scales (cosmological tests) and in strong gravitational fields (compact objects)~\cite{t1,t2}.

Astrophysical black holes are ideal laboratories for testing Einstein's Theory of General Relativity in the strong field regime because they are the systems with the strongest gravitational fields that can be found today in the Universe. In General Relativity, uncharged black holes are described by the Kerr solution~\cite{kerr} and are relatively simple objects, in the sense that they are completely specified by only two parameters, representing, respectively, the mass $M$ and the spin angular momentum $J$ of the black hole. This is the conclusion of the celebrated ``no-hair'' theorems~\cite{nh1,nh2,nh3}, which hold under specific assumptions. It is also remarkable that the spacetime around an astrophysical black hole formed from gravitational collapse should be well approximated by the ideal Kerr solution. We can quantify deviations from the Kerr background due to initial conditions before the creation of the black hole, the gravitational field of nearby stars or of the accretion disk, non-vanishing electric charges, etc., but it turns out that all these effects generally have an extremely weak impact on the spacetime geometry and can be safely ignored; see, for instance, Refs.~\cite{k1,k2,k3,k4,book} for more details. In conclusion, the spacetime metric around an object generated by the complete gravitational collapse of an astronomical system should be described by the Kerr metric and the detection of macroscopic deviations from the Kerr solutions could be a signature of new physics~\cite{new1,new2}.

Astrophysical black holes can be tested with electromagnetic techniques~\cite{e1,e2,e3,e4,e5,e6,e7,e8} and gravitational waves~\cite{g1,g2,g3,g4}. The two methods are complementary because they test different sectors of the theory. Electromagnetic techniques, strictly speaking, can test the motion of massive and massless particles in the strong gravitational field of a black hole. The gravitational wave signal emitted by a system with a black hole depends instead on the evolution of the gravitational field in response to a variation of the distribution of energy/momentum in the spacetime. For example, deviations from geodesic motion due to a new coupling between the matter and gravity sectors may produce an effect in the electromagnetic spectrum without affecting the gravitational wave signal. Modified theories of gravity in which black holes are still described by the Kerr solution~\cite{psaltis} may be tested with gravitational waves and have instead an electromagnetic spectrum compatible with that expected in General Relativity~\cite{barausse}.

In this paper, we will review our current efforts to test Einstein's Theory of General Relativity in the strong field regime using supermassive black holes and the electromagnetic technique called X-ray reflection spectroscopy. As of now, our results are the only tests of the Kerr metric in the strong field regime with an electromagnetic method. The paper is organized as follows. In Section~\ref{s-ref}, we briefly review the origin of the reflection spectrum of the accretion disk around black holes and the reflection model {\sc relxill}. In Section~\ref{s-gr}, we present our method to test General Relativity using X-ray reflection spectroscopy and our reflection model {\sc relxill\_nk} specifically designed to test the Kerr metric. In Section~\ref{s-results}, we review current constraints from the 11~supermassive black holes that we have analyzed so far. In Section~\ref{s-sys}, we list the systematic uncertainties in our model and, in Section~\ref{s-flav}, we present a preliminary study on the impact of different model choices on our tests of the Kerr metric. Summary and future plans are presented in Section~\ref{s-con}. Throughout the paper, we adopt units in which $G_{\rm N} = c = 1$ and the convention of a metric with signature $(-+++)$.

\section{X-ray reflection spectroscopy \label{s-ref}}

\begin{figure}[t]
\begin{center}
\includegraphics[type=pdf,ext=.pdf,read=.pdf,width=8.7cm]{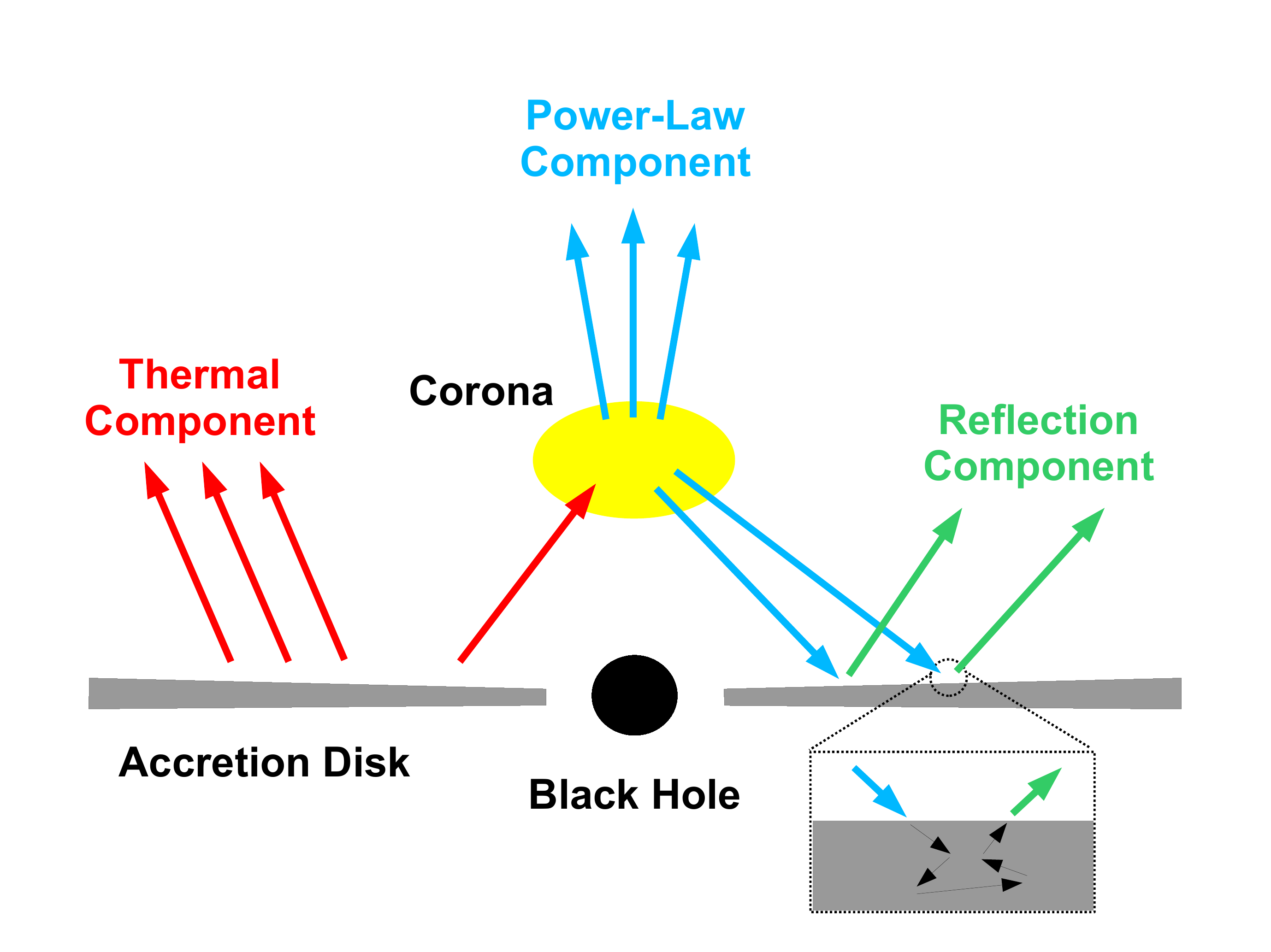}
\end{center}
\vspace{-0.3cm}
\caption{A black hole is accreting from a geometrically thin and optically thick disk. The disk has a multi-temperature blackbody spectrum (red arrows). Thermal photons from the disk can inverse Compton scatter off free electrons in the corona, producing a power-law component (blue arrows). Comptonized photons can illuminate the disk, generating a reflection component (green arrows). \label{f-corona}}
\end{figure}

Our astrophysical system is sketched in Fig.~\ref{f-corona}. A black hole is accreting from a geometrically thin and optically thick accretion disk. Since the disk is in thermal equilibrium, at every point the emission is like that of a blackbody, and the spectrum of the whole disk is a multi-temperature blackbody spectrum (red arrows in Fig.~\ref{f-corona}). The temperature of the disk depends on the black hole mass and mass accretion rate, and increases as the gas falls onto the gravitational potential of the black hole. For a black hole accreting at 10\% of its Eddington limit, the thermal spectrum of the inner part of the accretion disk is peaked in the soft X-ray band ($\sim 1$~keV) for stellar mass black holes and in the optical/UV band ($1-100$~eV) for supermassive black holes.

Thermal photons of the accretion disk can inverse Compton scatter off free electrons in the so-called corona, which is a generic name to call a hotter ($\sim 100$~keV) cloud of gas in the vicinity of the black hole. The corona may be represented, for instance, by the accretion flow between the inner edge of the accretion disk and the black hole, the atmosphere above the accretion disk, or the base of the black hole jet. Typically, at any given time only one of these coronas provides the main contribution for the inverse Compton scattering of the photons from the disk. The process makes in the corona produces a power-law component with an exponential cut-off (blue arrows in Fig.~\ref{f-corona}). Comptonized photons can illuminate the disk, generating a reflection component (green arrows in Fig.~\ref{f-corona}). The most prominent features of such a reflection component are some narrow (in the rest-frame of the gas) fluorescent emission lines, notably the iron K$\alpha$ complex at $6.4-6.97$~keV (depending on the ionization of iron ions), and the Compton hump at $10-30$~keV. X-ray reflection spectroscopy refers to the study of this reflection component.

The reflection spectrum in the rest frame of the gas depends on atomic physics only and is independent of the gravitational field (assuming that the Einstein Equivalence Principle holds, which may not be the case in theories in which matter does not universally couple to gravity~\cite{will}). The reflection spectrum that we observe far from the source is the result of relativistic effects (Doppler boosting, gravitational redshift, light bending) that photons experience traveling from the emission point on the disk to the detection point far from the source.

As of now, {\sc relxill} is the most advanced relativistic reflection model for the Kerr spacetime~\cite{relxill1,relxill2}. It is the result of the merger between {\sc xillver}~\cite{xillver} and {\sc relline} (later called {\sc relconv})~\cite{relline1,relline2}. {\sc xillver} is a pure atomic physics model calculating the reflection spectrum of a cold accretion disk illuminated by X-ray radiation. {\sc relconv} is a convolution model: from the spectrum at any point in the disk in the rest-frame of the gas, it calculates the total spectrum observed far from the source assuming that the spacetime metric is described by the Kerr solution and that the accretion flow is described by an infinitesimally thin Novikov-Thorne disk~\cite{ntm}. From the analysis of the reflection spectrum of astrophysical black holes with {\sc relxill}, we can estimate the parameters of the model. Note that X-ray reflection spectroscopy is currently the only method for measuring the spin of supermassive black holes~\cite{book,s1,s2}.

\section{Testing the Kerr hypothesis \label{s-gr}}

There are two approaches to test the Kerr nature of astrophysical black holes (Kerr hypothesis) with electromagnetic radiation. They are usually referred to as, respectively, top-down and bottom-up methods.

The top-down approach is the most natural and logical one. We want to test the predictions of General Relativity, according to which astrophysical black holes should be described by the Kerr solution, against another theory of gravity in which uncharged black holes are not described by the Kerr solution. To do this, we have to construct two models, one in which the spacetime is described by the Kerr metric and the other one in which the spacetime is described by the black hole solution of the other theory of gravity. We then fit the data with the Kerr and the non-Kerr models and we can check whether astronomical data prefer one of the two models and can rule out the other one. Since there are many theories of gravity, we should repeat this analysis for each of them. However, the main problem is that we do not know the rotating black hole solutions in most theories of gravity. We often know the non-rotating black hole solutions and sometimes we have some approximated solutions valid in the slow-rotating limit, while the complete rotating solutions are unknown. This is just a technical problem to solve the field equations of the corresponding gravity theory. Even in General Relativity, the non-rotating (Schwarzschild) solution was found by Schwarzschild in 1916, shortly after Einstein had proposed his theory. On the contrary, the rotating (Kerr) solution was found only in 1964 by Kerr. Since the spin plays an important rule in particle motion and our tests require fast-rotating black holes to break parameter degeneracy, without the complete rotating solution it is impossible to test the black holes of certain theories of gravity.

The bottom-up approach is a phenomenological method in which we want to test the Kerr metric with a null-experiment, but we are not considering any particular theory of gravity beyond General Relativity. The spacetime is described by a parametric black hole metric in which the Kerr solution is deformed by adding {\it ad hoc} ``deformation parameters''. The latter are introduced to quantify possible deviations from the Kerr metric and when all deformation parameters vanish we should recover the Kerr solution of General Relativity. There is no theory behind such a parametric black hole metric, but the idea is that these deformation parameters can capture possible non-Kerr features. As in any null-experiment, we expect that the Kerr hypothesis is correct and we want to verify it. If the analysis of astronomical data required some non-vanishing deformation parameter, this would point out that the spacetime metric around astrophysical black holes may not be described by the Kerr solution (assuming that the astrophysical model is correct).

There are several parametric black hole spacetimes in the literature that can be adopted for the bottom-up approach. A possible choice is the Johannsen metric~\cite{tj}. In Boyer-Lindquist-like coordinates, the line element of the Johannsen metric reads
\be\label{eq-jm}
ds^2 &=&-\frac{\tilde{\Sigma}\left(\Delta-a^2A_2^2\sin^2\theta\right)}{B^2}dt^2
+\frac{\tilde{\Sigma}}{\Delta A_5}dr^2+\tilde{\Sigma} d\theta^2 \nonumber\\
&&-\frac{2a\left[\left(r^2+a^2\right)A_1A_2-\Delta\right]\tilde{\Sigma}\sin^2\theta}{B^2}dtd\phi \nonumber\\
&&+\frac{\left[\left(r^2+a^2\right)^2A_1^2-a^2\Delta\sin^2\theta\right]\tilde{\Sigma}\sin^2\theta}{B^2}d\phi^2 \, ,
\ee
where $M$ is the black hole mass, $a = J/M$, $J$ is the black hole spin angular momentum, $\tilde{\Sigma} = \Sigma + f$, and
\be
\Sigma &=& r^2 + a^2 \cos^2\theta \, , \\
\Delta &=& r^2 - 2 M r + a^2 \, , \\
B &=& \left(r^2+a^2\right)A_1-a^2A_2\sin^2\theta \, .
\ee
The functions $f$, $A_1$, $A_2$, and $A_5$ are defined as
\be
f &=& \sum^\infty_{n=3} \epsilon_n \frac{M^n}{r^{n-2}} \, , \\
A_1 &=& 1 + \sum^\infty_{n=3} \alpha_{1n} \left(\frac{M}{r}\right)^n \, , \\
A_2 &=& 1 + \sum^\infty_{n=2} \alpha_{2n} \left(\frac{M}{r}\right)^n \, , \\
A_5 &=& 1 + \sum^\infty_{n=2} \alpha_{5n} \left(\frac{M}{r}\right)^n \, ,
\ee
where $\{ \epsilon_n \}$, $\{ \alpha_{1n} \}$, $\{ \alpha_{2n} \}$, and $\{ \alpha_{5n} \}$ are four infinite sets of deformation parameters. We note that this form of the Johannsen metric recovers the correct Newtonian limit and passes all Solar System experiments without fine-tuning.

In Refs.~\cite{noi-nk1,noi-nk2}, we have extended the {\sc relxill} model to the Johannsen metric in which we allow that one of the deformation parameters among $\epsilon_3$, $\alpha_{13}$, and $\alpha_{22}$ is non-vanishing. This new reflection model is called {\sc relxill\_nk}\footnote{The public version can be downloaded from the following URLs:\\
{\tiny \url{http://www.physics.fudan.edu.cn/tps/people/bambi/Site/RELXILL_NK.html}}\\
{\tiny \url{http://www.tat.physik.uni-tuebingen.de/~nampalliwar/relxill_nk/}}}, where NK stands for Non-Kerr. We assume standard atomic physics (Einstein Equivalence Principle) and therefore we keep {\sc xillver} without modifications. We construct the new convolution model {\sc relconv\_nk} that takes all relativistic effects of the Johannsen metric into account. We note that the new model can be easily extended to any stationary, axisymmetric, asymptotically flat spacetime without pathological properties (see, for instance, Refs.~\cite{noi-ext1,noi-ext2,noi-ext3}). The deformation parameters in {\sc relxill\_nk} are just model parameters like the ones already present in {\sc relxill} and their value can be estimated by fitting the observational data of astrophysical black holes with the theoretical predictions in the Johannsen spacetime of {\sc relxill\_nk}.

In order to avoid a spacetime with pathological properties (spacetime singularities, regions closed time-like curves, etc.), it is necessary to limit the parameter space by imposing constraints on the spin parameter $a_* = a/M$ and the deformation parameters. As in the Kerr metric, the constraint on the spin parameter is
\be
- 1 < a_* < 1 \, .
\ee 
For $|a_*| > 1$, there is no event horizon and the central singularity is naked. The constraints on the deformation parameters $\alpha_{13}$, $\alpha_{22}$, and $\epsilon_3$ are 
\be
\label{eq-c}
&&\alpha_{13} > - \frac{1}{2} \left( 1 + \sqrt{1 - a^2_*} \right)^4 \, , \nonumber\\
&&\frac{1}{a_*^2} \left( 1 + \sqrt{1 - a^2_*} \right)^4 
> \alpha_{22} > - \left( 1 + \sqrt{1 - a^2_*} \right)^2 \, , \qquad \nonumber\\
&&\epsilon_3 > - \left( 1 + \sqrt{1 - a_*^2} \right)^3 \, .
\ee
For more details on these constraints, see Refs.~\cite{tj,noi-ark}.

\section{Present results \label{s-results}}

In the past one and a half years, we have analyzed X-ray observations of 4~stellar-mass black holes~\cite{noi-gx,noi-gs,noi-grs,noi-cyg} and 11~supermassive black holes~\cite{noi-ark,noi-1h0707,noi-univ,noi-mrk,noi-mcg,noi-bare,noi-e3} with {\sc relxill\_nk}. Supermassive black holes are probably more suitable than the stellar-mass ones for our tests, because their spectrum is easier to model (as the temperature of the accretion disk is lower) and they typically rotate very fast (so the inner edge of the accretion disk is closer to the black hole event horizon and the radiation emitted from the inner part of the accretion disk is more strongly affected by relativistic effects). In this article, we will only review our results for supermassive black holes.

\begin{table*}
 \centering
 \caption{List of the sources and of the observations analyzed as of now with {\sc relxill\_nk} to test the Kerr nature of the associated supermassive black hole. \label{t-obs}}
\vspace{0.3cm}
\begin{tabular}{cccccc}
\hspace{1.2cm} Source \hspace{1.2cm} & \hspace{0.2cm} Mission \hspace{0.2cm} & \hspace{0.2cm} Observation ID \hspace{0.2cm} & \hspace{0.2cm} Year \hspace{0.2cm} & \hspace{0.2cm} Exposure (ks) \hspace{0.2cm} & \hspace{0.2cm} References \hspace{0.2cm} \\ 
\hline\hline
1H0419--577 &\textsl{Suzaku}&702041010&2007&179&\cite{noi-bare,noi-e3}\\
\hline
1H0707--495 &\textsl{XMM-Newton}&0554710801&2011&98&\cite{noi-1h0707,noi-univ}\\
&\textsl{NuSTAR}&60001102002&2014&144&\\
&\textsl{NuSTAR}&60001102004&2014&49&\\
&\textsl{NuSTAR}&60001102006&2014&47&\\
&\textsl{Swift}&00080720001&2014&20&\\
&\textsl{Swift}&00080720004&2014&17&\\
\hline
Ark~120 &\textsl{Suzaku}&702014010&2007&91&\cite{noi-bare}\\
\hline
Ark~564 &\textsl{Suzaku}&702117010&2007&80&\cite{noi-ark,noi-e3}\\
\hline
Fairall~9 &\textsl{Suzaku}&702043010&2007&145&\cite{noi-bare}\\
\hline
MCG--6--30--15 &\textsl{NuSTAR}&60001047002&2013&23&\cite{noi-mcg,noi-e3}\\
&\textsl{NuSTAR}&60001047003&2013&127&\\
&\textsl{NuSTAR}&60001047005&2013&30&\\
&\textsl{XMM-Newton}&0693781201&2013&134&\\
&\textsl{XMM-Newton}&0693781301&2013&134&\\
&\textsl{XMM-Newton}&0693781401&2013&49&\\
\hline
Mrk~335 &\textsl{Suzaku}&701031010&2006&151&\cite{noi-mrk}\\
\hline
PKS~0558--504&\textsl{Suzaku}&701011010&2007&20&\cite{noi-bare,noi-e3}\\
&\textsl{Suzaku}&701011020&2007&19&\\
&\textsl{Suzaku}&701011030&2007&21&\\
&\textsl{Suzaku}&701011040&2007&20&\\
&\textsl{Suzaku}&701011050&2007&20&\\
\hline
RBS~1124 &\textsl{Suzaku}&702114010&2007&79&\cite{noi-bare}\\
\hline
Swift~J0501.9--3239 &\textsl{Suzaku}&703014010&2008&36&\cite{noi-bare,noi-e3}\\
\hline
Ton~S180 &\textsl{Suzaku}&701021010&2006&108&\cite{noi-bare}\\
\hline\hline
\end{tabular}
%\end{table*}
\vspace{0.8cm}
%\begin{table*}
 \centering
 \caption{Measurements of the deformation parameters $\alpha_{13}$, $\alpha_{22}$, and $\epsilon_3$ from the 11~supermassive black holes that we have studied until now. We report the uncertainties at 90\% confidence level for one relevant parameter. -- means that we have not measured that deformation parameter for the corresponding object. $\times$ indicates that we tried to measure the deformation parameter but it was not possible to get any clear constraint. See the references in the last column for more details. \label{t-con}}
\vspace{0.3cm}
\begin{tabular}{cccccc}
\hspace{1.2cm} Source \hspace{1.2cm} & \hspace{0.3cm} $L/L_{\rm Edd}$ \hspace{0.3cm} & \hspace{0.3cm} $\alpha_{13}$ \hspace{0.3cm} & \hspace{0.3cm} $\alpha_{22}$ \hspace{0.3cm} & \hspace{0.3cm} $\epsilon_3$ \hspace{0.3cm} & References \\ 
\hline
1H0419--577 & $\gtrsim 1$ & $0.00_{-0.14}^{+0.04}$ & $0.00_{-0.04}^{+0.13}$ & $-0.2_{-2.1}^{+0.5}$ &\cite{noi-bare,noi-e3}\\
1H0707--495 & $\gtrsim 1$ & $(-1.9, 0.5)$ & -- & -- &\cite{noi-1h0707,noi-univ}\\
Ark~120 & $\sim 0.04$ & $0.00_{-0.08}^{+0.01}$ & $0.01_{-0.03}^{+0.06}$ & -- &\cite{noi-bare}\\
Ark~564 & $\gtrsim 0.5$ & $-0.2_{-0.2}^{+0.3}$ & $0_{-0.3}^{+0.05}$ & $0.41_{-0.69}^{+0.11}$ &\cite{noi-ark,noi-e3}\\
Fairall~9 & $\sim 0.05$ & $\times$ & $1.3_{-0.4}^{+0.2}$ & -- &\cite{noi-ark}\\
MCG--6--30--15 & $\sim 0.4$ & $0.00_{-0.20}^{+0.07}$ & $0.0_{-0.1}^{+0.6}$ & $-0.05_{-0.17}^{+0.29}$ &\cite{noi-mcg,noi-e3}\\
Mrk~335 & $\sim 0.25$ & $\times$ & $\times$ & -- &\cite{noi-mrk}\\
PKS~0558--504& $\gtrsim 1$ & $0.03_{-0.20}^{+0.02}$ & $-0.03_{-0.02}^{+0.19}$ & $0.0_{-0.8}^{+0.1}$ &\cite{noi-bare,noi-e3}\\
RBS~1124 & ? & $\times$ & $\times$ & -- &\cite{noi-bare}\\
Swift~J0501.9--3239 & ? & $0.00_{-0.07}^{+0.03}$ & $0.11_{-0.18}^{+0.05}$ & -- &\cite{noi-bare}\\
Ton~S180 & $\gtrsim 1$ & $0.01_{-0.32}^{+0.02}$ & $-0.02_{-0.04}^{+0.30}$ & -- &\cite{noi-bare}\\
\hline
\end{tabular}
\end{table*}

The list of sources and observations analyzed as of now with {\sc relxill\_nk} is reported in Tab.~\ref{t-obs}. Details on the observations and the data reduction can be found in the references in the last column of the table.

Tab.~\ref{t-con} shows the summary of our measurements of the deformation parameters $\alpha_{13}$, $\alpha_{22}$, and $\epsilon_3$ from every source. For some sources, we do not get a clear measurement (either the constraint is very weak or we find multiple measurements) and in such a case we report $\times$ in Tab.~\ref{t-con}. More details can be found in the corresponding reference. For other sources, we have not tried to measure a certain deformation parameter, and in such a case we report -- in the table. The best-fit tables of every source and the constraints on the plane spin parameter vs deformation parameter can be found in the corresponding reference in the last column of Tab.~\ref{t-con}. The second column of Tab.~\ref{t-con} show the accretion luminosity of the source in Eddington units. The Novikov-Thorne model for the description of geometrically thin and optically thick accretion disk is normally thought to hold when the black hole is accreting between 5\% to 30\% of its Eddington limit~\cite{kulk}. However, as it is common in X-ray astronomy, we have applied our model even for sources in which the accretion luminosity is higher than 30\%.

As we will discuss in the next section, {\sc relxill\_nk} has a number of simplifications that introduce systematic uncertainties in the final estimates of the model parameters, and therefore even in the measurements of the deformation parameters $\alpha_{13}$, $\alpha_{22}$, and $\epsilon_3$. Unfortunately, it is difficult to estimate all the systematic uncertainties in the final measurement, so Tab.~\ref{t-con} only shows the statistical ones, but work is underway to have a better understanding of the actual accuracy of these measurements. However, it is surely remarkable that our tests are consistent with the Kerr hypothesis. Fairall~9 is the only source for which we do not recover the Kerr metric at a 90\% confidence level, but still the Kerr solution is recovered at a slightly higher confidence level (see Ref.~\cite{noi-bare} for more details). We would like to stress that our results are currently the only tests of the Kerr metric in the strong field regime with electromagnetic radiation.

\section{Systematic uncertainties from the model\label{s-sys}}

Systematic uncertainties due to approximations in our theoretical model can be grouped into three classes: $i)$ approximations in {\sc xillver}, $ii)$ simplifications in the disk model in {\sc relconv\_nk}, $iii)$ relativistic effects neglected in {\sc relconv\_nk}.

\subsection{Approximations in {\sc xillver}}

{\sc xillver} computes the reflection spectrum of a cold accretion disk illuminated by an X-ray source by solving some radiative transfer equations. The main simplifications in these calculations are:
\begin{enumerate}
\item The accretion disk is supposed to be cold and we ignore the thermal X-ray photons from the disk itself. This is presumably a reasonable approximation for supermassive black holes, in which the temperature of the inner edge of the disk is $1-100$~eV, but it is not for stellar-mass black holes in the soft state, where the disk temperature is around 1~keV.
\item The density of the disk is supposed to be constant in height and over radii.
\item Compton scattering is treated non-relativistically.
\item Elemental abundance in the disk is supposed to be the Solar one, with the exception of the iron one, which is allowed to vary. 
\item The incident radiation from the X-ray source is supposed to illuminate the disk at a constant angle. The correct illumination angle could only be derived assuming a specific coronal geometry.
\end{enumerate}

\subsection{Simplifications in the disk model in {\sc relconv\_nk}}

{\sc relconv\_nk} takes into account the structure of the accretion disk and the spacetime metric. The limitations of the accretion disk model are:
\begin{enumerate}
\item The disk is approximated as infinitesimally thin on the equatorial plane. In reality, the disk should have a finite thickness, which should increase as the mass accretion rate increases.
\item The reflection component emitted by particles in the plunging region, between the inner edge of the disk and the black hole event horizon, is completely neglected.
\item The ionization parameter of the disk is supposed to be constant over the whole disk, while it should increase as we move to smaller radii.
\end{enumerate}

\subsection{Relativistic effects neglected in {\sc relconv\_nk}}

Some relativistic effects are ignored in {\sc relconv\_nk}:
\begin{enumerate}
\item We ignore any radiation crossing the equatorial plane between the inner edge of the disk and the black hole event horizon. Such radiation can originate from $i)$ multiple-images of the accretion disk, and $ii)$ the reflection process on the other side of the disk. 
\item We ignore the effect of returning radiation (which would also affect the spectrum of the radiation illuminating the disk) as well as the direct radiation from the X-ray source on the other side of the disk (assuming the system is symmetric with respect to the equatorial plane).
\item The spectrum of the corona is a power-law with an exponential cut-off $E_{\rm cut}$. At every point of the disk, the spectrum of the incident radiation has a different $E_{\rm cut}$ because of the difference in the gravitational field. Within the lamppost model, it is possible to calculate the exact $E_{\rm cut}$ at every point of the disk and this is indeed done in {\sc relxilllp} (see next section). Without knowing the geometry of the corona, it is impossible to calculate the exact spectrum of the incident radiation at every point of the disk.
\end{enumerate}

\begin{table*}
\centering
%\vspace{0.2cm} 
\begin{tabular}{lcccc}
& \hspace{0.2cm} {\sc relxill\_nk} \hspace{0.2cm} & \hspace{0.2cm} {\sc relxillCp\_nk} \hspace{0.2cm} & \hspace{0.2cm} {\sc relxillD\_nk} \hspace{0.2cm} & \hspace{0.2cm} {\sc relxilllp\_nk} \hspace{0.2cm} \\
\hline
{\sc tbabs} &&&& \\
$N_{\rm H}/10^{22}$~cm$^{-2}$ & $0.0136^\star$ & $0.0136^\star$ & $0.0136^\star$ & $0.0136^\star$ \\
\hline
{\sc zpowerlaw}/{\sc nthcomp} &&&& \\
%{\sc nthcomp} &&&& \\
$\Gamma$ & $2.43_{-0.03}^{+0.03}$ & $2.403_{-0.022}^{+0.036}$ & $2.43_{-0.03}^{+0.03}$ & $2.439_{-0.012}^{+0.010}$ \\
Norm~$(10^{-3})$ & $0.5_{\rm (P)}^{+2.3}$ & $0.3_{\rm (P)}^{+1.3}$ & $0.5_{\rm (P)}^{+1.0}$ & $3.25_{-0.16}^{+0.67}$ \\
\hline
{\sc relxill} &&&& \\
$q_{\rm in}$ & $> 9.77$ & $> 9.76$ & $> 9.67$ & -- \\
$q_{\rm out}$ & $3^\star$ & $3^\star$ & $3^\star$ & -- \\
$R_{\rm br}$ [$M$] & $3.14_{-0.44}^{+0.19}$ & $3.14_{-0.46}^{+0.19}$ & $3.03_{-0.16}^{+0.25}$ & -- \\
$h$ [$M$] & -- & -- & -- & $< 2.5$ \\
$i$ [deg] & $37.1_{-3.2}^{+2.2}$ & $37.2_{-1.6}^{+2.6}$ & $37.8_{-3.6}^{+1.8}$ & $39.3_{-1.1}^{+1.0}$ \\
$a_*$ & $0.995_{-0.005}^{\rm (P)}$ & $0.995_{-0.005}^{\rm (P)}$ & $0.996_{-0.005}^{\rm (P)}$ & $0.996_{-0.007}^{\rm (P)}$ \\
$\alpha_{13}$ & $0.00_{-0.32}^{+0.02}$ & $-0.01_{-0.35}^{+0.07}$ & $-0.08_{-0.17}^{+0.15}$ & $0.00_{-0.02}^{+0.06}$ \\
$z$ & $0.062^\star$ & $0.062^\star$ & $0.062^\star$ & $0.062^\star$ \\
$\log\xi$ & $3.28_{-0.04}^{+0.03}$ & $3.12_{-0.05}^{+0.04}$ & $3.27_{-0.12}^{+0.03}$ & $2.994_{-0.052}^{+0.013}$ \\
$A_{\rm Fe}$  & $3.0_{-0.8}^{+1.0}$ & $2.8_{-0.7}^{+1.0}$ & $3.0_{-0.8}^{+1.4}$ & $4.6_{-0.4}^{+0.9}$ \\
$\log ( n_{\rm e}/10^{15}$~cm$^{-3})$ & $15^\star$ & $15^\star$ & $< 15.6$ & $15^\star$ \\
$E_{\rm cut}$ [keV] & $300^\star$ & -- & $300^\star$ & $300^\star$ \\
$kT_{\rm e}$ [keV] & -- & $60^\star$ & -- & -- \\
Norm~$(10^{-3})$ & $0.119_{-0.022}^{+0.026}$ & $0.114_{-0.016}^{+0.021}$ & $0.119_{-0.024}^{+0.029}$ & $2.76_{-0.09}^{+0.10}$ \\ 
\hline
$\chi^2$/dof & 1352.62/1313 & 1350.80/1313 & 1353.04/1312 & 1448.77/1314 \\
& =1.030 & =1.029 & =1.031 & =1.103 \\
\hline
\end{tabular}
%\vspace{0.3cm}
\caption{Summary of the best-fit values for the supermassive black hole in Ton~S180.}
\label{t-ton}
\end{table*}

\begin{figure*}[t]
\begin{center}
\includegraphics[width=8.5cm,trim={1.0cm 0 3cm 17.5cm},clip]{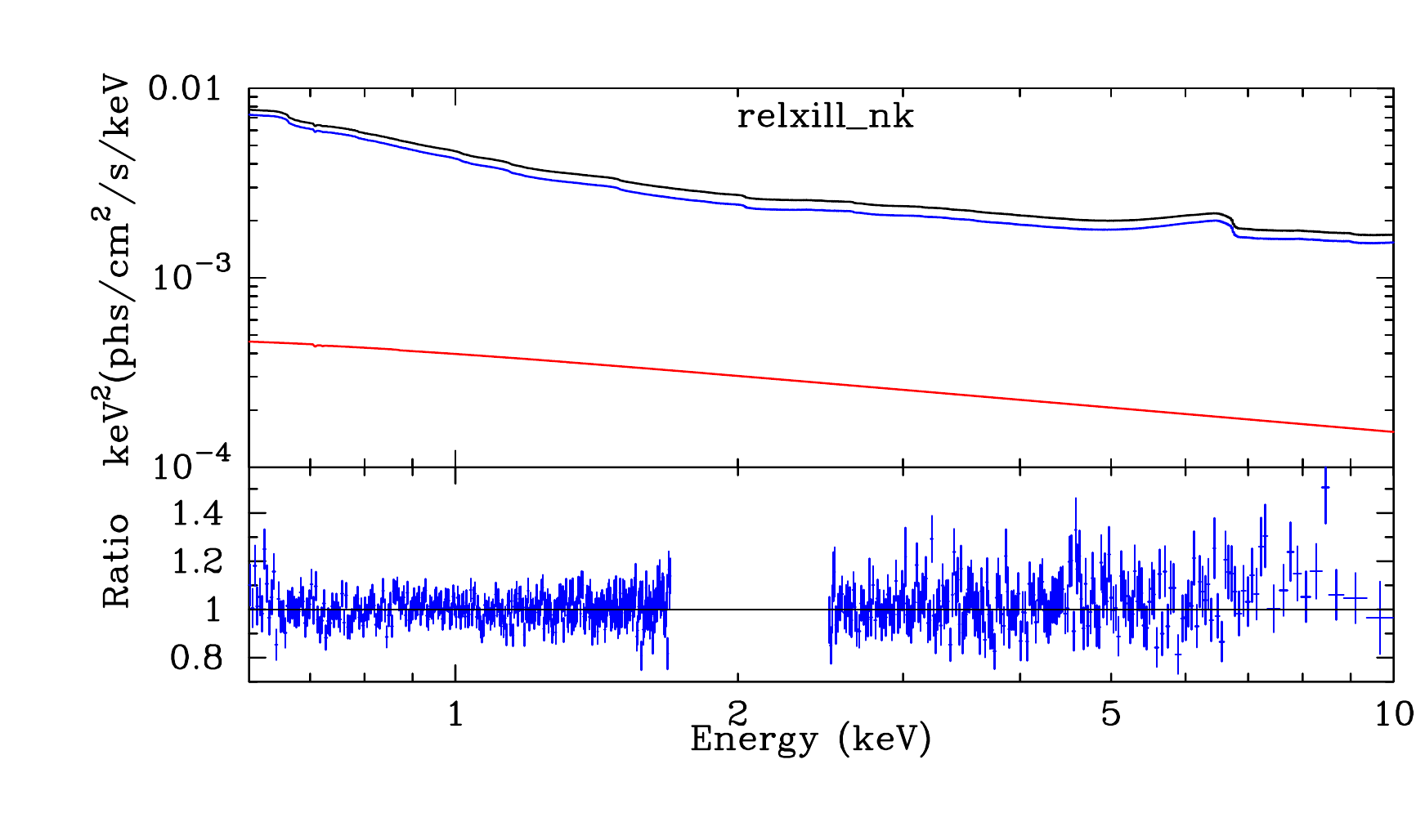}
\includegraphics[width=8.5cm,trim={1.0cm 0 3cm 17.5cm},clip]{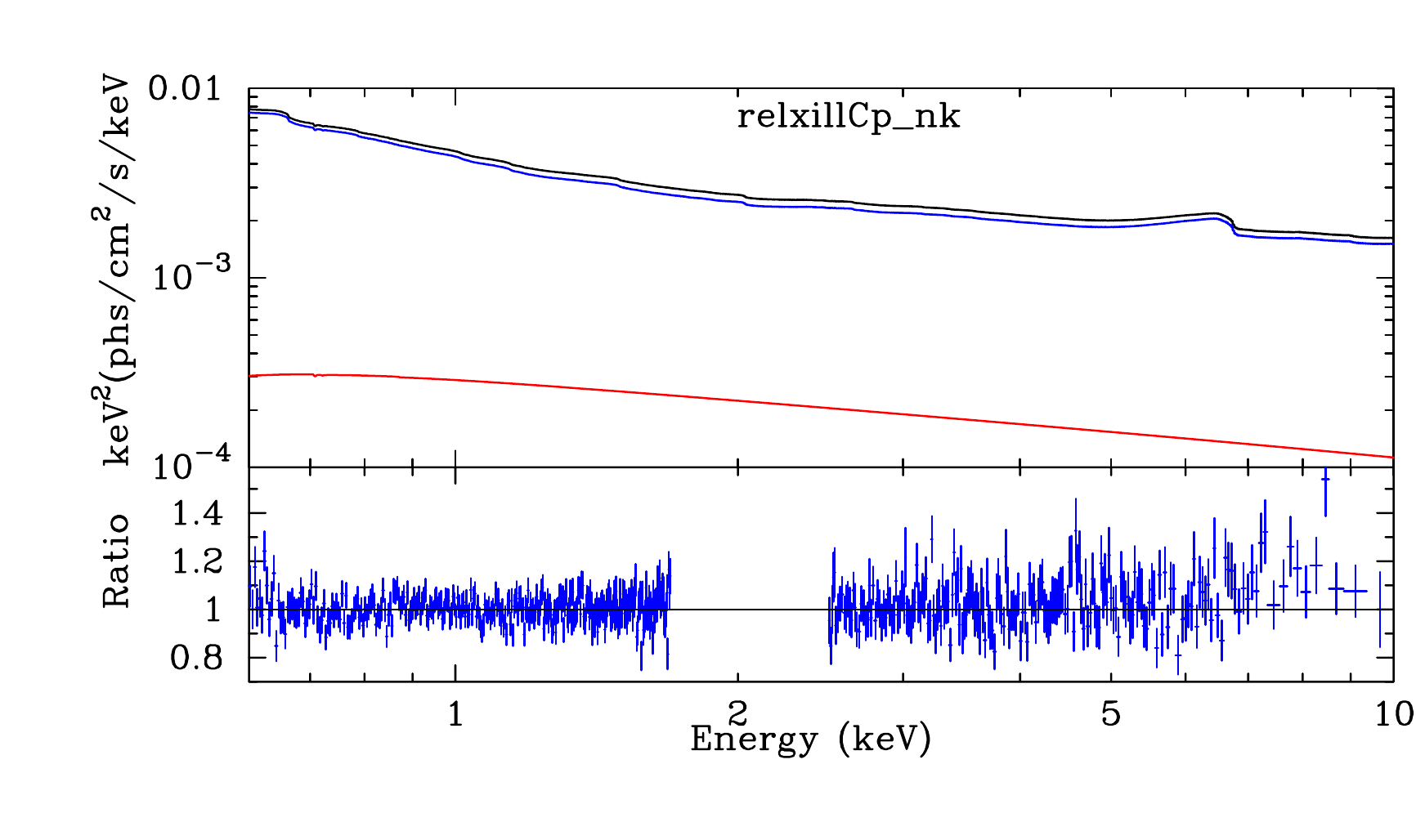} \\
\includegraphics[width=8.5cm,trim={1.0cm 0 3cm 17.5cm},clip]{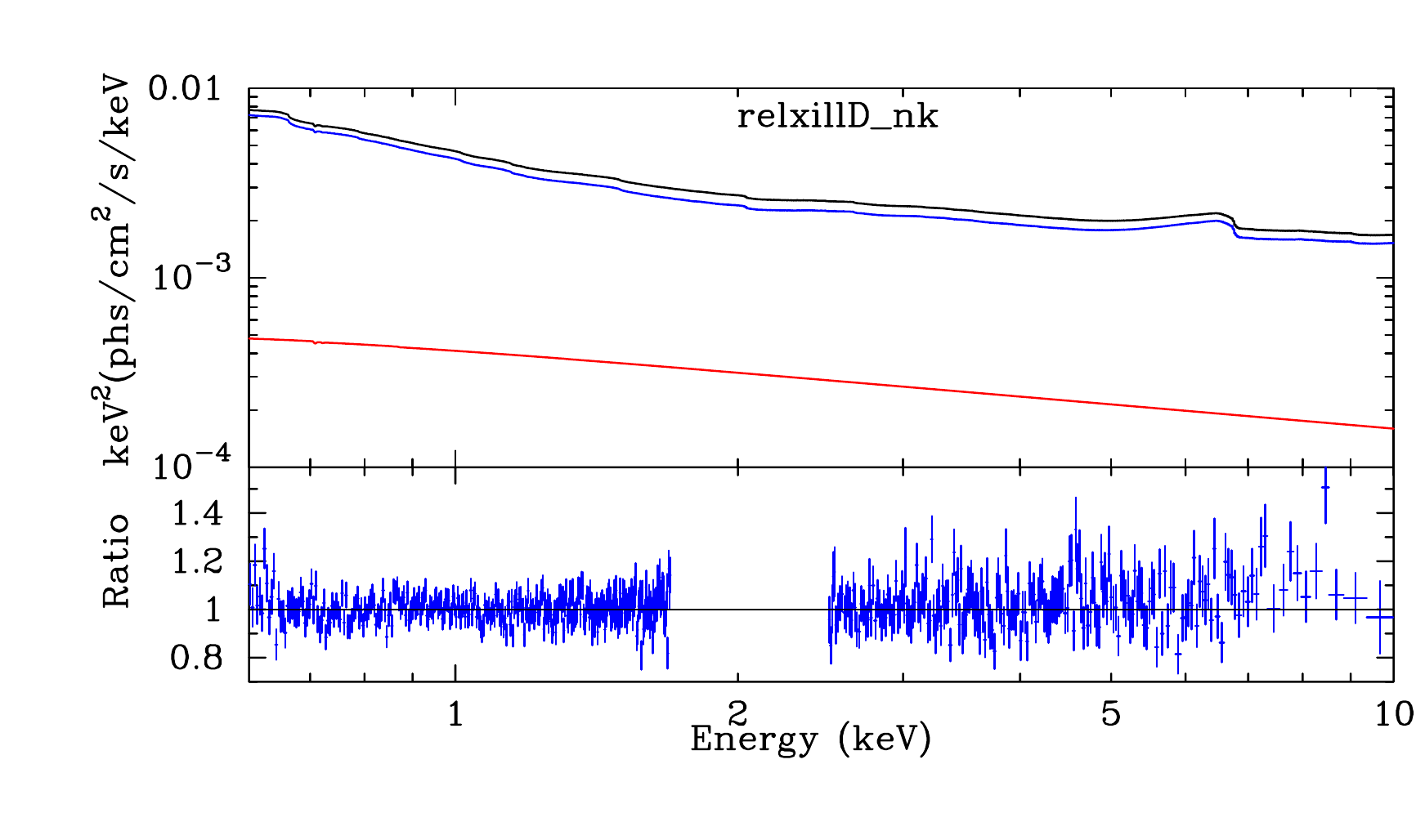}
\includegraphics[width=8.5cm,trim={1.0cm 0 3cm 17.5cm},clip]{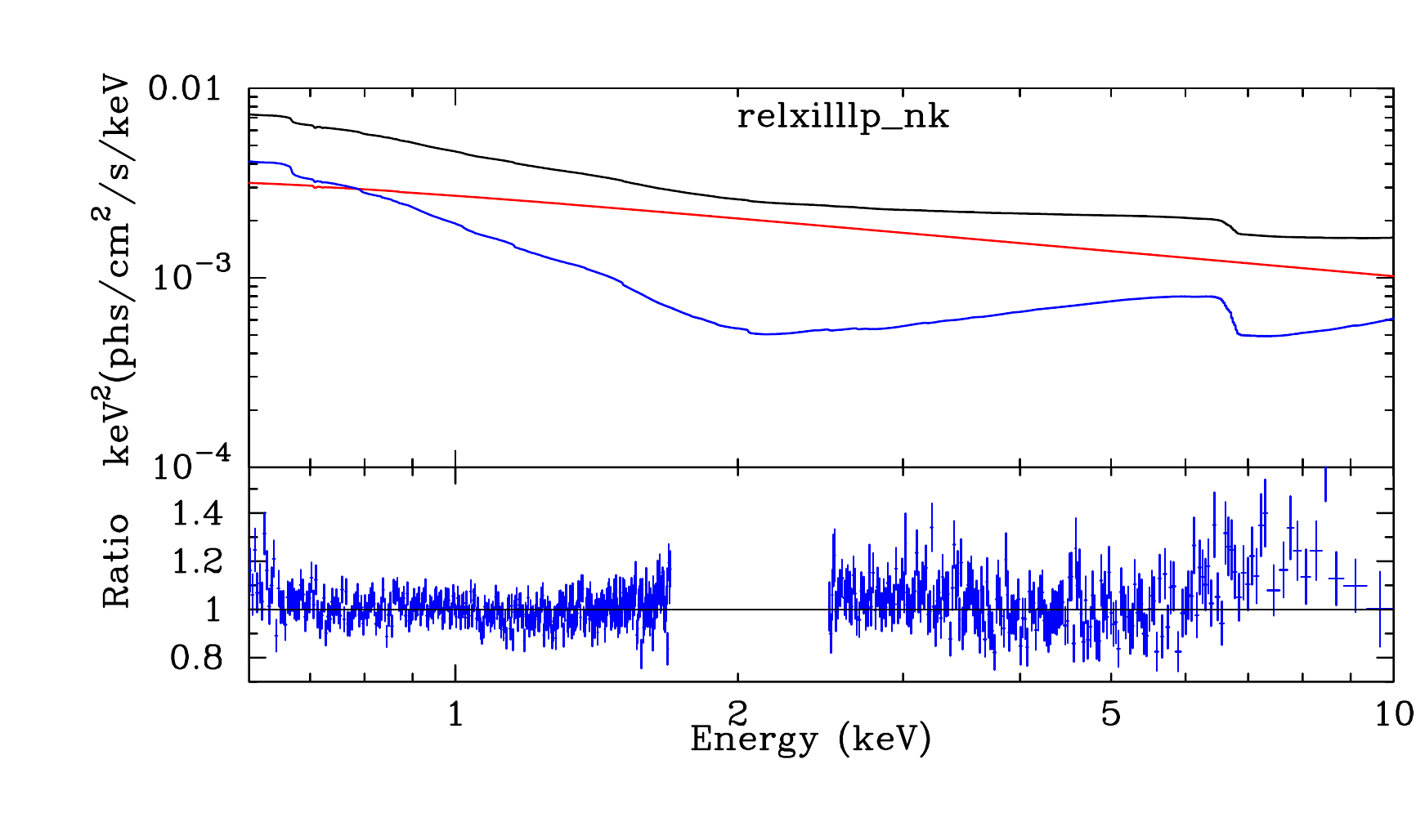}
\end{center}
\vspace{-0.7cm}
\caption{Spectra of the best fit models with the corresponding components (upper panels) and data to best-fit model ratios (lower panels) for Ton~S180 when the reflection component is modeled by {\sc relxill\_nk} (top left panel), {\sc relxillCp\_nk} (top right panel), {\sc relxillD\_nk} (bottom left panel), and {\sc relxilllp\_nk} (bottom right panel). The total spectra are in black, power law components from the coronas are in red, and the relativistic reflection components from the disk are in blue. \label{r-ton}}
\end{figure*}

\begin{figure*}[t]
\begin{center}
\includegraphics[width=8.5cm,trim={0cm 2cm 0cm 1cm},clip]{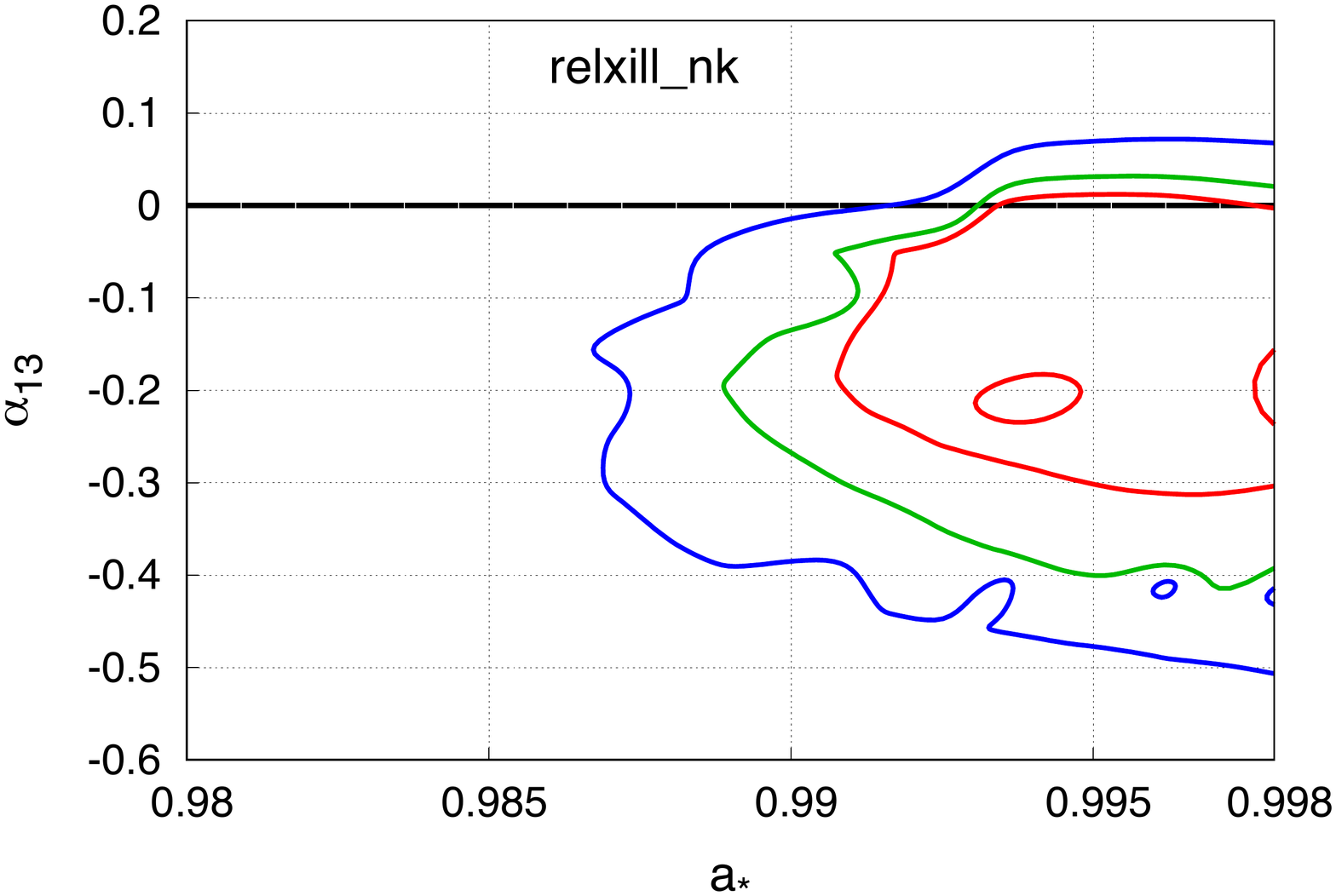}
\includegraphics[width=8.5cm,trim={0cm 2cm 0cm 1cm},clip]{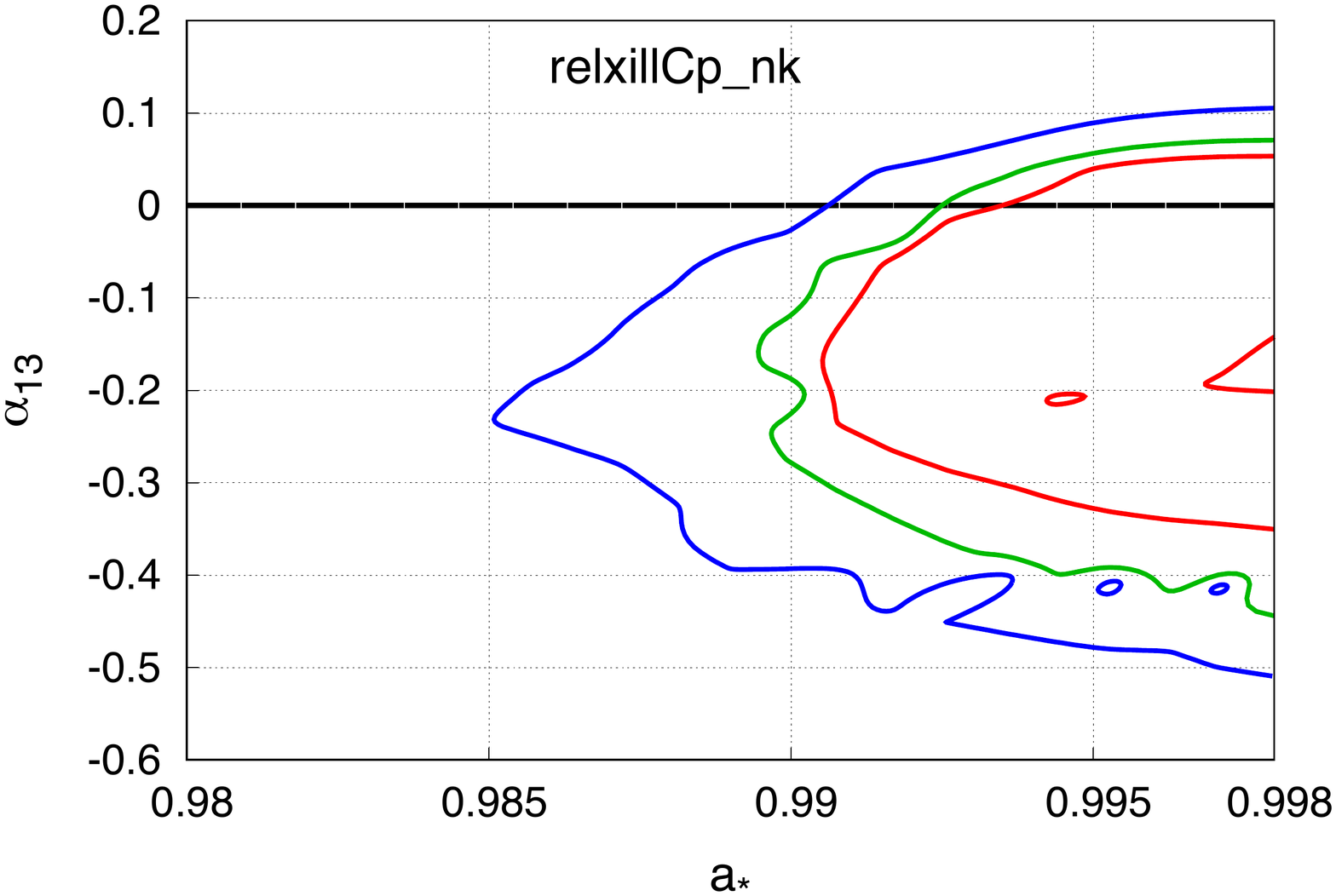} \\
\includegraphics[width=8.5cm,trim={0cm 2cm 0cm 1cm},clip]{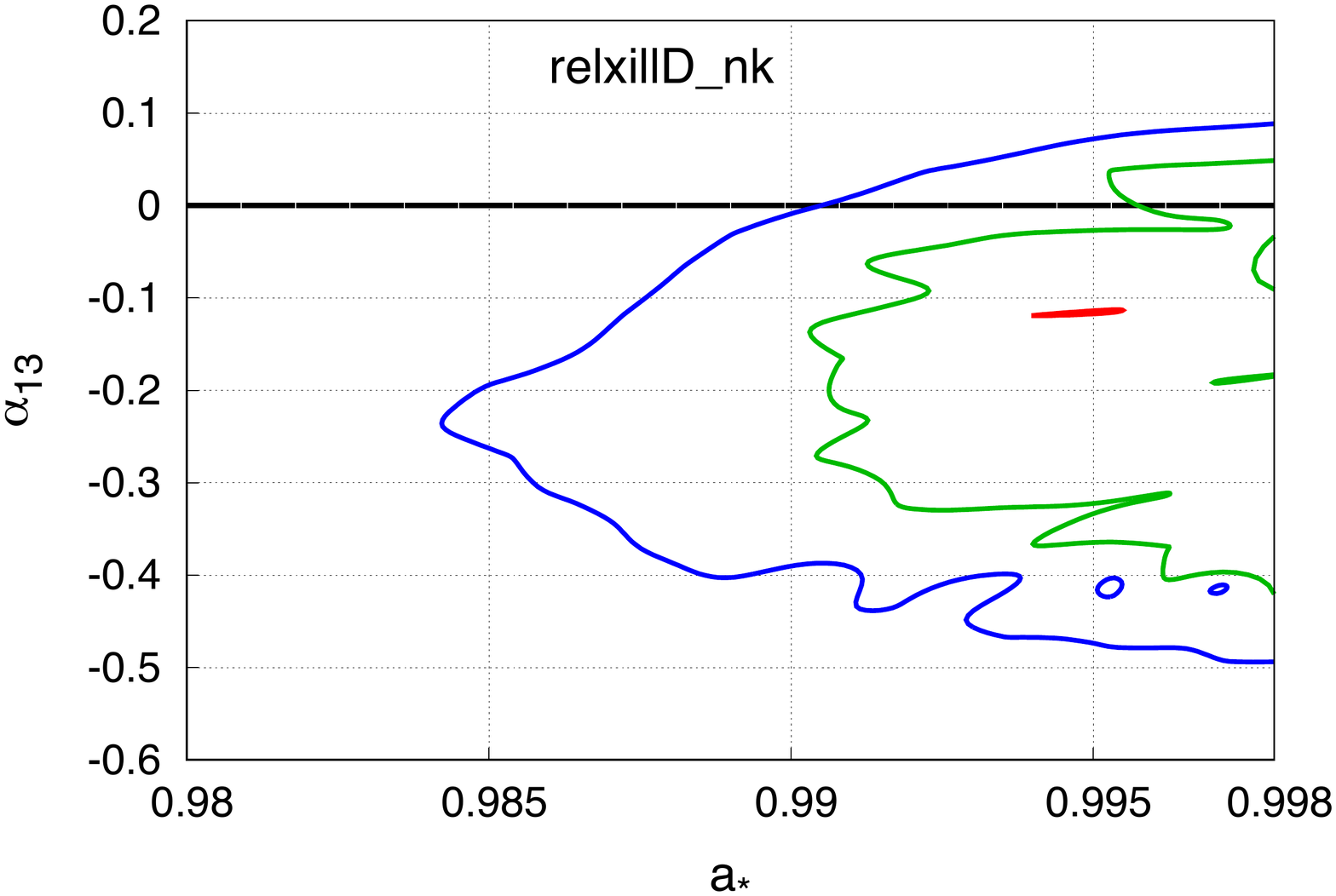}
\includegraphics[width=8.5cm,trim={0cm 2cm 0cm 1cm},clip]{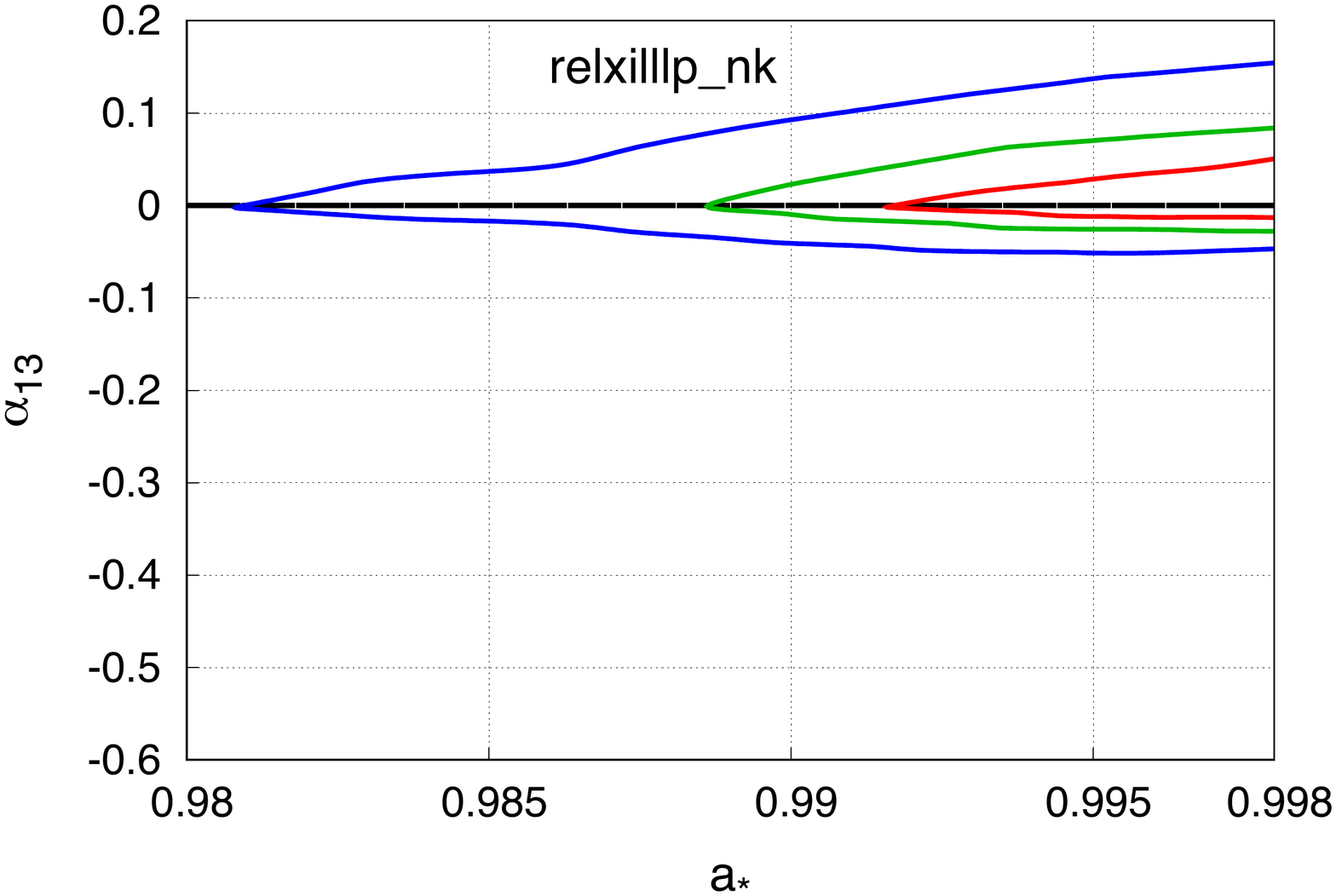}
\end{center}
\vspace{-0.7cm}
\caption{Constraints on the spin parameter $a_*$ and the Johannsen deformation parameters $\alpha_{13}$ for Ton~S180 when the reflection component is modeled by {\sc relxill\_nk} (top left panel), {\sc relxillCp\_nk} (top right panel), {\sc relxillD\_nk} (bottom left panel), and {\sc relxilllp\_nk} (bottom right panel). The red, green, and blue curves are, respectively, the 68\%, 90\%, and 99\% confidence level boundaries for two relevant parameters. \label{c-ton}}
\end{figure*}

\begin{table*}
\centering
%\vspace{0.2cm} 
\begin{tabular}{lcccc}
& \hspace{0.2cm} {\sc relxill\_nk} \hspace{0.2cm} & \hspace{0.2cm} {\sc relxillCp\_nk} \hspace{0.2cm} & \hspace{0.2cm} {\sc relxillD\_nk} \hspace{0.2cm} & \hspace{0.2cm} {\sc relxilllp\_nk} \hspace{0.2cm} \\
\hline
{\sc tbabs} &&&& \\
$N_{\rm H}/10^{22}$~cm$^{-2}$ & $0.145^\star$ & $0.145^\star$ & $0.145^\star$ & $0.145^\star$ \\
\hline
{\sc zpowerlaw}/{\sc nthcomp} &&&& \\
$\Gamma$ & $2.41_{-0.03}^{+0.06}$ & $2.315_{-0.017}^{+0.033}$ & $2.175_{-0.019}^{+0.032}$ & $2.279_{-0.004}^{+0.004}$ \\
Norm~$(10^{-3})$ & $4.1_{-0.7}^{+1.4}$ & $1.5_{\rm (P)}^{+2.0}$ & $9.1_{-0.9}^{+0.6}$ & $14.84_{-0.08}^{+0.06}$ \\
\hline
{\sc relxill} &&&& \\
$q_{\rm in}$ & $8.9_{-0.6}^{+0.6}$ & $8.5_{-0.7}^{+0.7}$ & $5.30_{-0.27}^{+0.05}$ & -- \\
$q_{\rm out}$ & $3^\star$ & $3^\star$ & $3^\star$ & -- \\
$R_{\rm br}$ [$M$] & $4.6_{-0.6}^{+0.5}$ & $4.4_{-0.4}^{+0.3}$ & $4.96_{-0.60}^{+0.11}$ & -- \\
$h$ [$M$] & -- & -- & -- & $24.7_{-1.3}^{+1.2}$ \\
$i$ [deg] & $21_{-3}^{+8}$ & $21_{-6}^{+3}$ & $16_{-3}^{+3}$ & $88.65_{-0.22}^{\rm (P)}$ \\
$a_*$ & $> 0.996$ & $> 0.996$ & $0.9897_{-0.0092}^{+0.0013}$ & $> 0.990$ \\
$\alpha_{13}$ & $0.01_{-0.08}^{+0.01}$ & $0.00_{-0.09}^{+0.04}$ & $-0.8_{-0.3}^{+0.9}$ & $0.21_{-0.04}^{+0.09}$ \\
$z$ & $0.0327^\star$ & $0.0327^\star$ & $0.0327^\star$ & $0.0327^\star$ \\
$\log\xi$ & $2.996_{-0.118}^{+0.025}$ & $2.998_{-0.100}^{+0.020}$ & $2.717_{-0.093}^{+0.023}$ & $1.00_{-0.04}^{+0.03}$ \\
$A_{\rm Fe}$  & $1.5_{-0.8}^{+0.5}$ & $1.7_{-0.5}^{+0.6}$ & $3.94_{-0.20}^{+0.45}$ & $1.10_{-0.13}^{+0.16}$ \\
$\log ( n_{\rm e}/10^{15}$~cm$^{-3})$ & $15^\star$ & $15^\star$ & $17.79_{-0.06}^{+0.13}$ & $15^\star$ \\
$E_{\rm cut}$ [keV] & $300^\star$ & -- & $300^\star$ & $300^\star$ \\
$kT_{\rm e}$ [keV] & -- & $60^\star$ & -- & -- \\
Norm~$(10^{-3})$ & $0.83_{-0.09}^{+0.08}$ & $0.64_{-0.09}^{+0.05}$ & $0.204_{-0.006}^{+0.021}$ & $2.91_{-0.08}^{+0.11}$ \\ 
\hline
{\sc xillver} &&&& \\
$\log\xi$ & $0^\star$ & $0^\star$ & $0^\star$ & $0^\star$ \\
Norm~$(10^{-3})$ & $0.17_{-0.04}^{+0.05}$ & $0.113_{-0.023}^{+0.024}$ & $0.053_{-0.011}^{+0.011}$ & $2.51_{-0.21}^{+2.03}$ \\ 
\hline
{\sc zgauss} &&&& \\
$E_{\rm line}$ & $6.95_{-0.03}^{+0.03}$ & $6.95_{-0.03}^{+0.03}$ & $6.948_{-0.025}^{+0.025}$ & $6.95_{-0.04}^{+0.03}$ \\
\hline
{\sc zgauss} &&&& \\
$E_{\rm line}$ & $6.087_{-0.014}^{+0.012}$ & $6.087_{-0.014}^{+0.011}$ & $6.088_{-0.014}^{+0.009}$ & $6.085_{-0.012}^{+0.012}$ \\
\hline
$\chi^2$/dof & 1408.88/1308 & 1402.54/1308 & 1393.93/1307 & 1477.62/1309 \\
& =1.077 & =1.072 & =1.067 & =1.129 \\
\hline
\end{tabular}
%\vspace{0.3cm}
\caption{Summary of the best-fit values for the supermassive black hole in Ark~120.}
\label{t-ark}
\end{table*}

\begin{figure*}[t]
\begin{center}
\includegraphics[width=8.5cm,trim={1.0cm 0 3cm 17.5cm},clip]{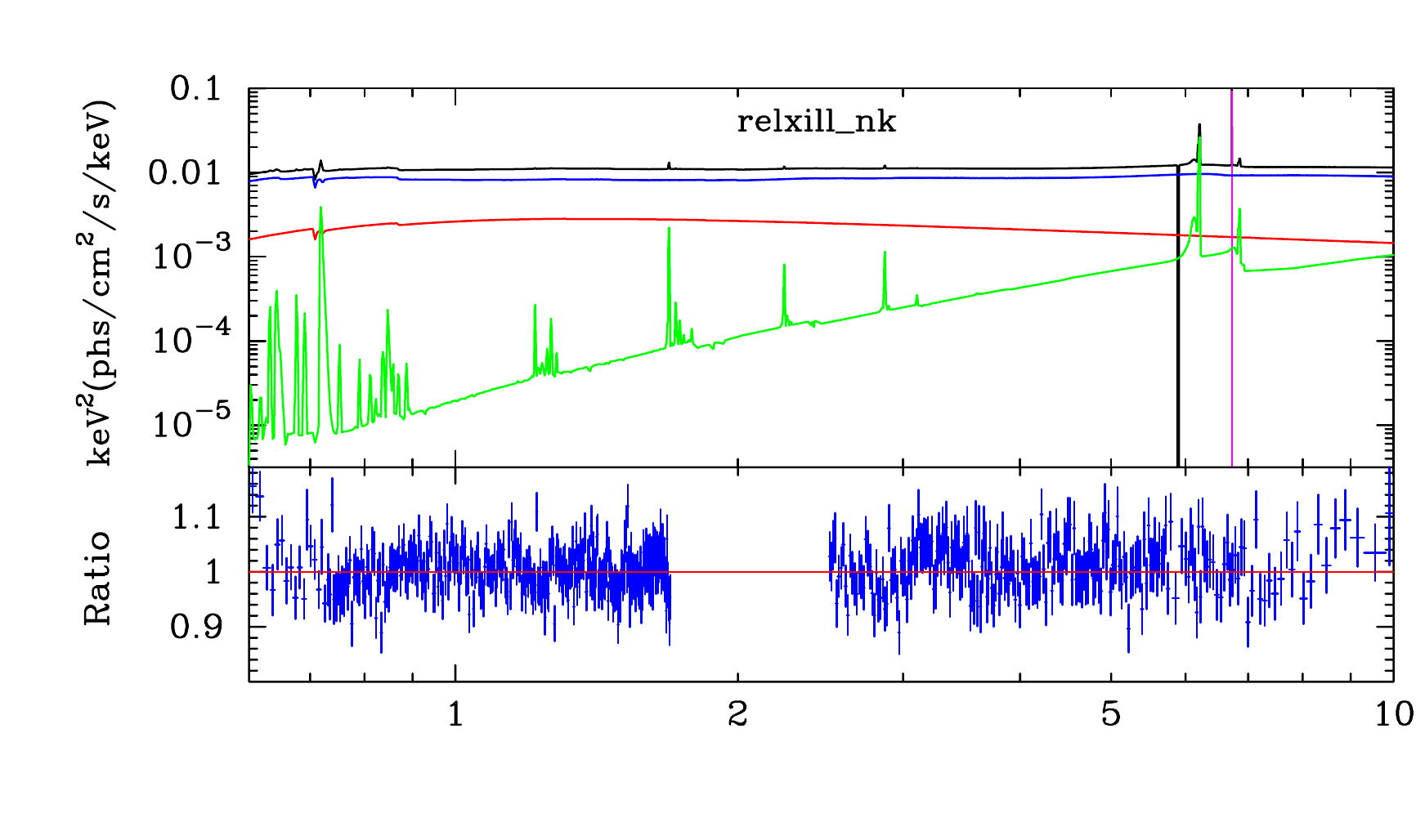}
\includegraphics[width=8.5cm,trim={1.0cm 0 3cm 17.5cm},clip]{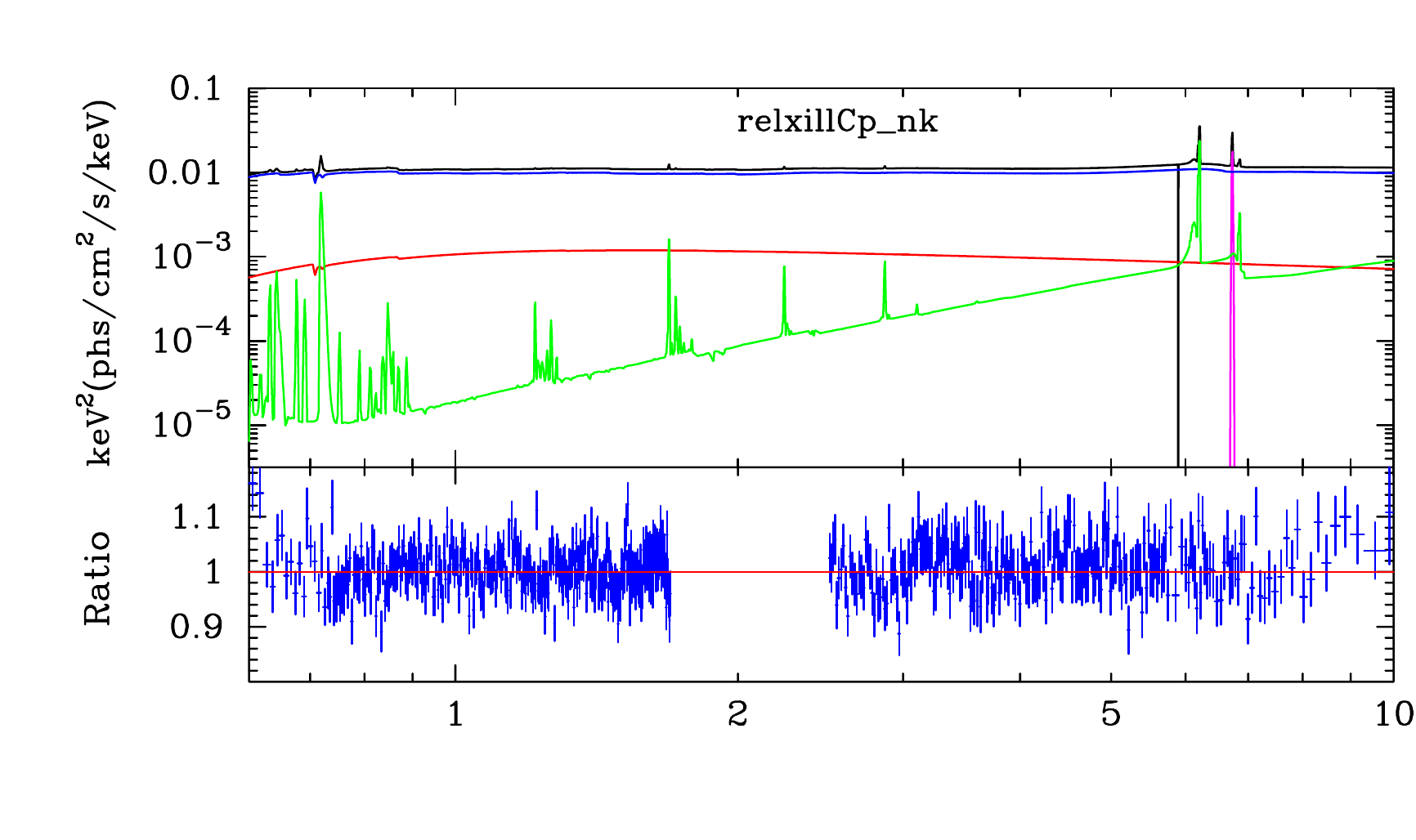} \\
\includegraphics[width=8.5cm,trim={1.0cm 0 3cm 17.5cm},clip]{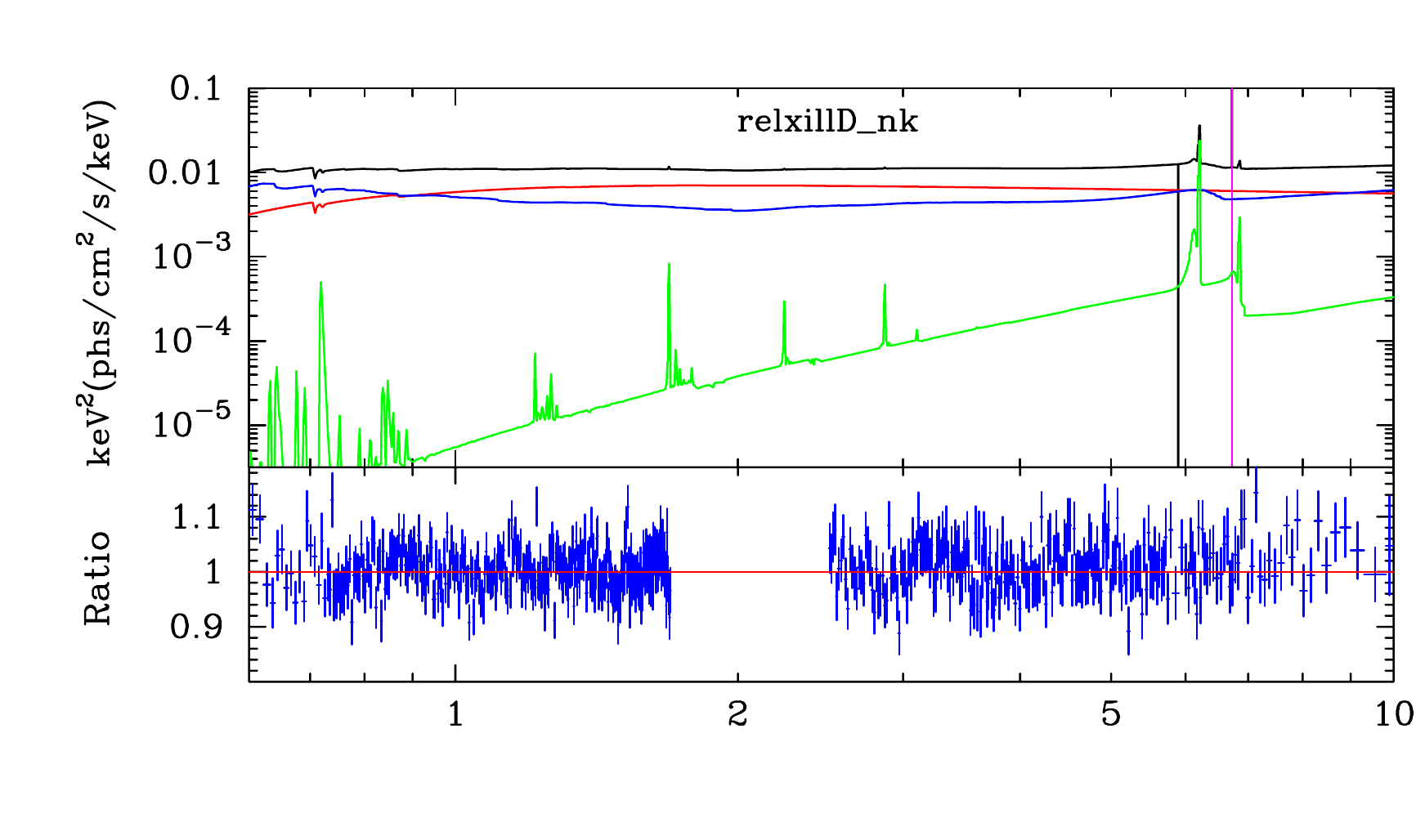}
\includegraphics[width=8.5cm,trim={1.0cm 0 3cm 17.5cm},clip]{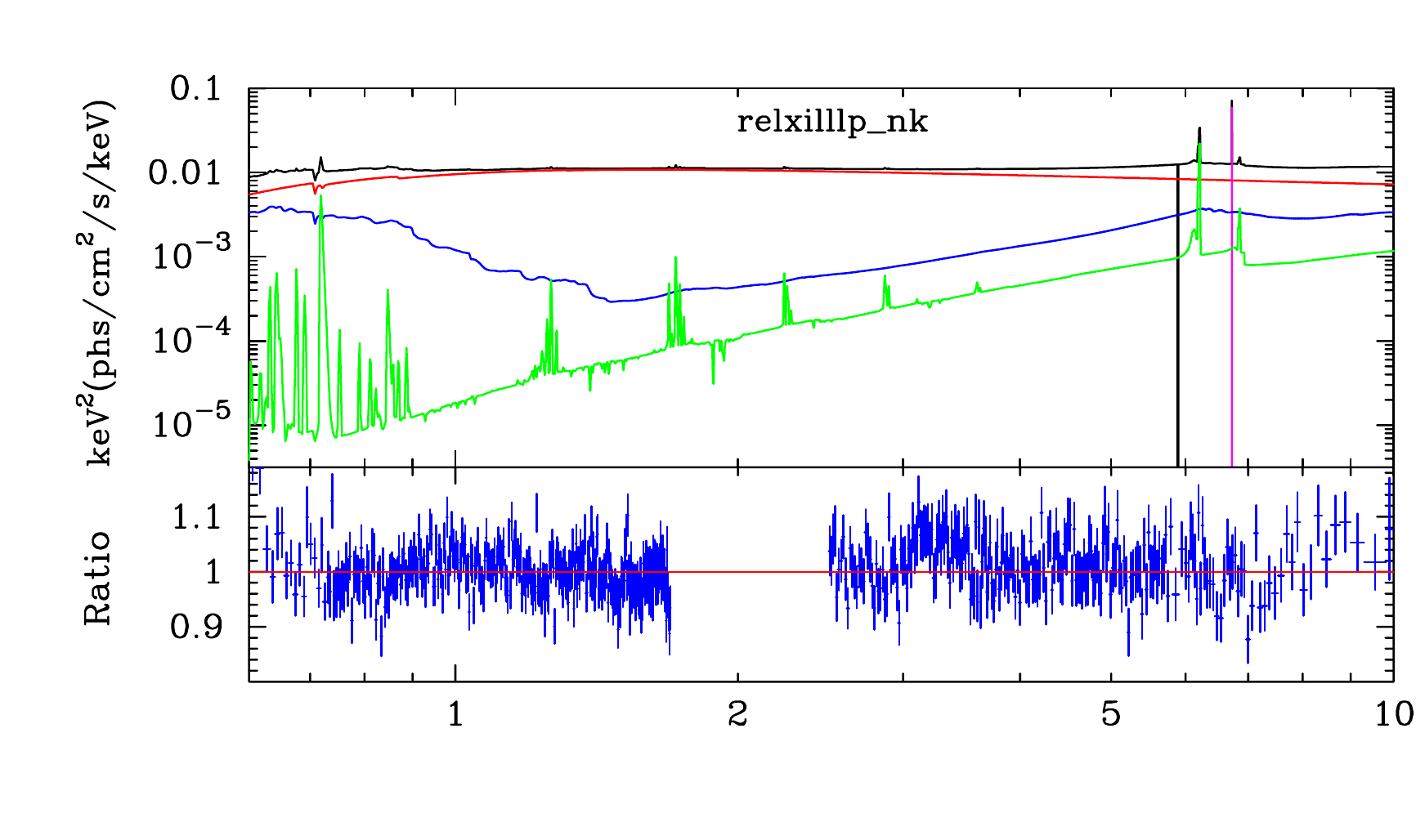}
\end{center}
\vspace{-0.7cm}
\caption{Spectra of the best fit models with the corresponding components (upper panels) and data to best-fit model ratios (lower panels) for Ark~120 when the reflection component is modeled by {\sc relxill\_nk} (top left panel), {\sc relxillCp\_nk} (top right panel), {\sc relxillD\_nk} (bottom left panel), and {\sc relxilllp\_nk} (bottom right panel). The total spectra are in black, power law components from the coronas are in red, the relativistic reflection components from the disk are in blue, the non-relativistic reflection components from cold material are in green, and the gaussian lines are in magenta. \label{r-ark}}
\end{figure*}

\begin{figure*}[t]
\begin{center}
\includegraphics[width=8.5cm,trim={0cm 2cm 0cm 1cm},clip]{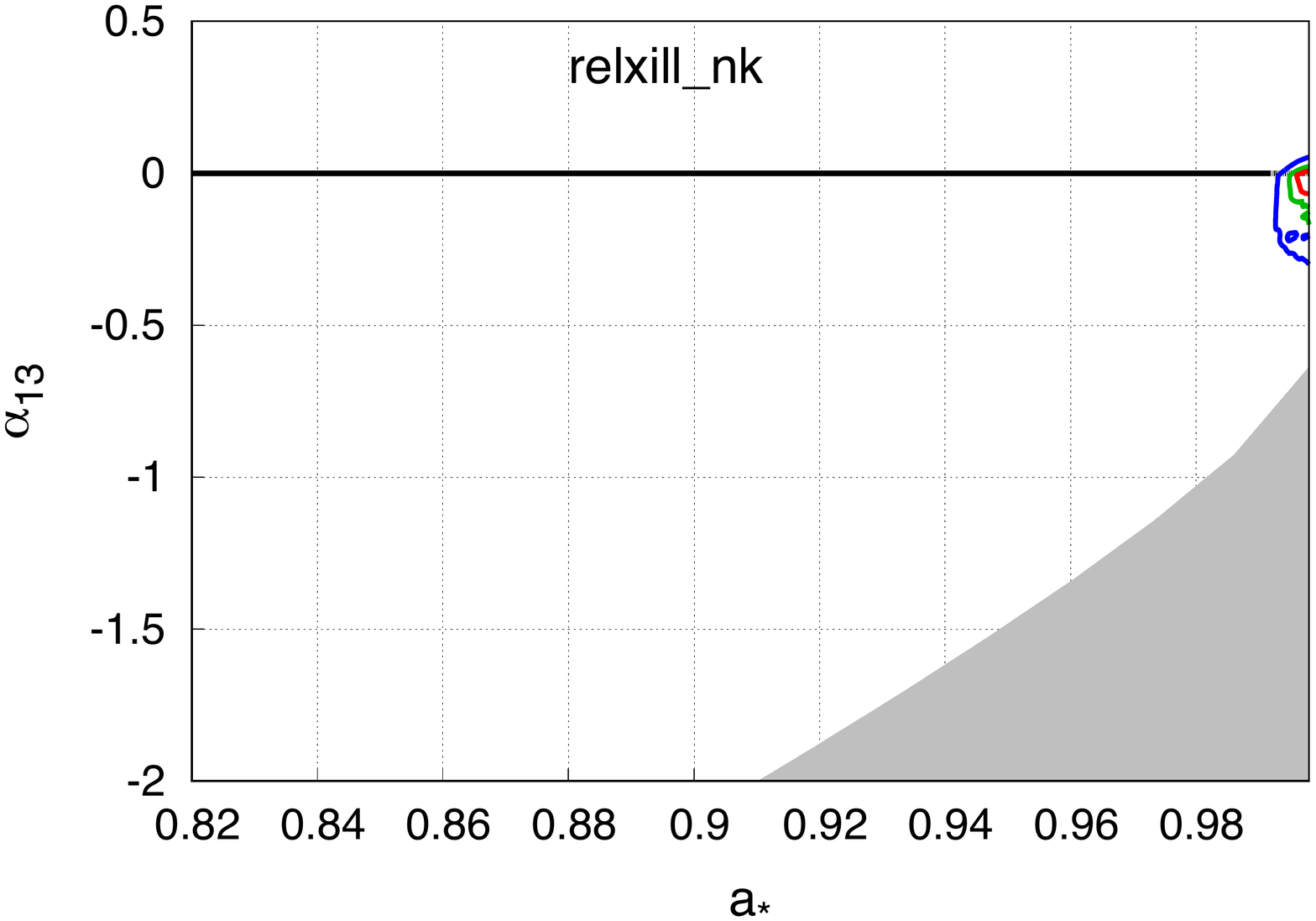}
\includegraphics[width=8.5cm,trim={0cm 2cm 0cm 1cm},clip]{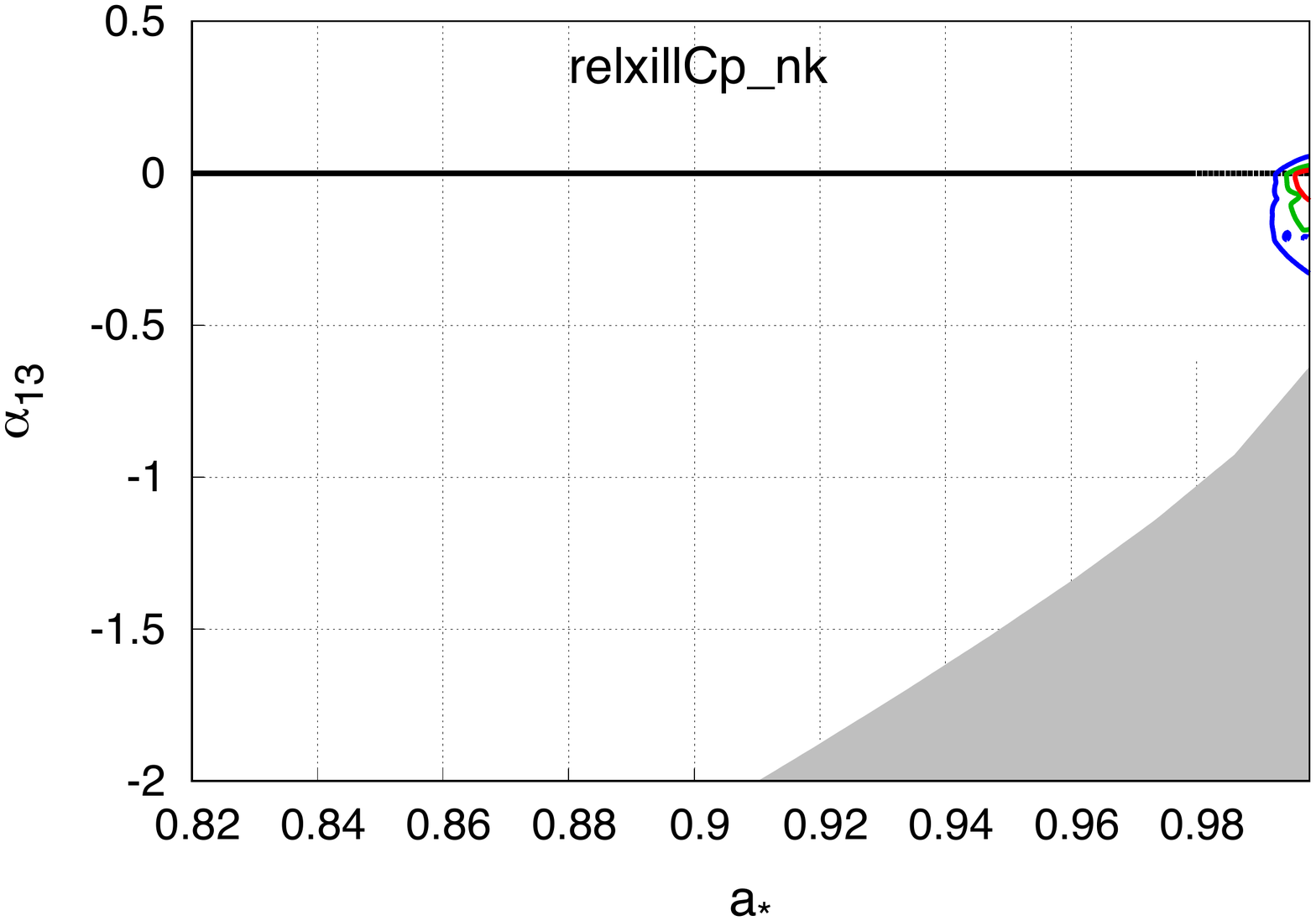} \\
\includegraphics[width=8.5cm,trim={0cm 2cm 0cm 1cm},clip]{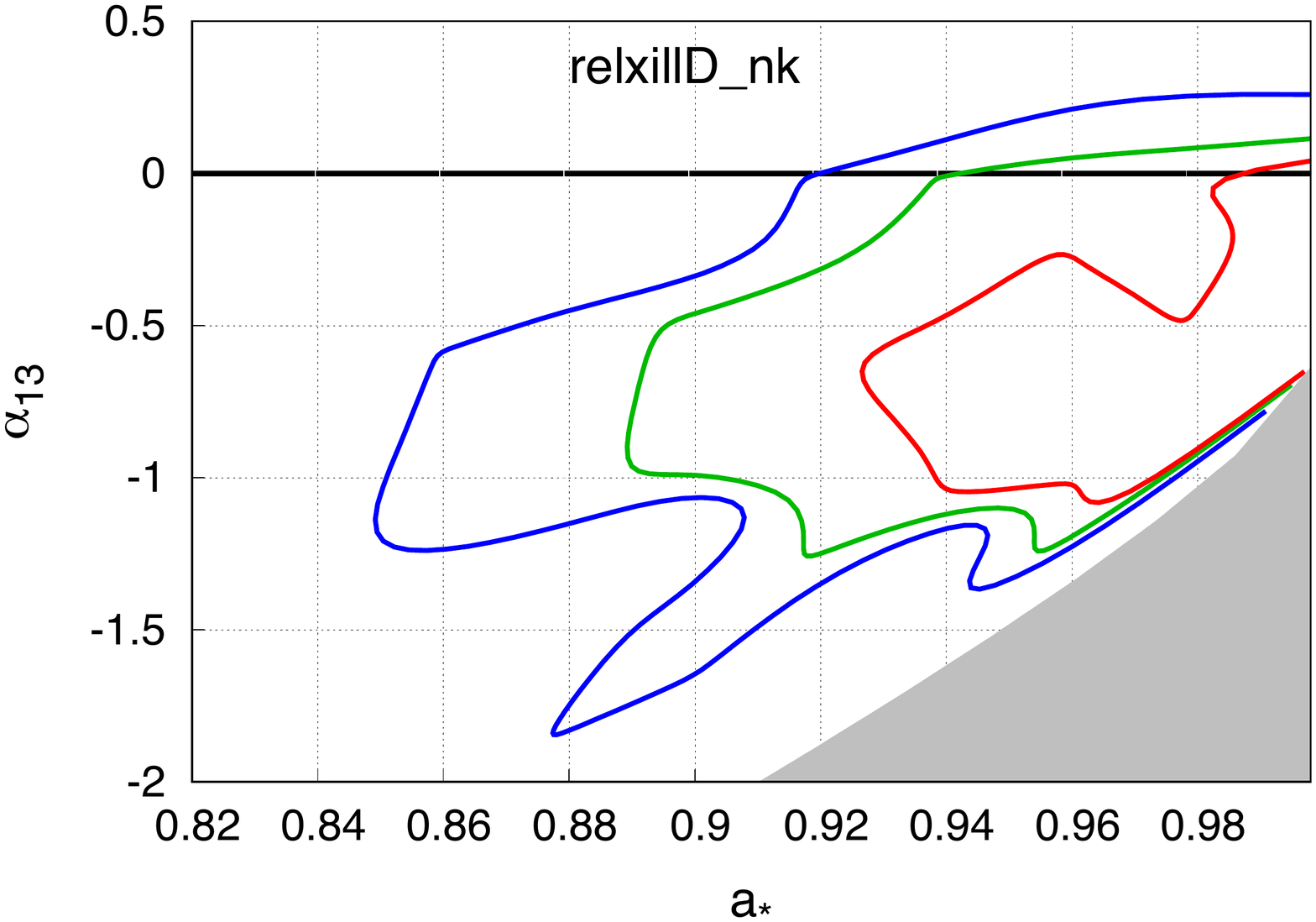}
\includegraphics[width=8.5cm,trim={0cm 2cm 0cm 1cm},clip]{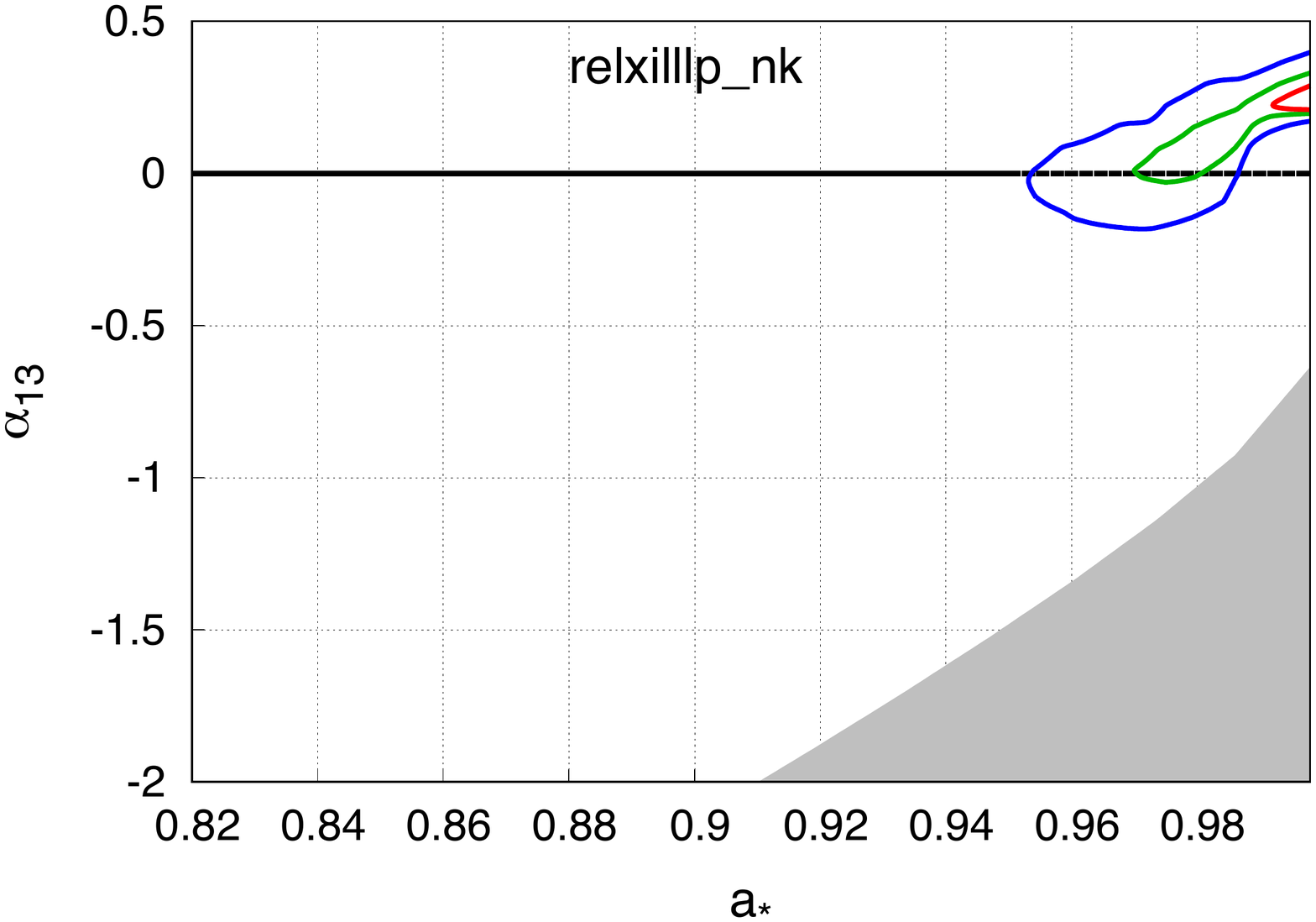}
\end{center}
\vspace{-0.7cm}
\caption{Constraints on the spin parameter $a_*$ and the Johannsen deformation parameters $\alpha_{13}$ for Ark~120 when the reflection component is modeled by {\sc relxill\_nk} (top left panel), {\sc relxillCp\_nk} (top right panel), {\sc relxillD\_nk} (bottom left panel), and {\sc relxilllp\_nk} (bottom right panel). The red, green, and blue curves are, respectively, the 68\%, 90\%, and 99\% confidence level boundaries for two relevant parameters. The grayed regions are ignored in our analysis because the do not meet the conditions in~(\ref{eq-c}). \label{c-ark}}
\end{figure*}

\begin{table*}
\centering
%\vspace{0.2cm} 
\begin{tabular}{lcccc}
& \hspace{0.2cm} {\sc relxill\_nk} \hspace{0.2cm} & \hspace{0.2cm} {\sc relxillCp\_nk} \hspace{0.2cm} & \hspace{0.2cm} {\sc relxillD\_nk} \hspace{0.2cm} & \hspace{0.2cm} {\sc relxilllp\_nk} \hspace{0.2cm} \\
\hline
{\sc tbabs} &&&& \\
$N_{\rm H}/10^{22}$~cm$^{-2}$ & $0.0134^\star$ & $0.0134^\star$ & $0.0134^\star$ & $0.0134^\star$ \\
\hline
{\sc zpowerlaw}/{\sc nthcomp} &&&& \\
$\Gamma$ & $2.16_{-0.04}^{+0.03}$ & $2.16_{-0.03}^{+0.03}$ & $2.15_{-0.03}^{+0.05}$ & $2.13_{-0.04}^{+0.05}$ \\
Norm~$(10^{-3})$ & $5.994_{-0.016}^{+0.058}$ & $4.83_{-0.05}^{+0.03}$ & $5.98_{-0.07}^{+0.07}$ & $5.96_{-0.08}^{+0.09}$ \\
\hline
{\sc relxill} &&&& \\
$q_{\rm in}$ & $7.4_{-1.0}^{+0.8}$ & $7.4_{-1.7}^{+1.1}$ & $4.9_{-0.8}^{+3.0}$ & -- \\
$q_{\rm out}$ & $= q_{\rm in}$ & $= q_{\rm in}$ & $= q_{\rm in}$ & -- \\
$R_{\rm br}$ [$M$] & -- & -- & -- & -- \\
$h$ [$M$] & -- & -- & -- & $< 2.6$ \\
$i$ [deg] & $71_{-5}^{+3}$ & $71_{-4}^{+3}$ & $65_{-3}^{+8}$ & $65.6_{-0.8}^{+3.9}$ \\
$a_*$ & $> 0.995$ & $> 0.995$ & $> 0.986$ & $> 0.990$ \\
$\alpha_{13}$ & $0.00_{-0.17}^{+0.04}$ & $0.01_{-0.15}^{+0.05}$ & $-0.26_{-0.11}^{+0.31}$ & $-0.27_{-0.03}^{+0.01}$ \\
$z$ & $0.104^\star$ & $0.104^\star$ & $0.104^\star$ & $0.104^\star$ \\
$\log\xi$ & $0.67_{-0.22}^{+0.19}$ & $0.70_{-0.23}^{+0.11}$ & $0.2_{\rm (P)}^{+0.9}$ & $0.81_{-0.23}^{+0.25}$ \\
$A_{\rm Fe}$  & $2.1_{-0.6}^{+0.5}$ & $2.0_{-0.5}^{+0.5}$ & $2.5_{-1.5}^{+1.0}$ & $1.8_{-0.5}^{+0.4}$ \\
$\log ( n_{\rm e}/10^{15}$~cm$^{-3})$ & $15^\star$ & $15^\star$ & $> 17.6$ & $15^\star$ \\
$E_{\rm cut}$ [keV] & $300^\star$ & -- & $300^\star$ & $300^\star$ \\
$kT_{\rm e}$ [keV] & -- & $60^\star$ & -- & -- \\
Norm~$(10^{-3})$ & $0.081_{-0.022}^{+0.012}$ & $0.082_{-0.010}^{+0.005}$ & $0.042_{-0.014}^{+0.019}$ & $1.91_{-0.07}^{+0.78}$ \\ 
\hline
{\sc xillver} &&&& \\
$\log\xi$ & $0^\star$ & $0^\star$ & $0^\star$ & $0^\star$ \\
Norm~$(10^{-3})$ & $0.017_{-0.006}^{+0.008}$ & $0.018_{-0.007}^{+0.007}$ & $0.017_{-0.006}^{+0.013}$ & $0.019_{-0.007}^{+0.008}$ \\ 
\hline
$\chi^2$/dof & 2489.24/2344 & 2490.78/2344 & 2488.98/2343 & 2495.58/2344 \\
& =1.062 & =1.063 & =1.062 & =1.065 \\
\hline
\end{tabular}
%\vspace{0.3cm}
\caption{Summary of the best-fit values for the supermassive black hole in 1H0419--577.}
\label{t-1h}
\end{table*}

\begin{figure*}[t]
\begin{center}
\includegraphics[width=8.5cm,trim={1.0cm 0 3cm 17.5cm},clip]{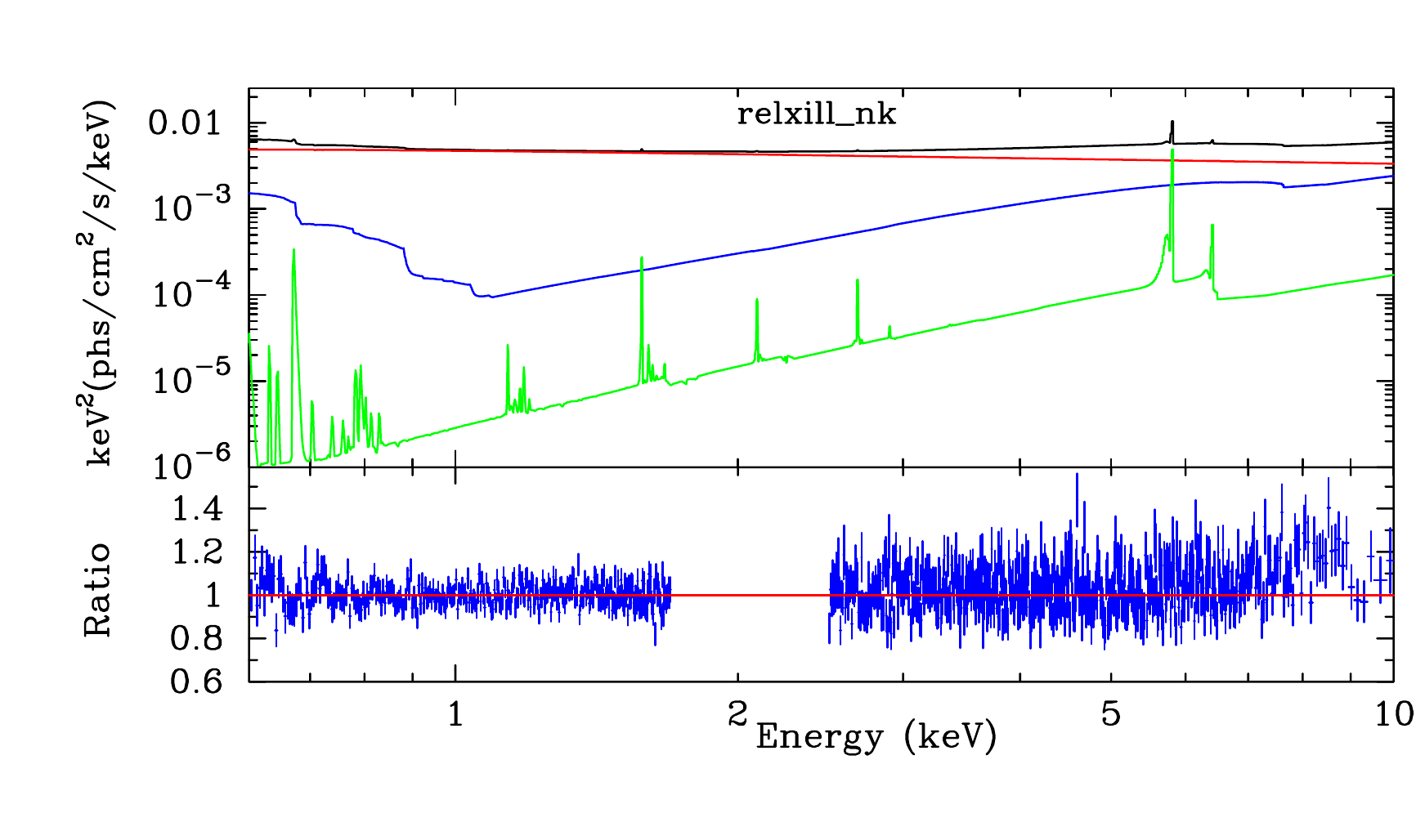}
\includegraphics[width=8.5cm,trim={1.0cm 0 3cm 17.5cm},clip]{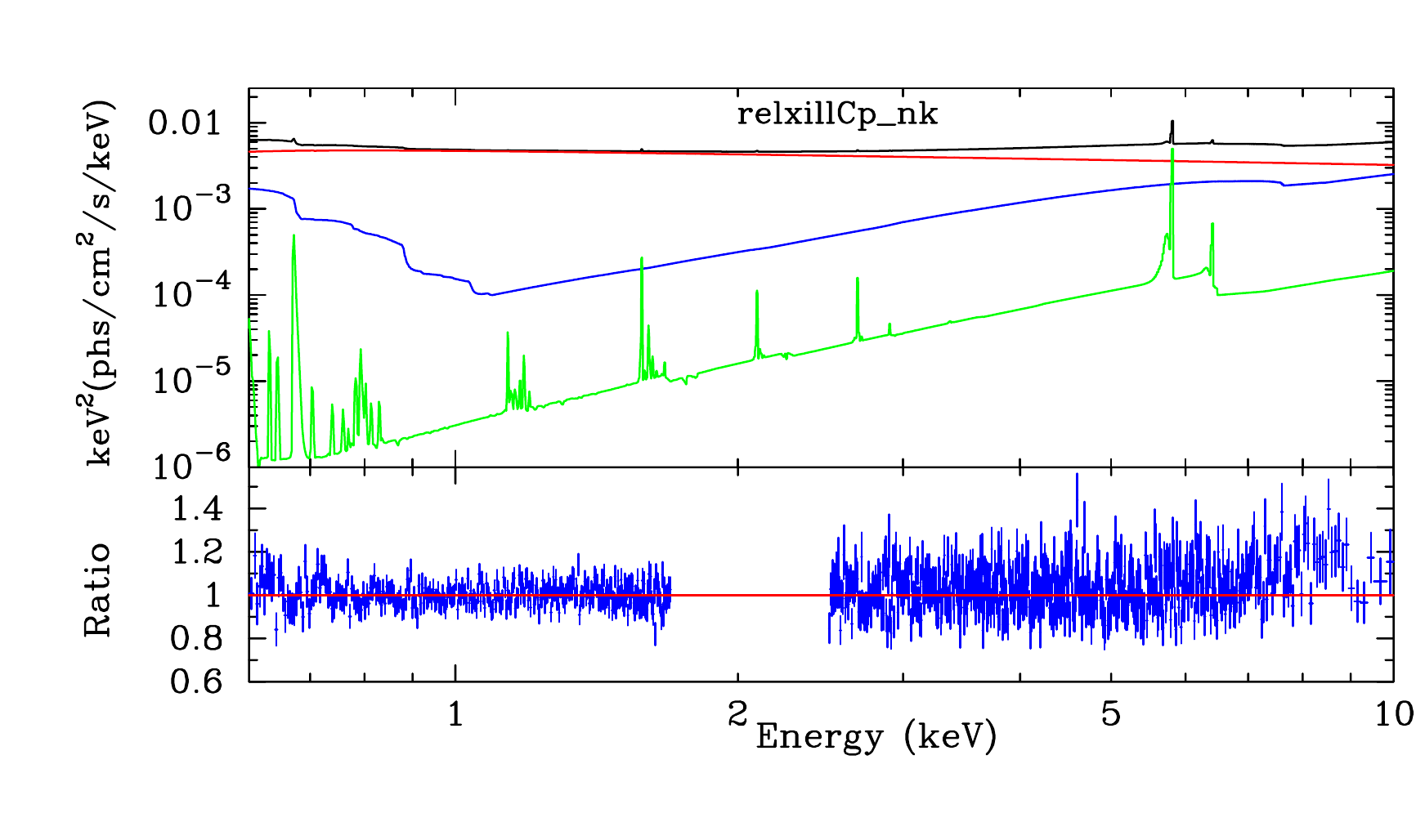} \\
\includegraphics[width=8.5cm,trim={1.0cm 0 3cm 17.5cm},clip]{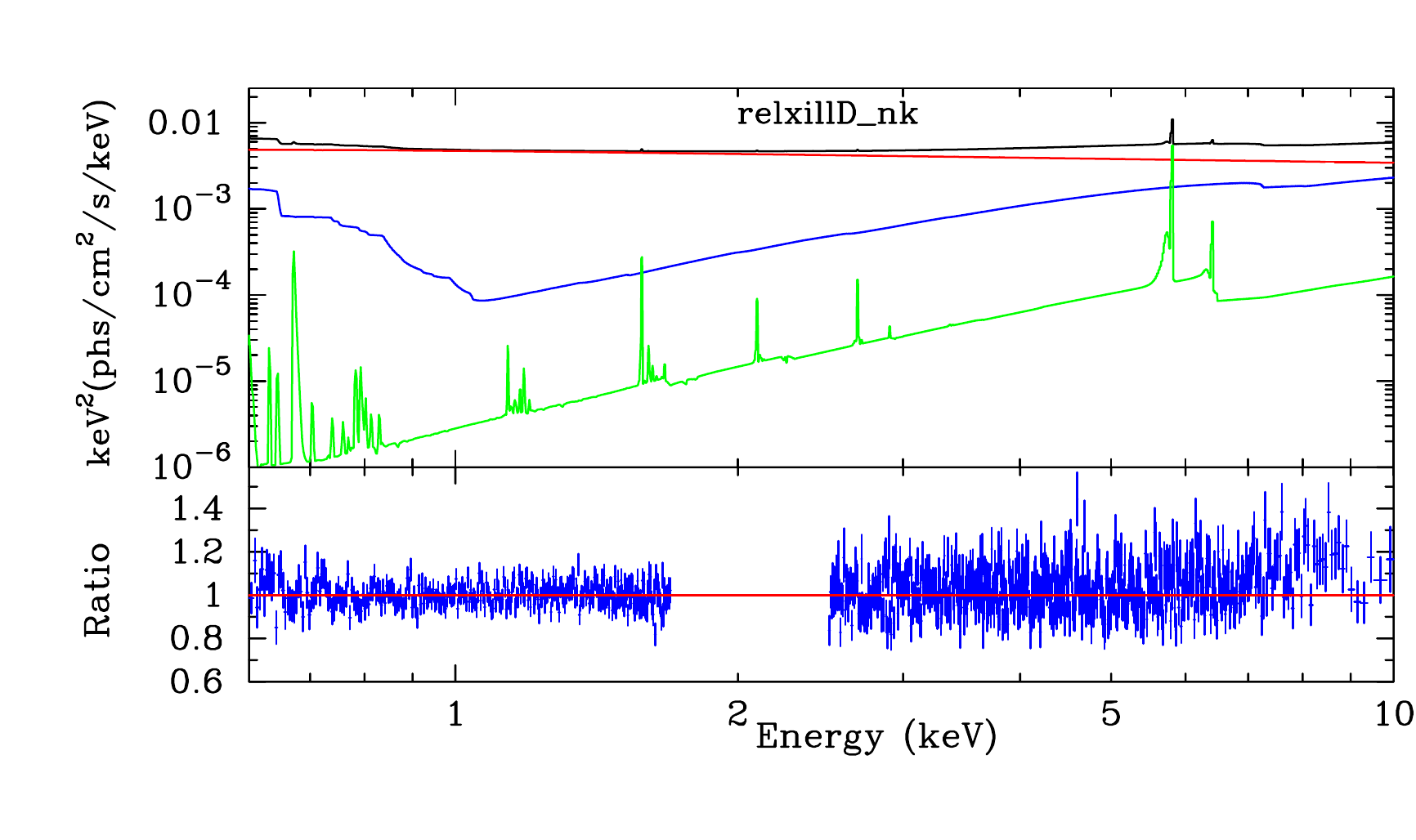}
\includegraphics[width=8.5cm,trim={1.0cm 0 3cm 17.5cm},clip]{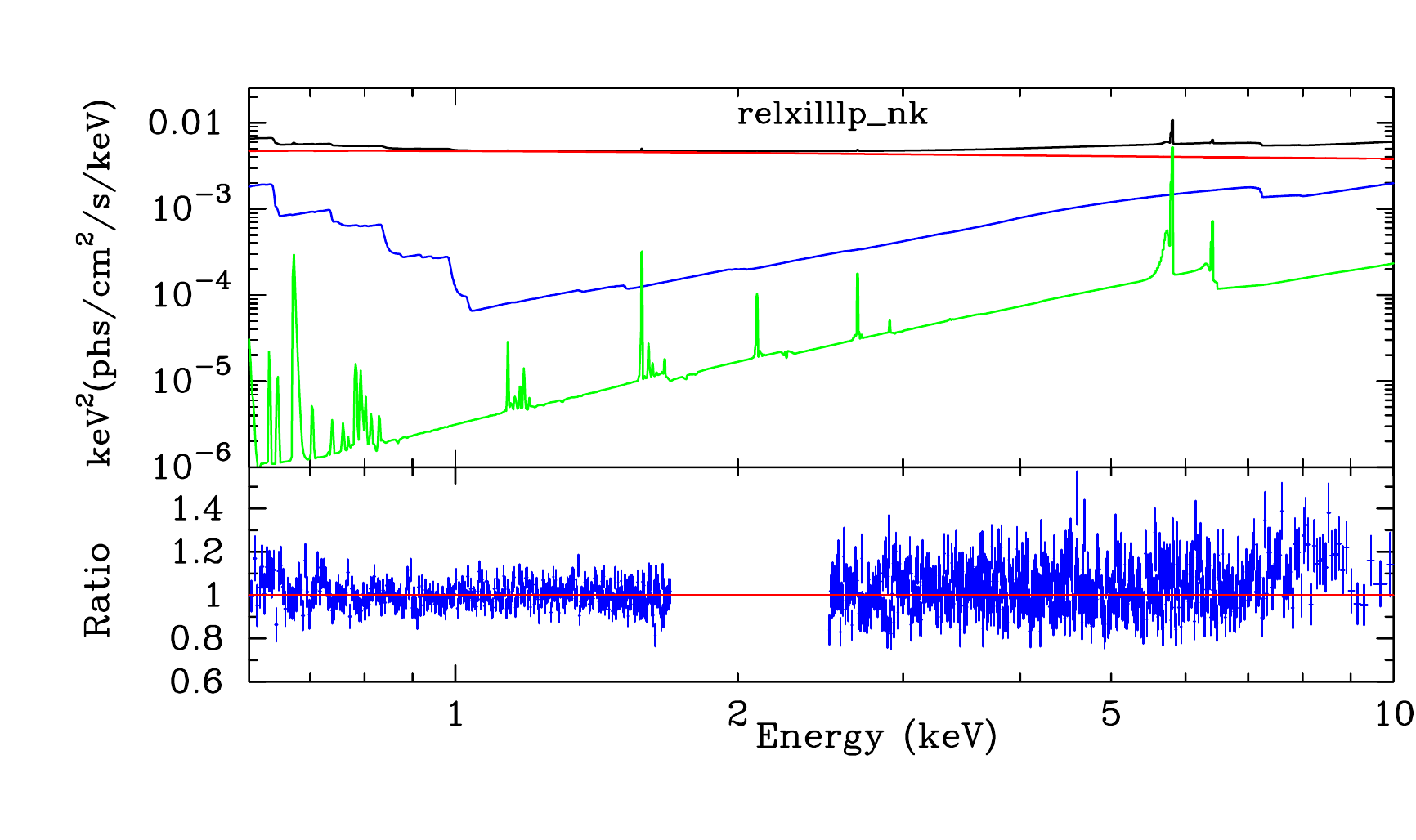}
\end{center}
\vspace{-0.7cm}
\caption{Spectra of the best fit models with the corresponding components (upper panels) and data to best-fit model ratios (lower panels) for 1H0419--577 when the reflection component is modeled by {\sc relxill\_nk} (top left panel), {\sc relxillCp\_nk} (top right panel), {\sc relxillD\_nk} (bottom left panel), and {\sc relxilllp\_nk} (bottom right panel). The total spectra are in black, power law components from the coronas are in red, the relativistic reflection components from the disk are in blue, and the non-relativistic reflection components from cold material are in green. \label{r-1h}}
\end{figure*}

\begin{figure*}[t]
\begin{center}
\includegraphics[width=8.5cm,trim={0cm 2cm 0cm 1cm},clip]{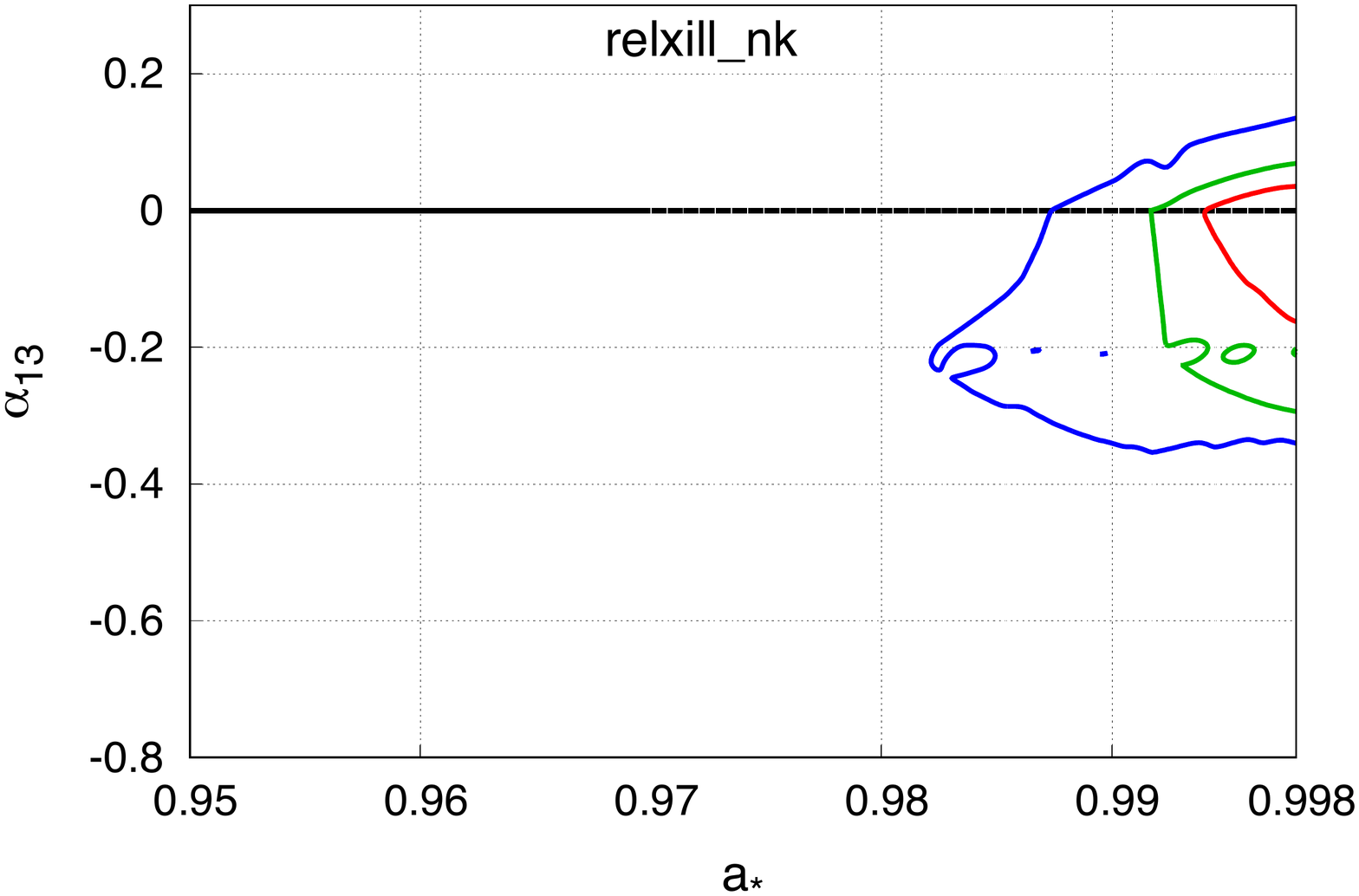}
\includegraphics[width=8.5cm,trim={0cm 2cm 0cm 1cm},clip]{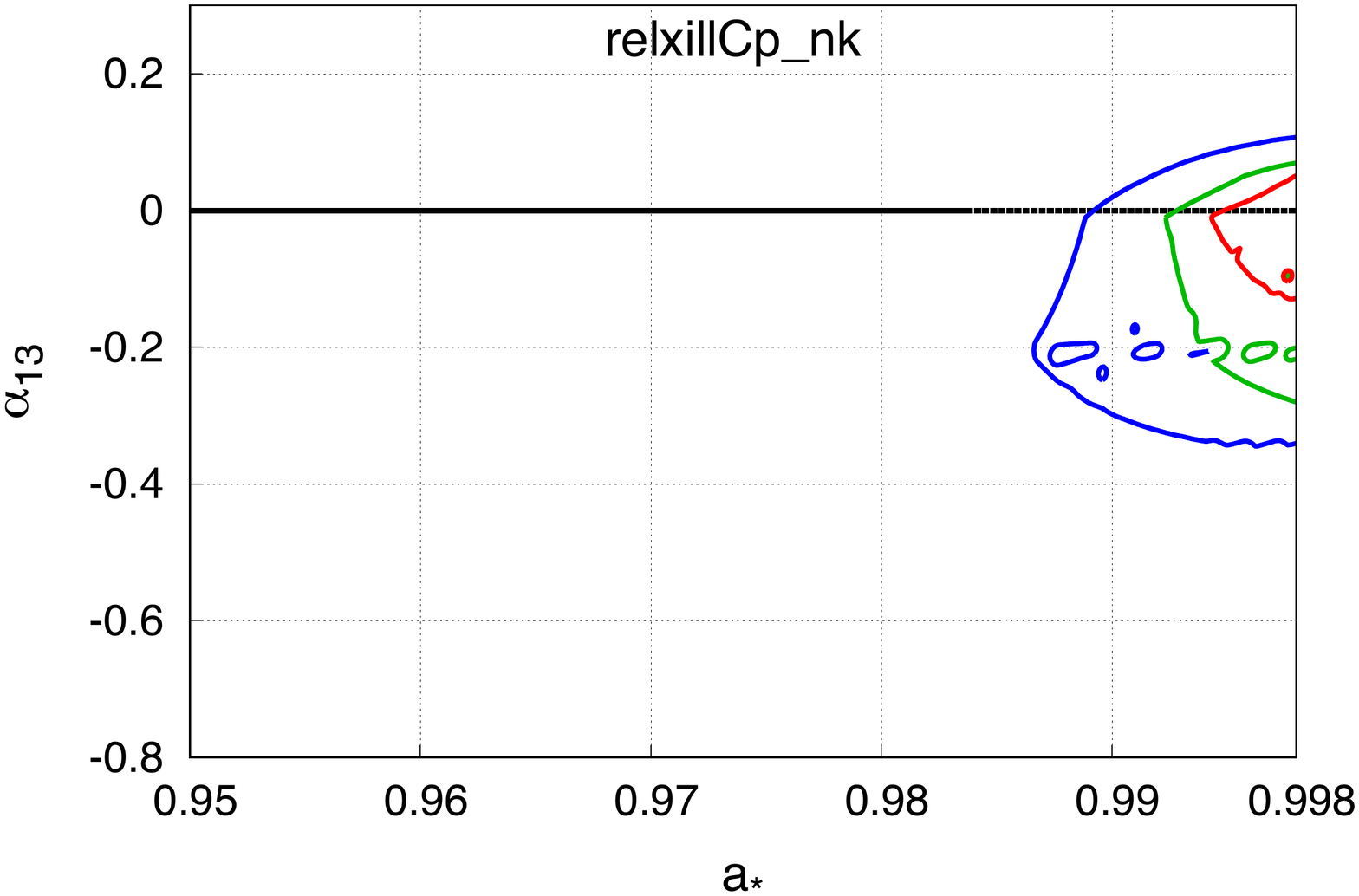} \\
\includegraphics[width=8.5cm,trim={0cm 2cm 0cm 1cm},clip]{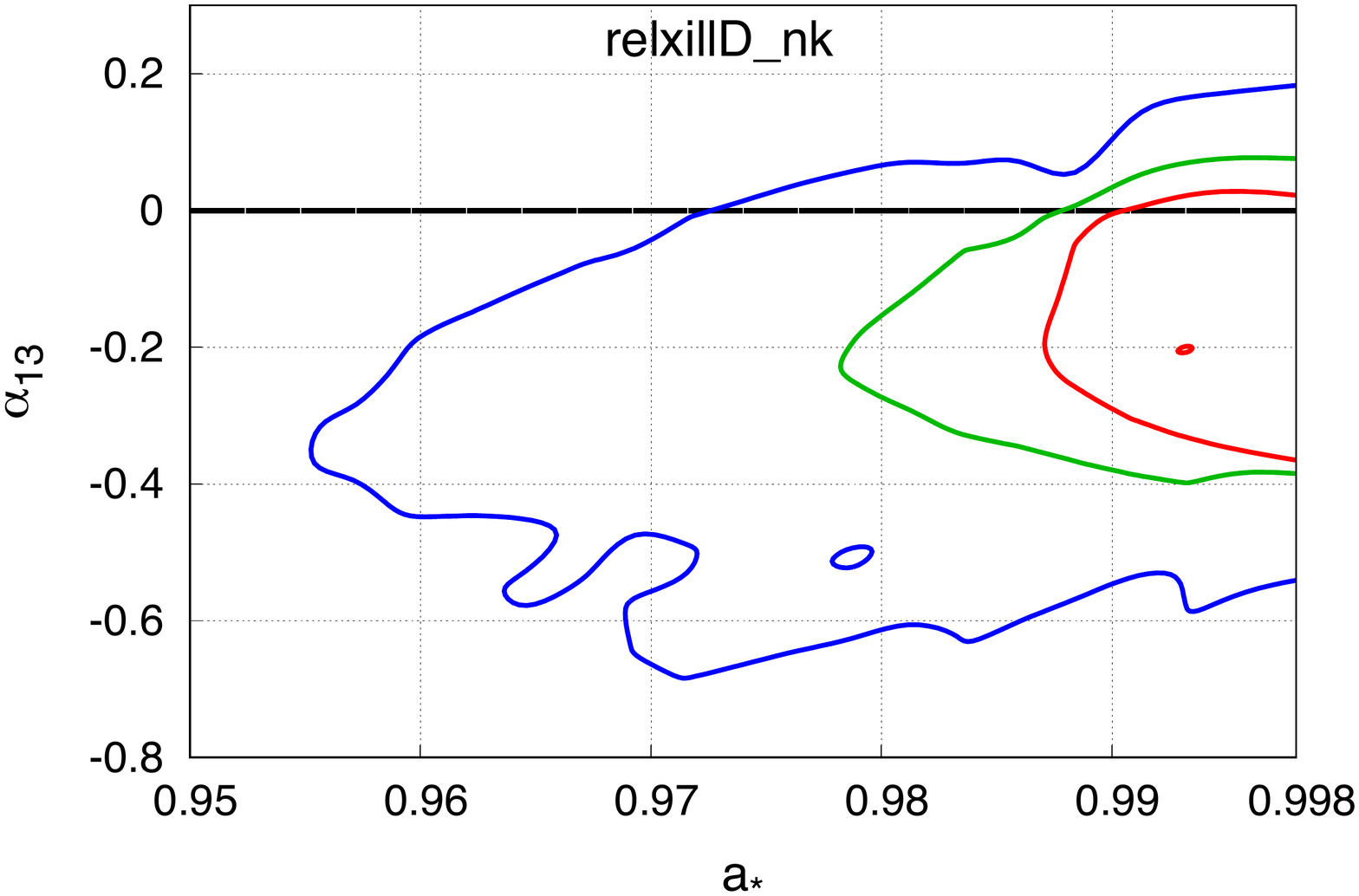}
\includegraphics[width=8.5cm,trim={0cm 2cm 0cm 1cm},clip]{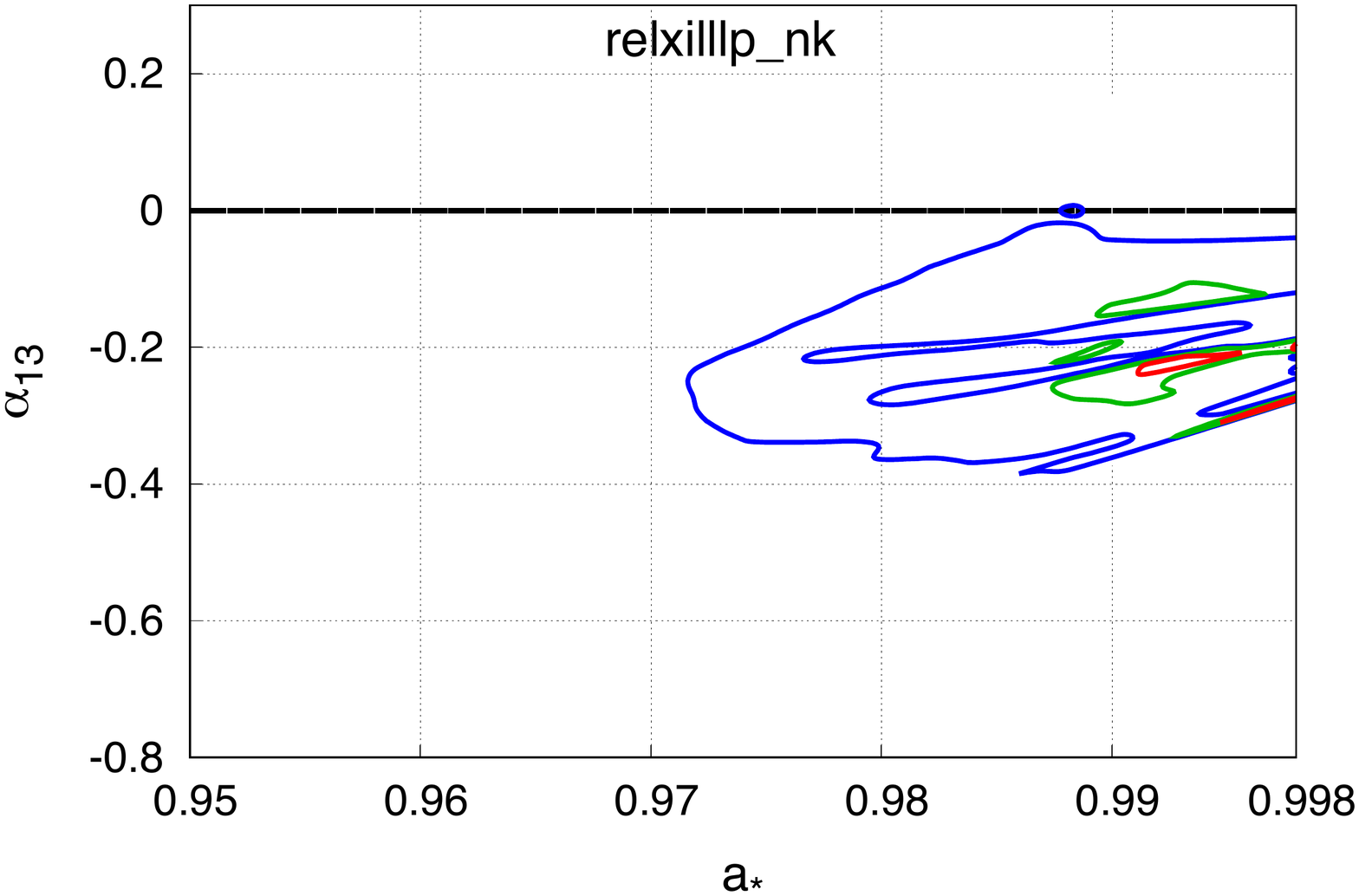}
\end{center}
\vspace{-0.7cm}
\caption{Constraints on the spin parameter $a_*$ and the Johannsen deformation parameters $\alpha_{13}$ for 1H0419--577 when the reflection component is modeled by {\sc relxill\_nk} (top left panel), {\sc relxillCp\_nk} (top right panel), {\sc relxillD\_nk} (bottom left panel), and {\sc relxilllp\_nk} (bottom right panel). The red, green, and blue curves are, respectively, the 68\%, 90\%, and 99\% confidence level boundaries for two relevant parameters. \label{c-1h}}
\end{figure*}

\begin{table*}
\centering
%\vspace{0.2cm} 
\begin{tabular}{lcccc}
& \hspace{0.2cm} {\sc relxill\_nk} \hspace{0.2cm} & \hspace{0.2cm} {\sc relxillCp\_nk} \hspace{0.2cm} & \hspace{0.2cm} {\sc relxillD\_nk} \hspace{0.2cm} & \hspace{0.2cm} {\sc relxilllp\_nk} \hspace{0.2cm} \\
\hline
{\sc tbabs} &&&& \\
$N_{\rm H}/10^{22}$~cm$^{-2}$ & $0.0184^\star$ & $0.0184^\star$ & $0.0184^\star$ & $0.0184^\star$ \\
\hline
{\sc zpowerlaw}/{\sc nthcomp} &&&& \\
$\Gamma$ & $2.308_{-0.026}^{+0.019}$ & $2.271_{-0.008}^{+0.057}$ & $2.306_{-0.025}^{+0.017}$ & $2.193_{-0.037}^{+0.007}$ \\
Norm~$(10^{-3})$ & $0.164_{\rm (P)}^{+0.007}$ & $3.03_{-0.25}^{+2.93}$ & $0.263_{\rm (P)}^{+0.008}$ & $11.92_{-0.12}^{+0.08}$ \\
\hline
{\sc relxill} &&&& \\
$q_{\rm in}$ & $> 9.9$ & $9.73_{-0.36}^{+0.18}$ & $> 9.9$ & -- \\
$q_{\rm out}$ & $= q_{\rm in}$ & $= q_{\rm in}$ & $= q_{\rm in}$ & -- \\
$R_{\rm br}$ [$M$] & -- & -- & -- & -- \\
$h$ [$M$] & -- & -- & -- & $2.50_{-0.04}^{+0.22}$ \\
$i$ [deg] & $< 13$ & $< 16$ & $< 21$ & $< 34$ \\
$a_*$ & $> 0.993$ & $> 0.990$ & $> 0.990$ & $0.990_{-0.063}^{+0.005}$ \\
$\alpha_{13}$ & $0.06_{-0.11}^{+0.02}$ & $0.11_{-0.14}^{+0.01}$ & $0.05_{-0.08}^{+0.03}$ & $0.05_{-0.11}^{+0.33}$ \\
$z$ & $0.0124^\star$ & $0.0124^\star$ & $0.0124^\star$ & $0.0124^\star$ \\
$\log\xi$ & $2.94_{-0.08}^{+0.04}$ & $2.770_{-0.007}^{+0.149}$ & $2.93_{-0.09}^{+0.04}$ & $1.00_{-0.12}^{+0.15}$ \\
$A_{\rm Fe}$  & $1.6_{-0.5}^{+0.4}$ & $1.4_{-0.3}^{+0.3}$ & $1.6_{-0.7}^{+0.5}$ & $1.8_{-0.5}^{+0.4}$ \\
$\log ( n_{\rm e}/10^{15}$~cm$^{-3})$ & $15^\star$ & $15^\star$ & $< 17.5$ & $15^\star$ \\
$E_{\rm cut}$ [keV] & $300^\star$ & -- & $300^\star$ & $300^\star$ \\
$kT_{\rm e}$ [keV] & -- & $60^\star$ & -- & -- \\
Norm~$(10^{-3})$ & $0.72_{-0.10}^{+0.09}$ & $0.503_{-0.093}^{+0.021}$ & $0.53_{-0.09}^{+0.06}$ & $7.3_{-0.3}^{+0.9}$ \\ 
\hline
{\sc xillver} &&&& \\
$\log\xi$ & $0^\star$ & $0^\star$ & $0^\star$ & $0^\star$ \\
Norm~$(10^{-3})$ & $0.130_{-0.024}^{+0.024}$ & $0.131_{-0.016}^{+0.036}$ & $0.130_{-0.026}^{+0.023}$ & $0.112_{-0.014}^{+0.022}$ \\ 
\hline
$\chi^2$/dof & 1352.52/1313 & 1353.55/1313 & 1352.61/1312 & 1415.54/1313 \\
& =1.030 & =1.031 & =1.031 & =1.078 \\
\hline
\end{tabular}
%\vspace{0.3cm}
\caption{Summary of the best-fit values for the supermassive black hole in Swift~J0501--3239.}
\label{t-swift}
\end{table*}

\begin{figure*}[t]
\begin{center}
\includegraphics[width=8.5cm,trim={1.0cm 0 3cm 17.5cm},clip]{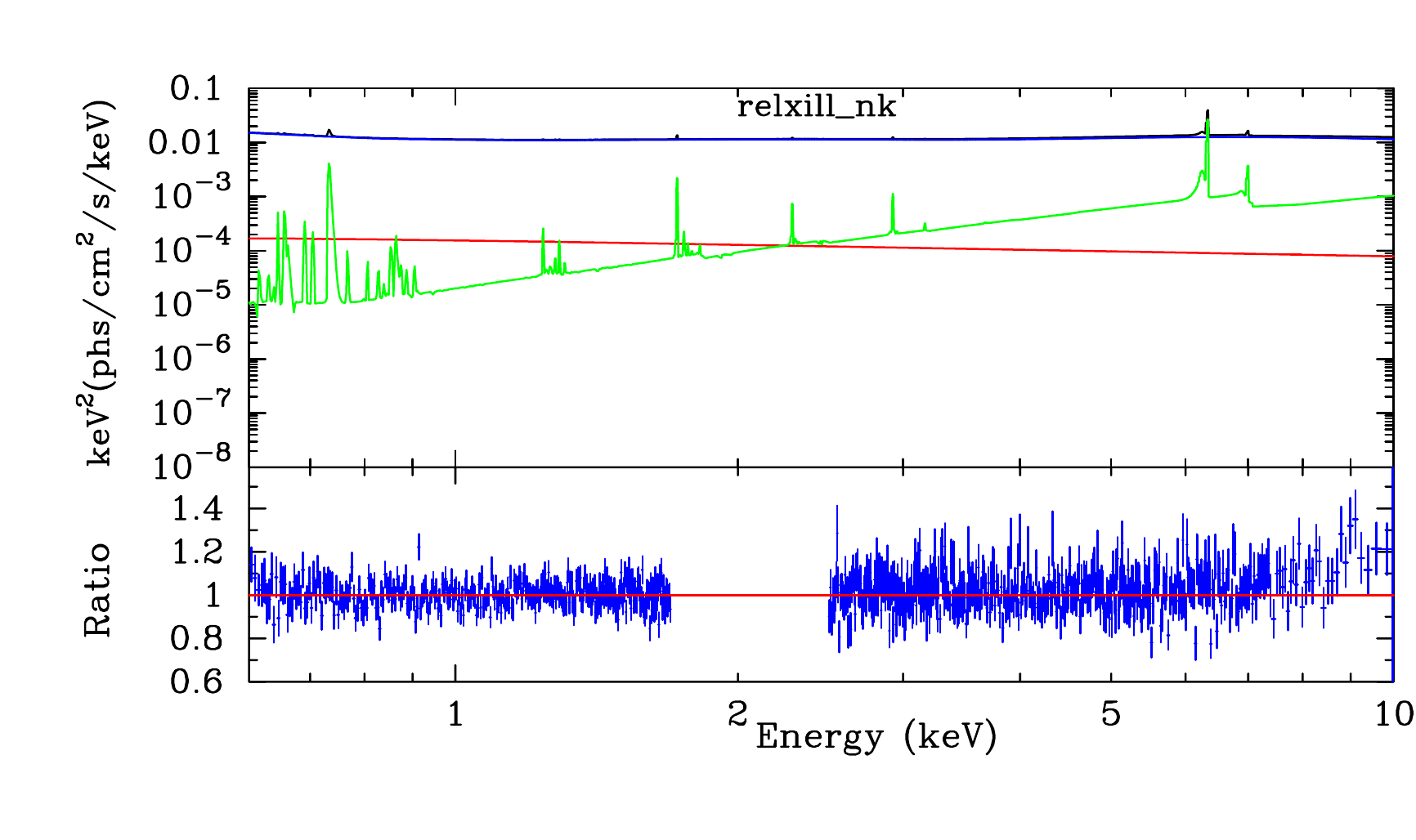}
\includegraphics[width=8.5cm,trim={1.0cm 0 3cm 17.5cm},clip]{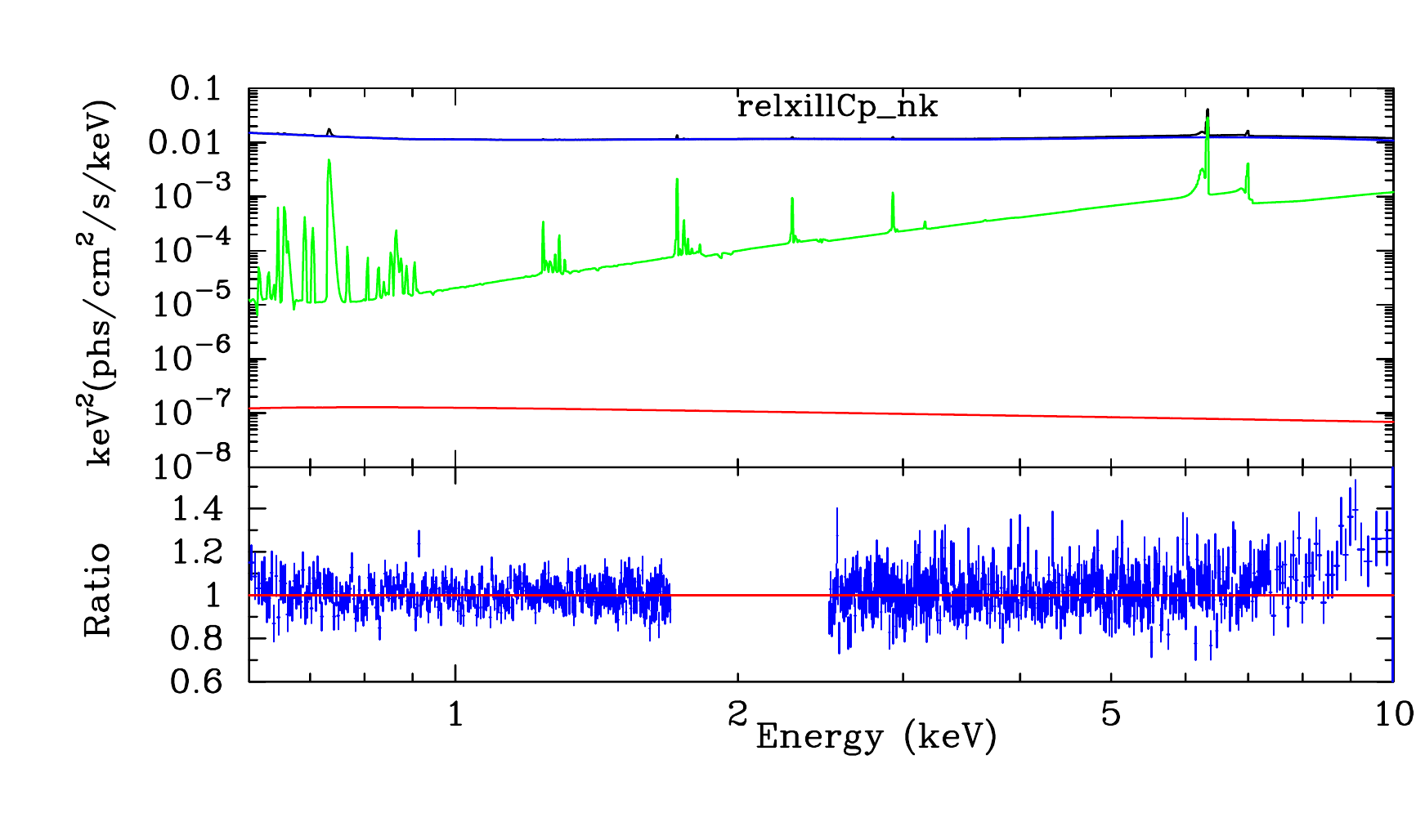} \\
\includegraphics[width=8.5cm,trim={1.0cm 0 3cm 17.5cm},clip]{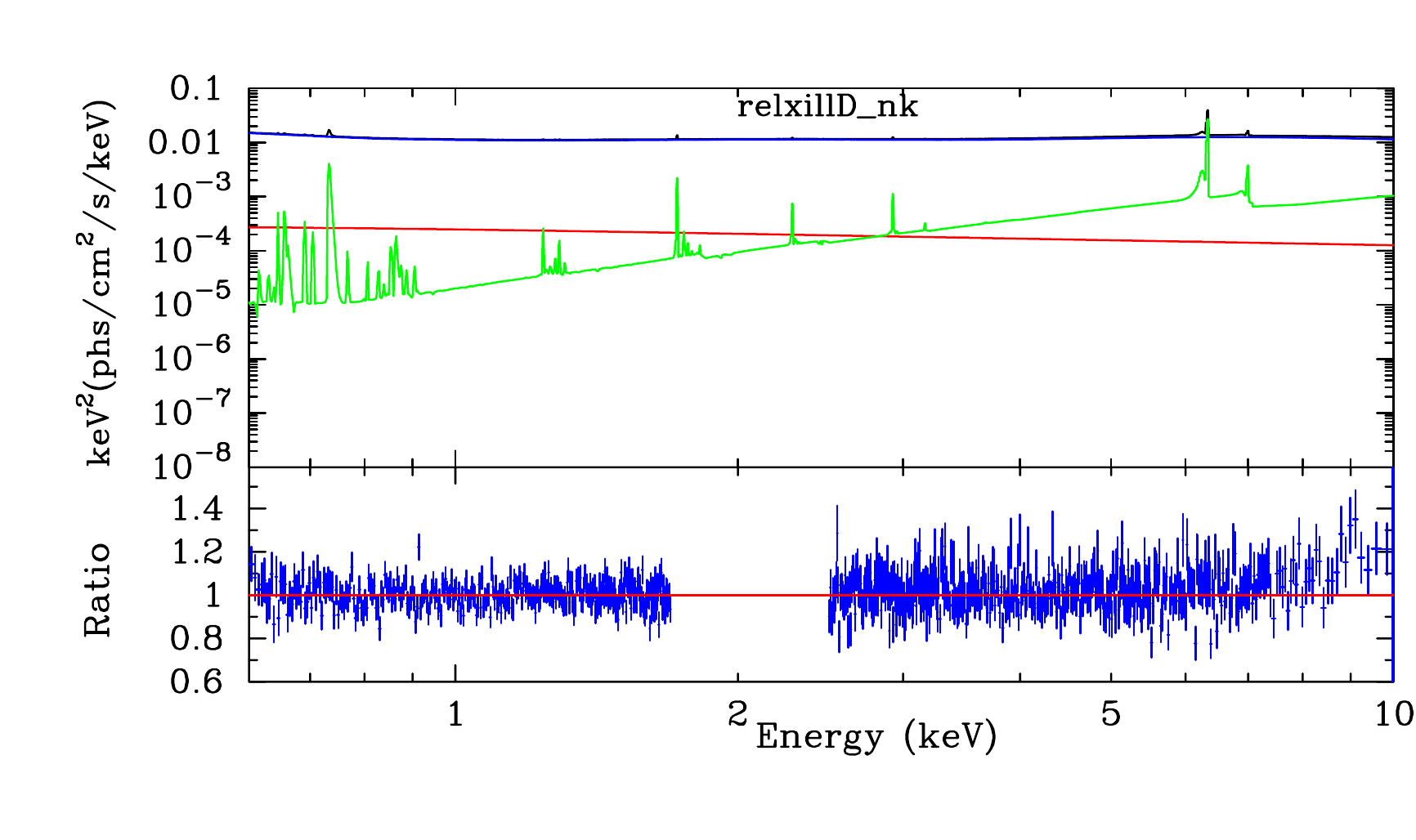}
\includegraphics[width=8.5cm,trim={1.0cm 0 3cm 17.5cm},clip]{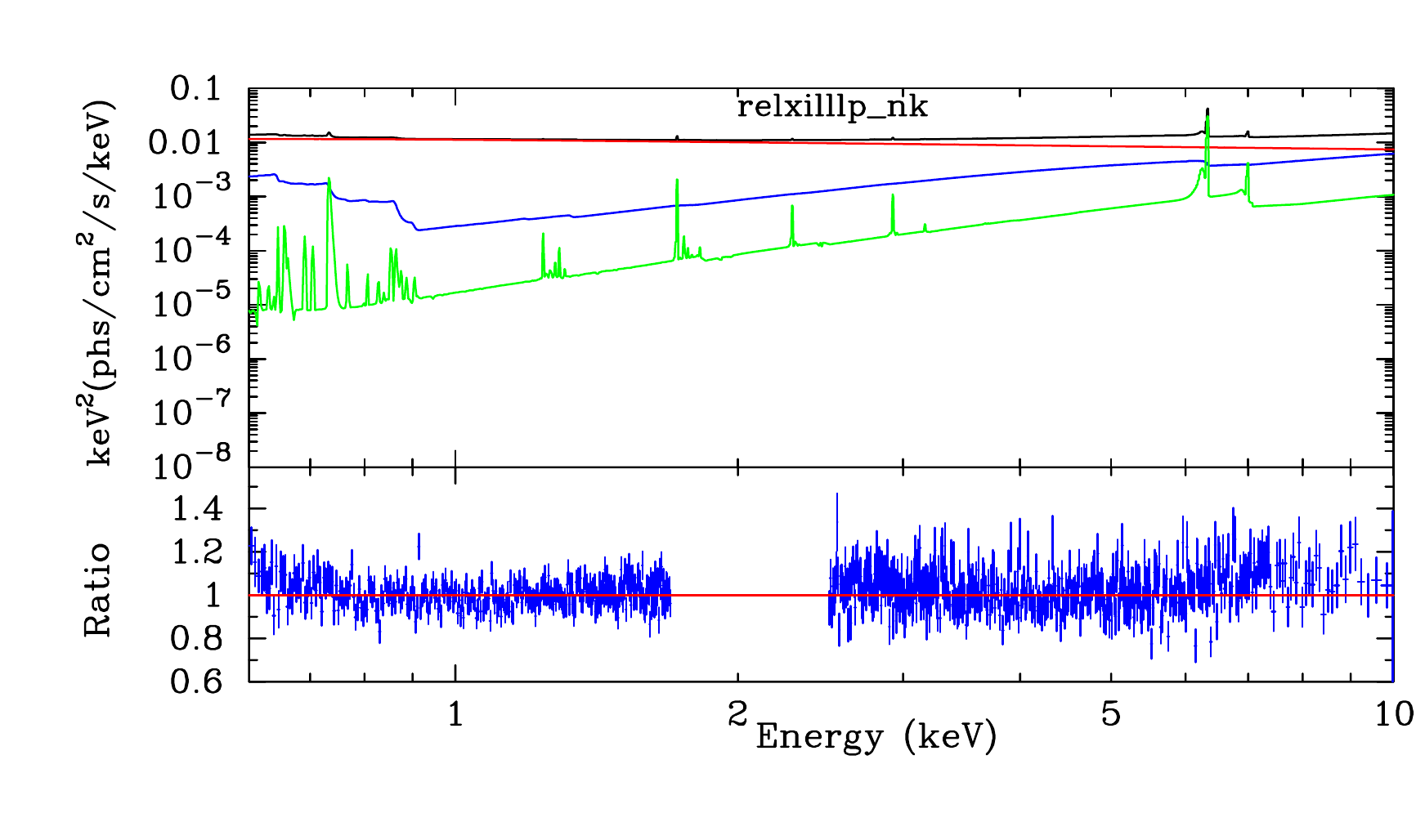}
\end{center}
\vspace{-0.7cm}
\caption{Spectra of the best fit models with the corresponding components (upper panels) and data to best-fit model ratios (lower panels) for Swift~J0501--3239 when the reflection component is modeled by {\sc relxill\_nk} (top left panel), {\sc relxillCp\_nk} (top right panel), {\sc relxillD\_nk} (bottom left panel), and {\sc relxilllp\_nk} (bottom right panel). The total spectra are in black, power law components from the coronas are in red, the relativistic reflection components from the disk are in blue, and the non-relativistic reflection components from cold material are in green. \label{r-swift}}
\end{figure*}

\begin{figure*}[t]
\begin{center}
\includegraphics[width=8.5cm,trim={0cm 2cm 0cm 1cm},clip]{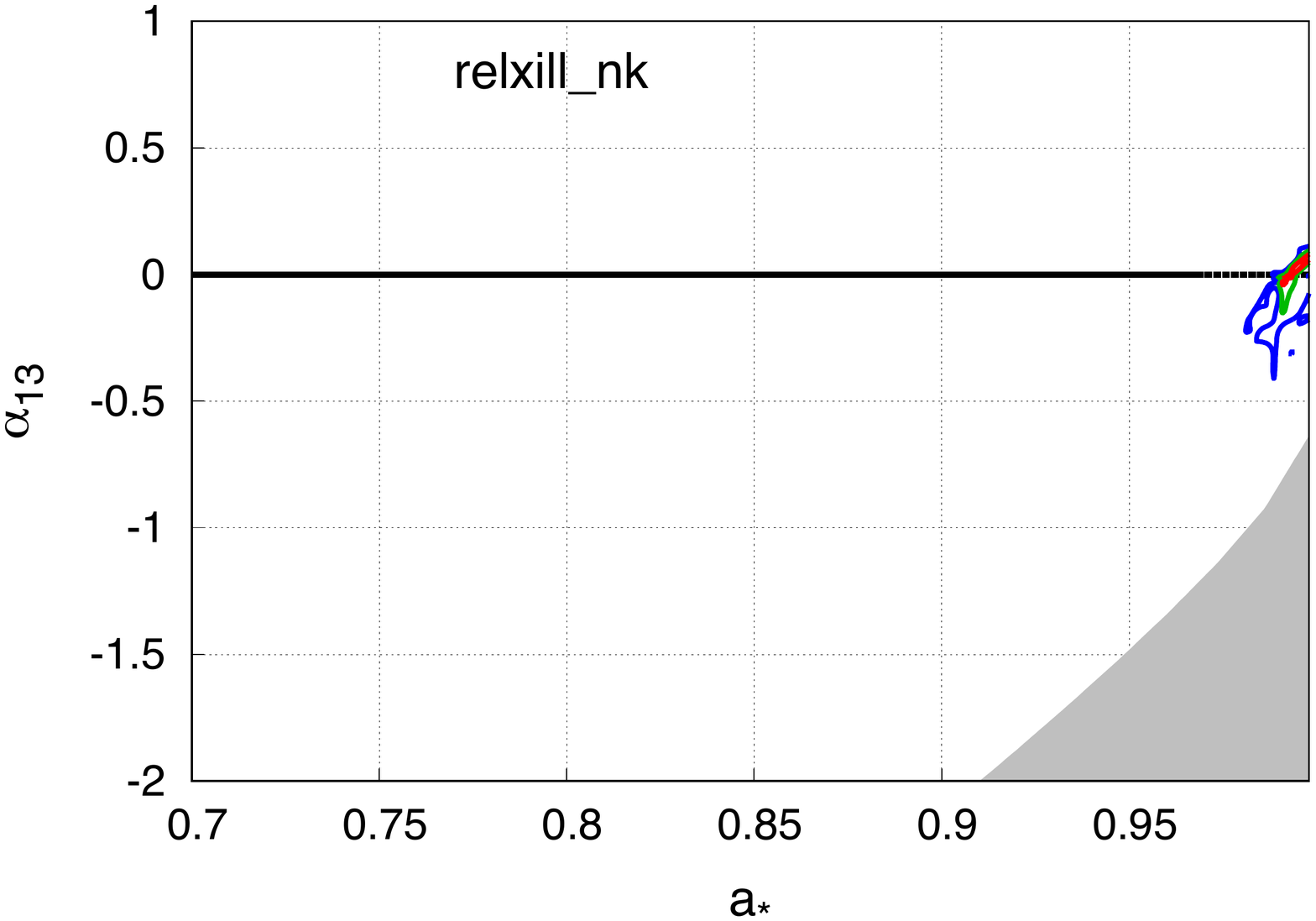}
\includegraphics[width=8.5cm,trim={0cm 2cm 0cm 1cm},clip]{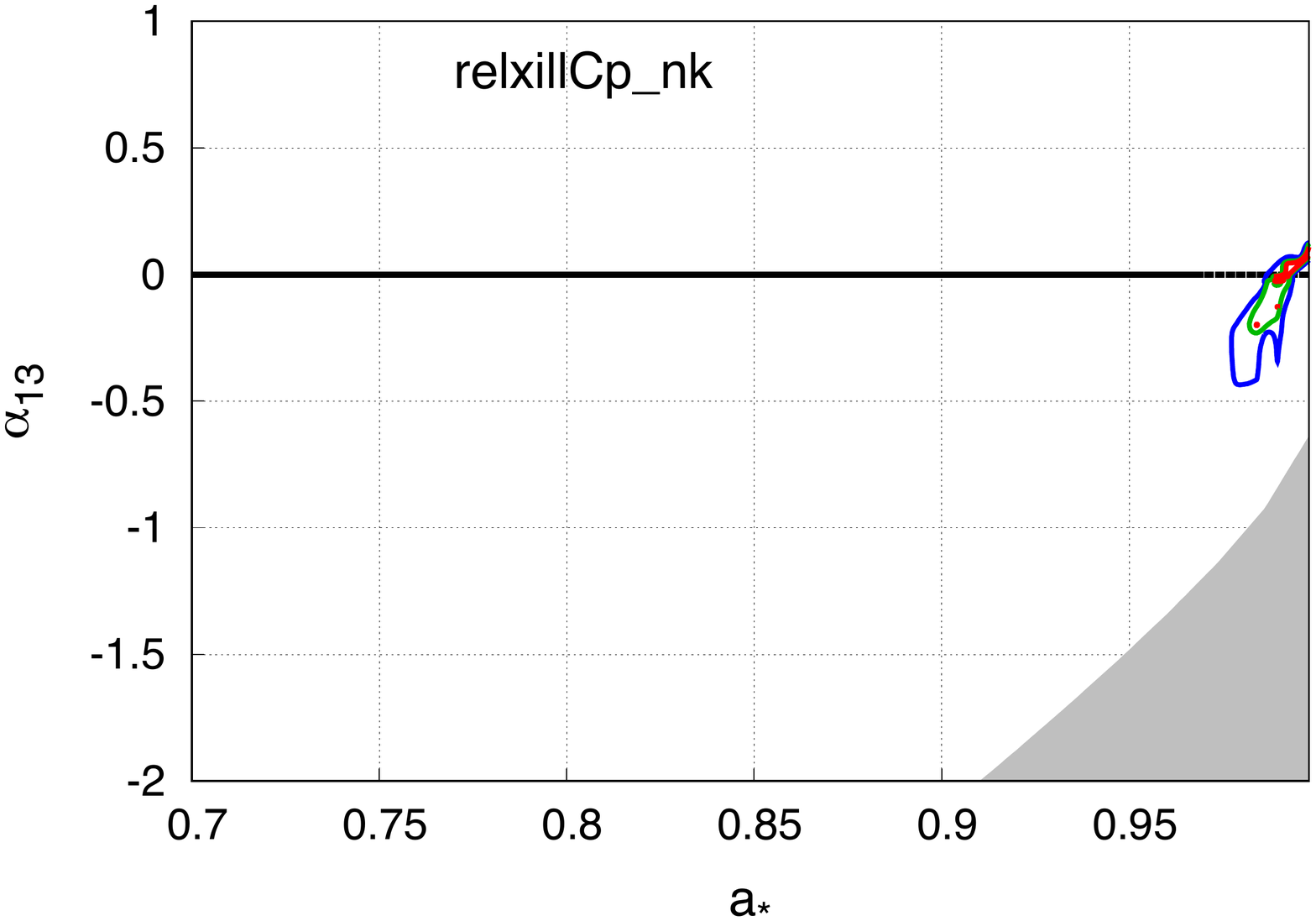} \\
\includegraphics[width=8.5cm,trim={0cm 2cm 0cm 1cm},clip]{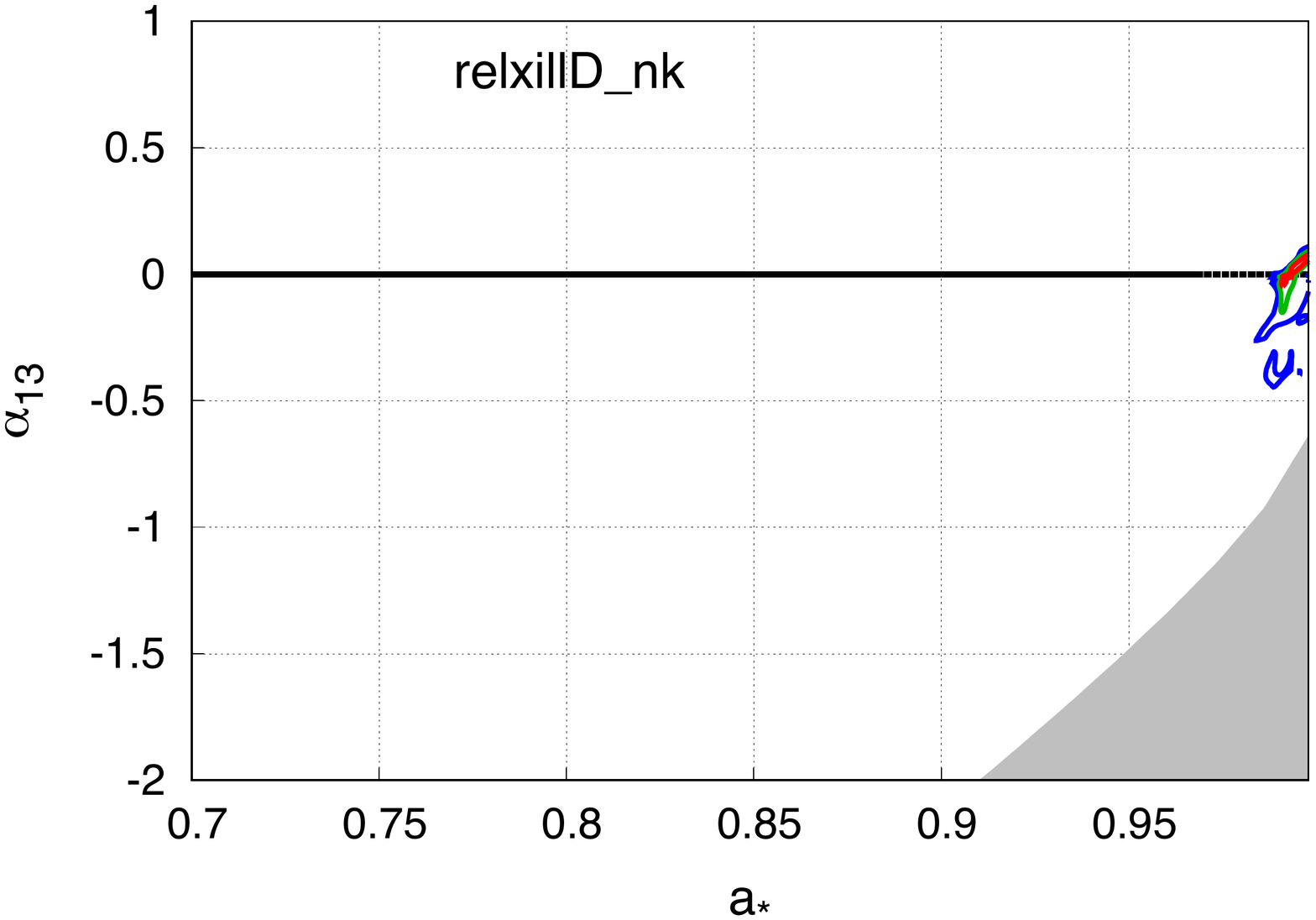}
\includegraphics[width=8.5cm,trim={0cm 2cm 0cm 1cm},clip]{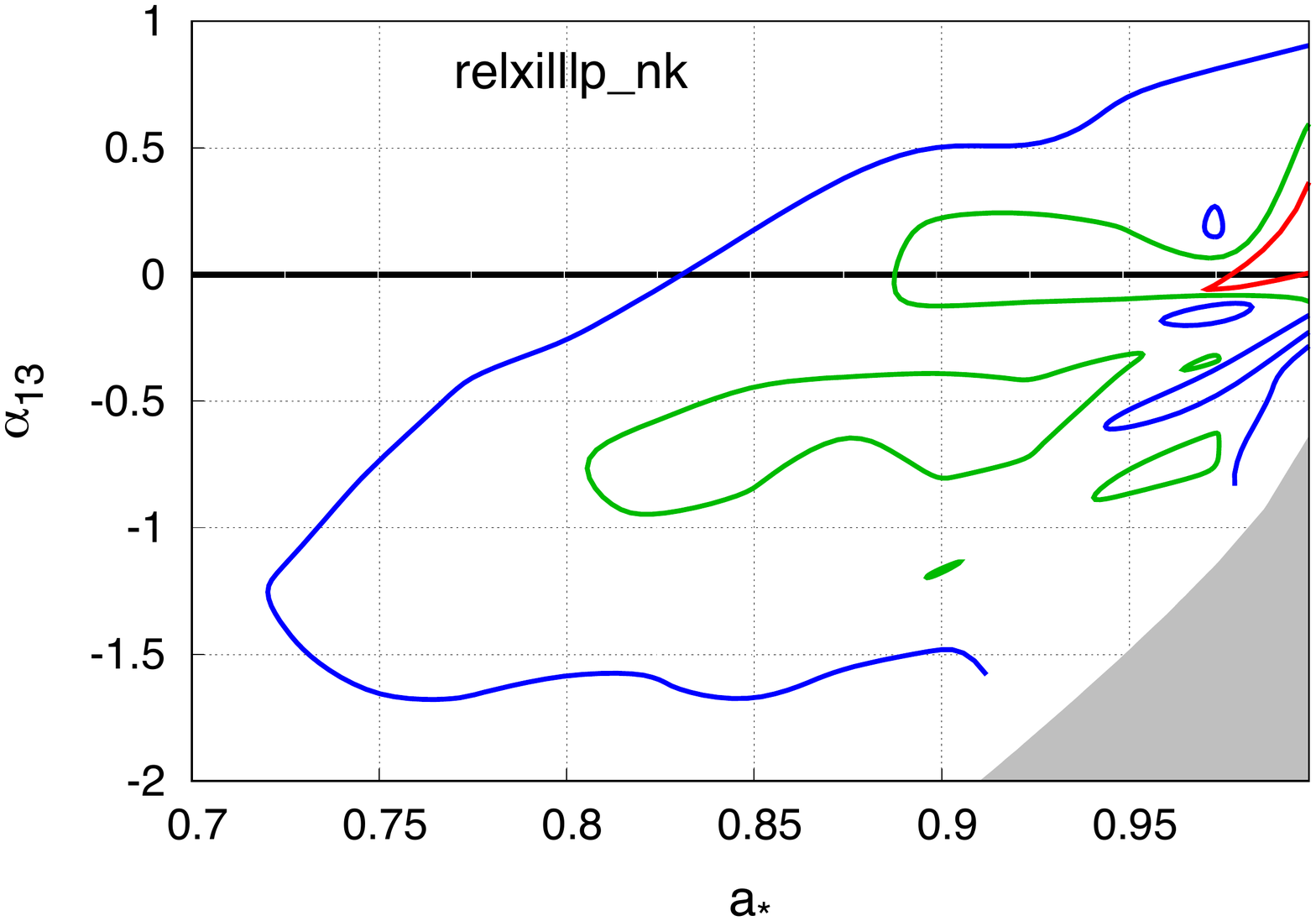}
\end{center}
\vspace{-0.7cm}
\caption{Constraints on the spin parameter $a_*$ and the Johannsen deformation parameters $\alpha_{13}$ for Swift~J0501--3239 when the reflection component is modeled by {\sc relxill\_nk} (top left panel), {\sc relxillCp\_nk} (top right panel), {\sc relxillD\_nk} (bottom left panel), and {\sc relxilllp\_nk} (bottom right panel). The red, green, and blue curves are, respectively, the 68\%, 90\%, and 99\% confidence level boundaries for two relevant parameters. The grayed regions are ignored in our analysis because the do not meet the conditions in~(\ref{eq-c}). \label{c-swift}}
\end{figure*}

\begin{figure*}[t]
\begin{center}
\includegraphics[width=8.5cm,trim={0cm 2cm 0cm 1cm},clip]{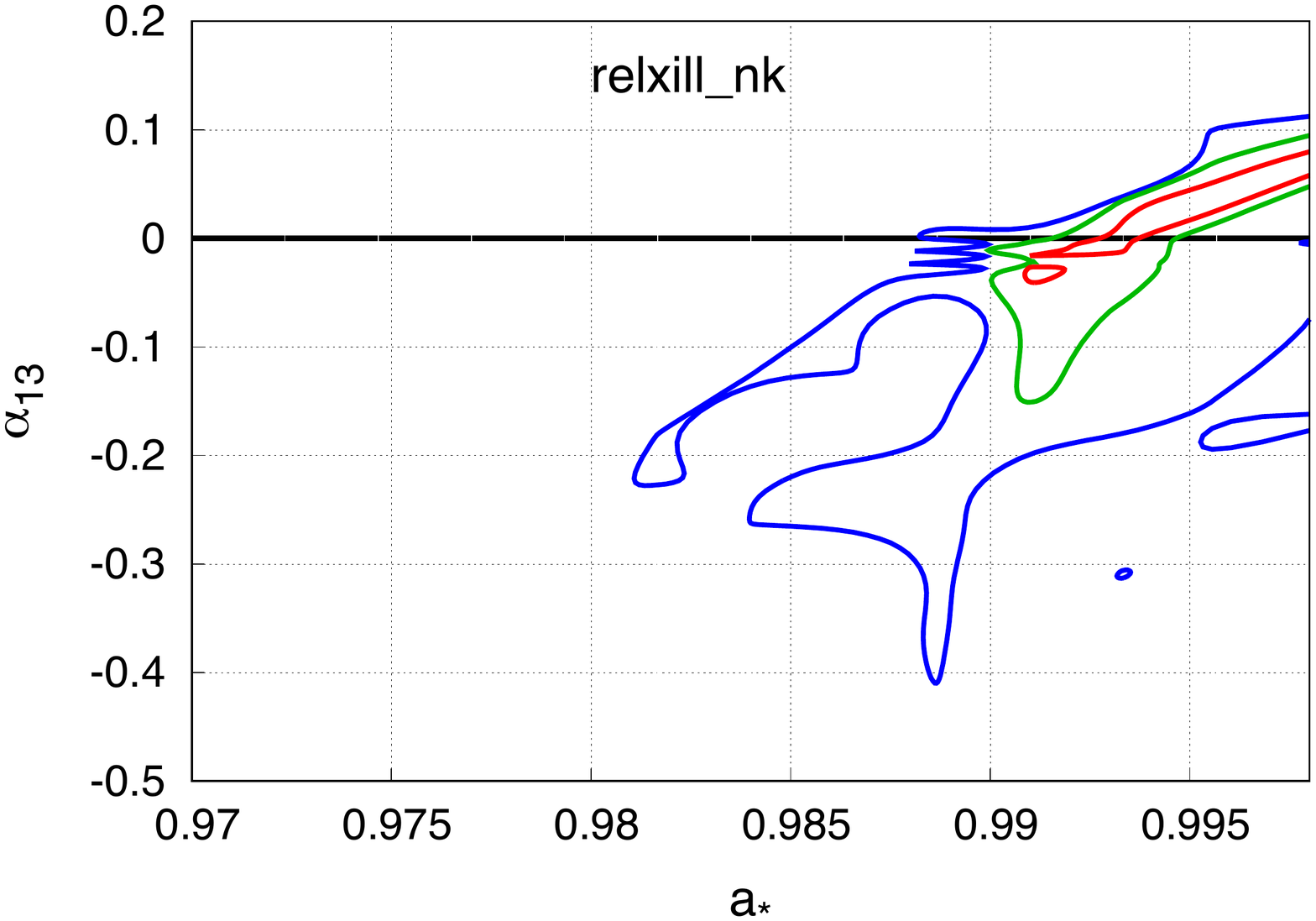}
\includegraphics[width=8.5cm,trim={0cm 2cm 0cm 1cm},clip]{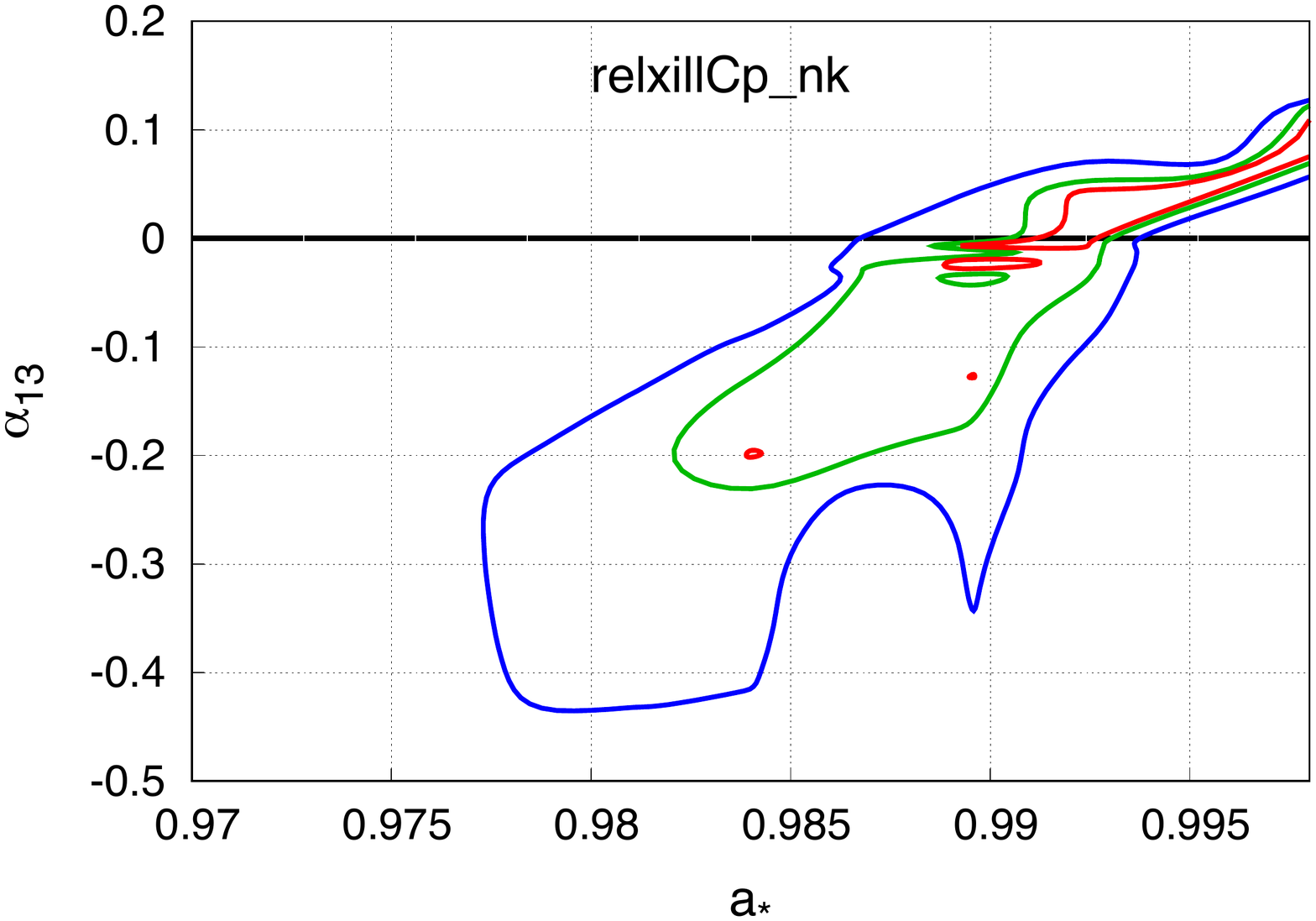} \\
\includegraphics[width=8.5cm,trim={0cm 2cm 0cm 1cm},clip]{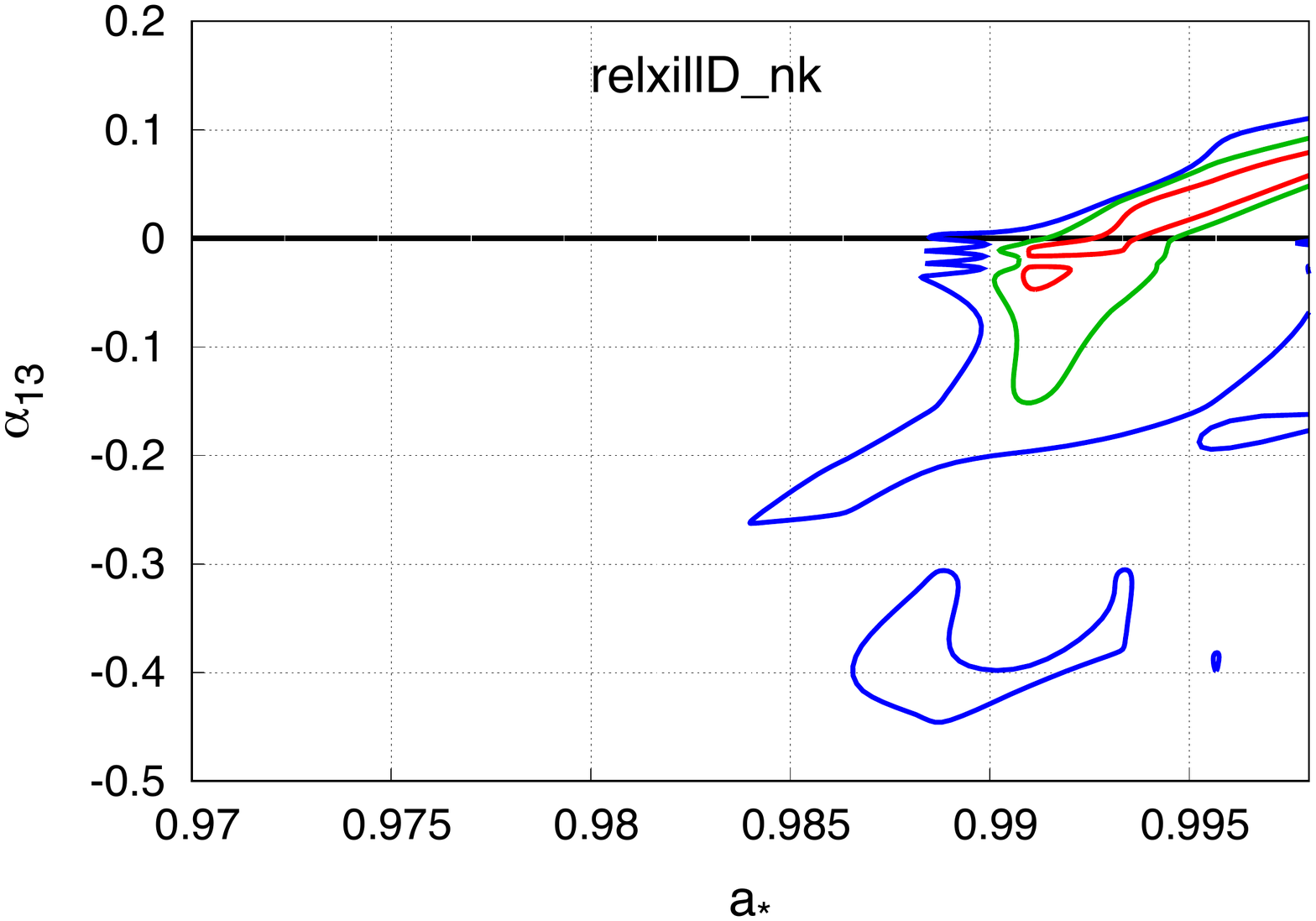}
\includegraphics[width=8.5cm,trim={0cm 2cm 0cm 1cm},clip]{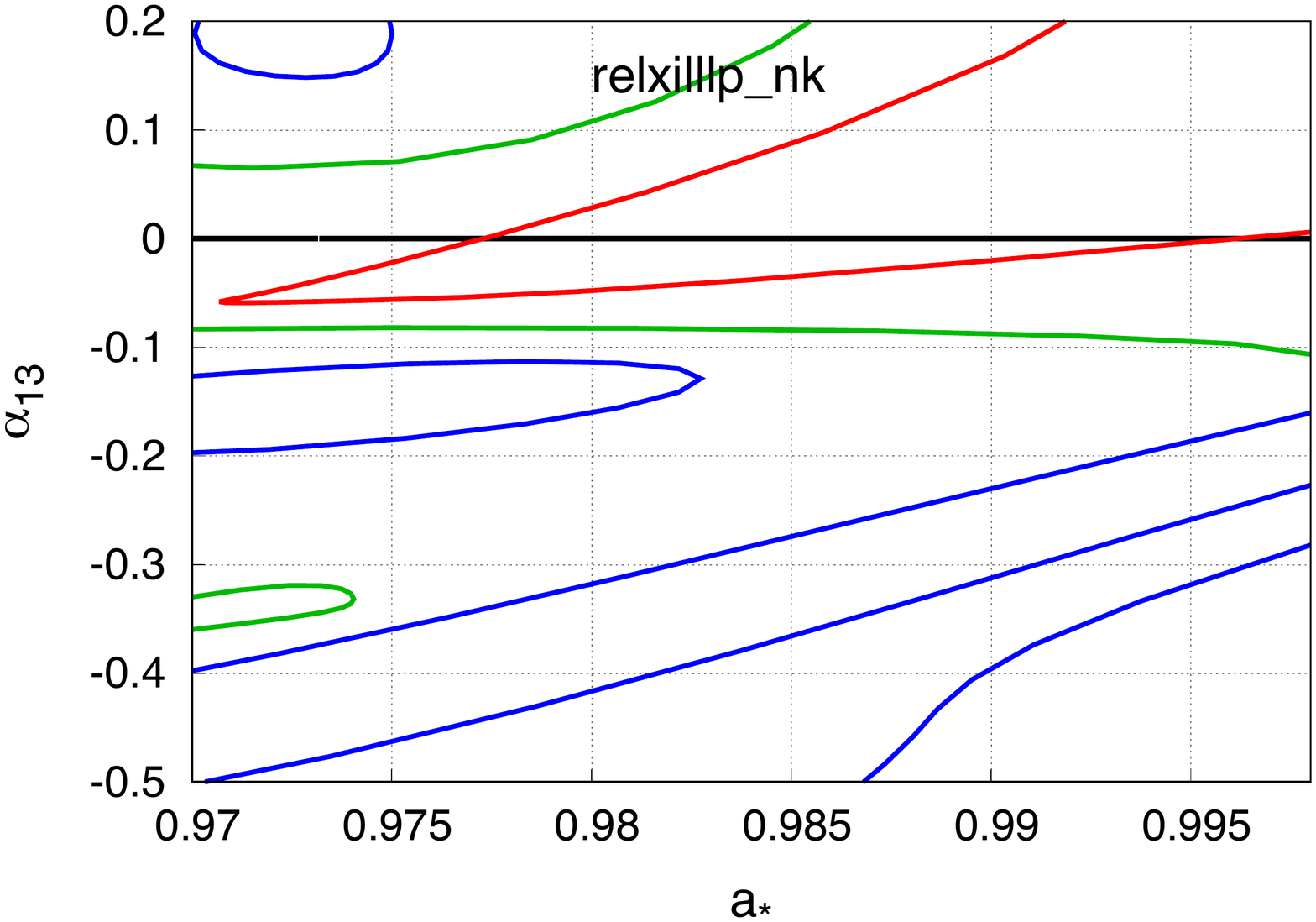}
\end{center}
\vspace{-0.7cm}
\caption{As in Fig.~\ref{c-swift}, but zooming the relevant part of the plane. \label{c-swift2}}
\end{figure*}

\begin{table*}
\centering
%\vspace{0.2cm} 
\begin{tabular}{lcccc}
& \hspace{0.2cm} {\sc relxill\_nk} \hspace{0.2cm} & \hspace{0.2cm} {\sc relxillCp\_nk} \hspace{0.2cm} & \hspace{0.2cm} {\sc relxillD\_nk} \hspace{0.2cm} & \hspace{0.2cm} {\sc relxilllp\_nk} \hspace{0.2cm} \\
\hline
{\sc tbabs} &&&& \\
$N_{\rm H}/10^{22}$~cm$^{-2}$ & $0.039^\star$ & $0.039^\star$ & $0.039^\star$ & $0.039^\star$ \\
\hline
{\sc zpowerlaw}/{\sc nthcomp} &&&& \\
$\Gamma$ & $2.322_{-0.011}^{+0.013}$ & $2.310_{-0.010}^{+0.013}$ & $2.242_{-0.009}^{+0.012}$ & $2.299_{-0.005}^{+0.009}$ \\
Norm~$(10^{-3})$ & $8.94_{-0.40}^{+0.21}$ & $5.8_{-0.3}^{+0.7}$ & $9.51_{-0.20}^{+0.16}$ & $8.6_{-0.4}^{+2.6}$ \\
\hline
{\sc relxill} &&&& \\
$q_{\rm in}$ & $> 9.7$ & $> 9.81$ & $> 8.7$ & -- \\
$q_{\rm out}$ & $3^\star$ & $3^\star$ & $2.24_{-0.21}^{+0.46}$ & -- \\
$R_{\rm br}$ [$M$] & $2.79_{-0.23}^{+0.35}$ & $2.81_{-0.06}^{+0.12}$ & $2.28_{-0.24}^{+0.23}$ & -- \\
$h$ [$M$] & -- & -- & -- & $< 2.7$ \\
$i$ [deg] & $44.2_{-2.7}^{+2.4}$ & $44.7_{-2.3}^{+2.3}$ & $71.5_{-2.1}^{+1.0}$ & $48.6_{-0.8}^{+1.1}$ \\
$a_*$ & $0.996_{-0.003}^{\rm (P)}$ & $0.996_{-0.003}^{\rm (P)}$ & $> 0.995$ & $> 0.993$ \\
$\alpha_{13}$ & $0.00_{-0.24}^{+0.06}$ & $0.00_{-0.23}^{+0.05}$ & $0.00_{-0.17}^{+0.03}$ & $-0.40_{-0.02}^{+0.02}$ \\
$z$ & $0.1372^\star$ & $0.1372^\star$ & $0.1372^\star$ & $0.1372^\star$ \\
$\log\xi$ & $2.995_{-0.105}^{+0.023}$ & $2.990_{-0.135}^{+0.022}$ & $2.69_{-0.12}^{+0.03}$ & $2.997_{-0.067}^{+0.009}$ \\
$A_{\rm Fe}$  & $5.3_{-1.1}^{+2.2}$ & $3.8_{-0.6}^{+0.8}$ & $> 8.6$ & $3.6_{-0.3}^{+0.8}$ \\
$\log ( n_{\rm e}/10^{15}$~cm$^{-3})$ & $15^\star$ & $15^\star$ & $18.00_{-0.16}^{+0.04}$ & $15^\star$ \\
$E_{\rm cut}$ [keV] & $300^\star$ & -- & $300^\star$ & $300^\star$ \\
$kT_{\rm e}$ [keV] & -- & $60^\star$ & -- & -- \\
Norm~$(10^{-3})$ & $0.100_{-0.014}^{+0.010}$ & $0.101_{-0.011}^{+0.014}$ & $0.0177_{-0.0036}^{+0.0020}$ & $1.89_{-0.07}^{+0.37}$ \\ 
\hline
{\sc zgauss} &&&& \\
$E_{\rm line}$ & $6.94_{-0.07}^{+0.09}$ & $6.95_{-0.07}^{+0.08}$ & $6.96_{-0.06}^{+0.06}$ & $6.96_{-0.05}^{+0.04}$ \\
\hline
$\chi^2$/dof & 1379.35/1311 & 1379.47/1311 & 1345.36/1309 & 1423.18/1312 \\
& =1.052 & =1.052 & =1.028 & =1.085 \\
\hline
\end{tabular}
%\vspace{0.3cm}
\caption{Summary of the best-fit values for the supermassive black hole in PKS~0558--504.}
\label{t-pks}
\end{table*}

\begin{figure*}[t]
\begin{center}
\includegraphics[width=8.5cm,trim={1.0cm 0 3cm 17.5cm},clip]{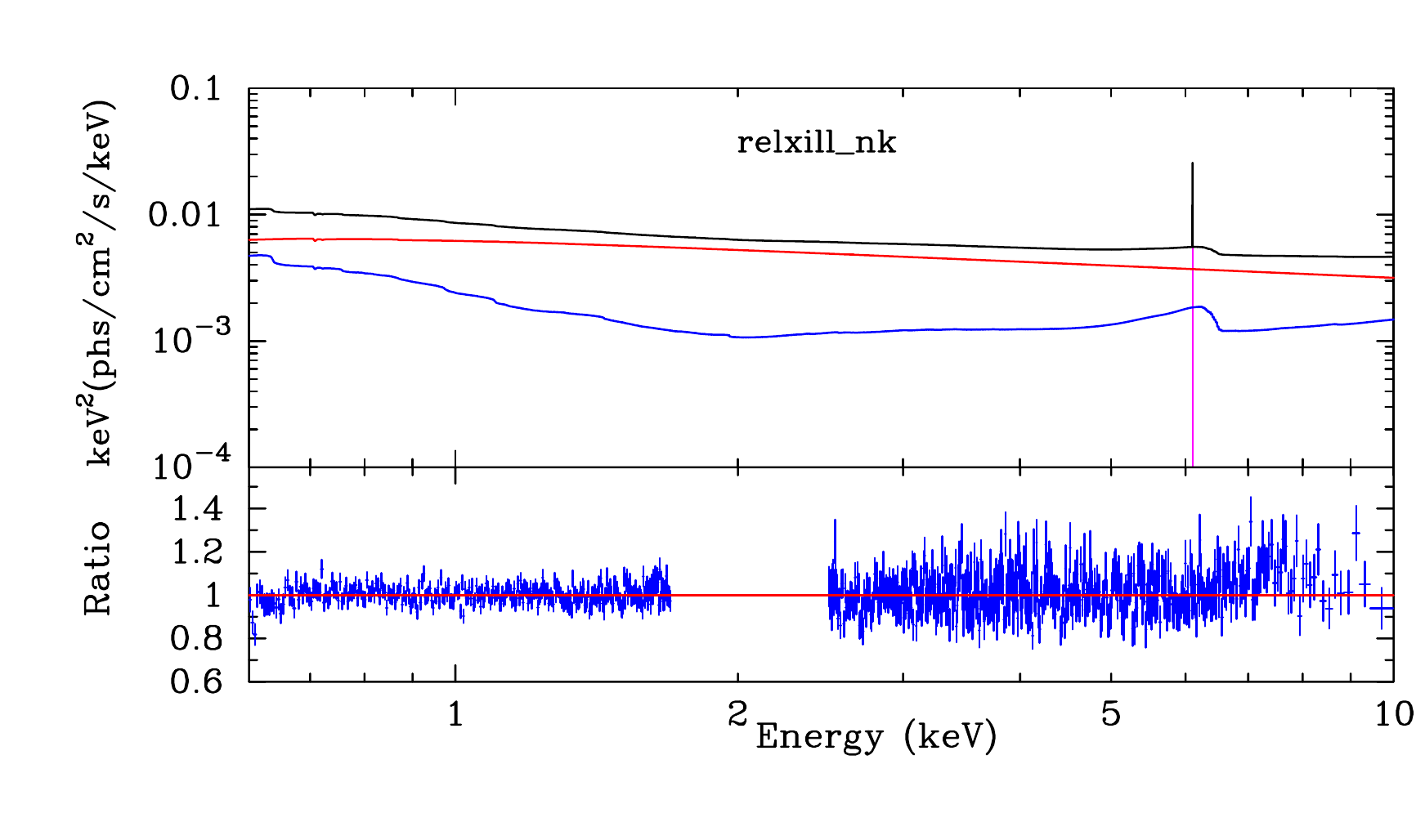}
\includegraphics[width=8.5cm,trim={1.0cm 0 3cm 17.5cm},clip]{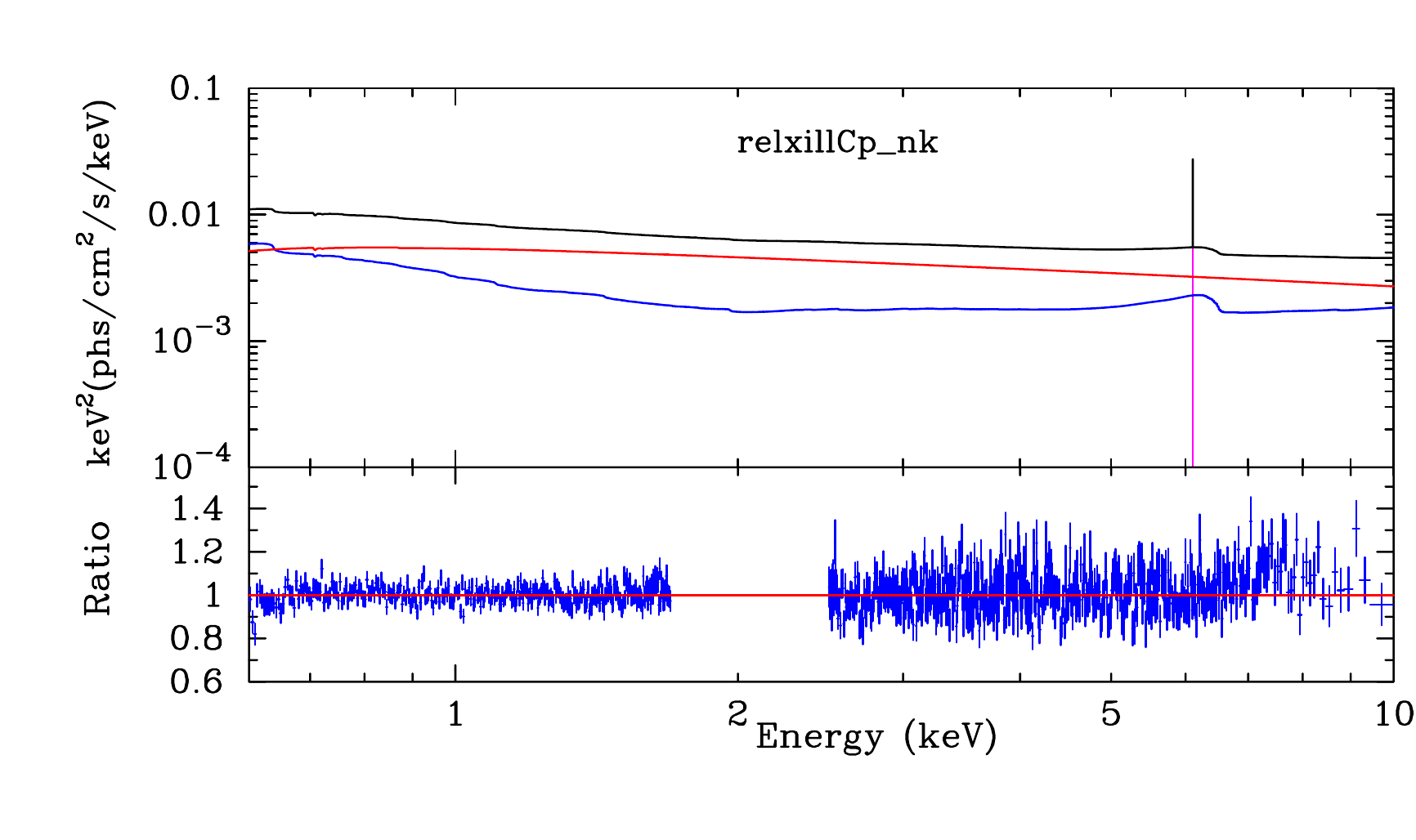} \\
\includegraphics[width=8.5cm,trim={1.0cm 0 3cm 17.5cm},clip]{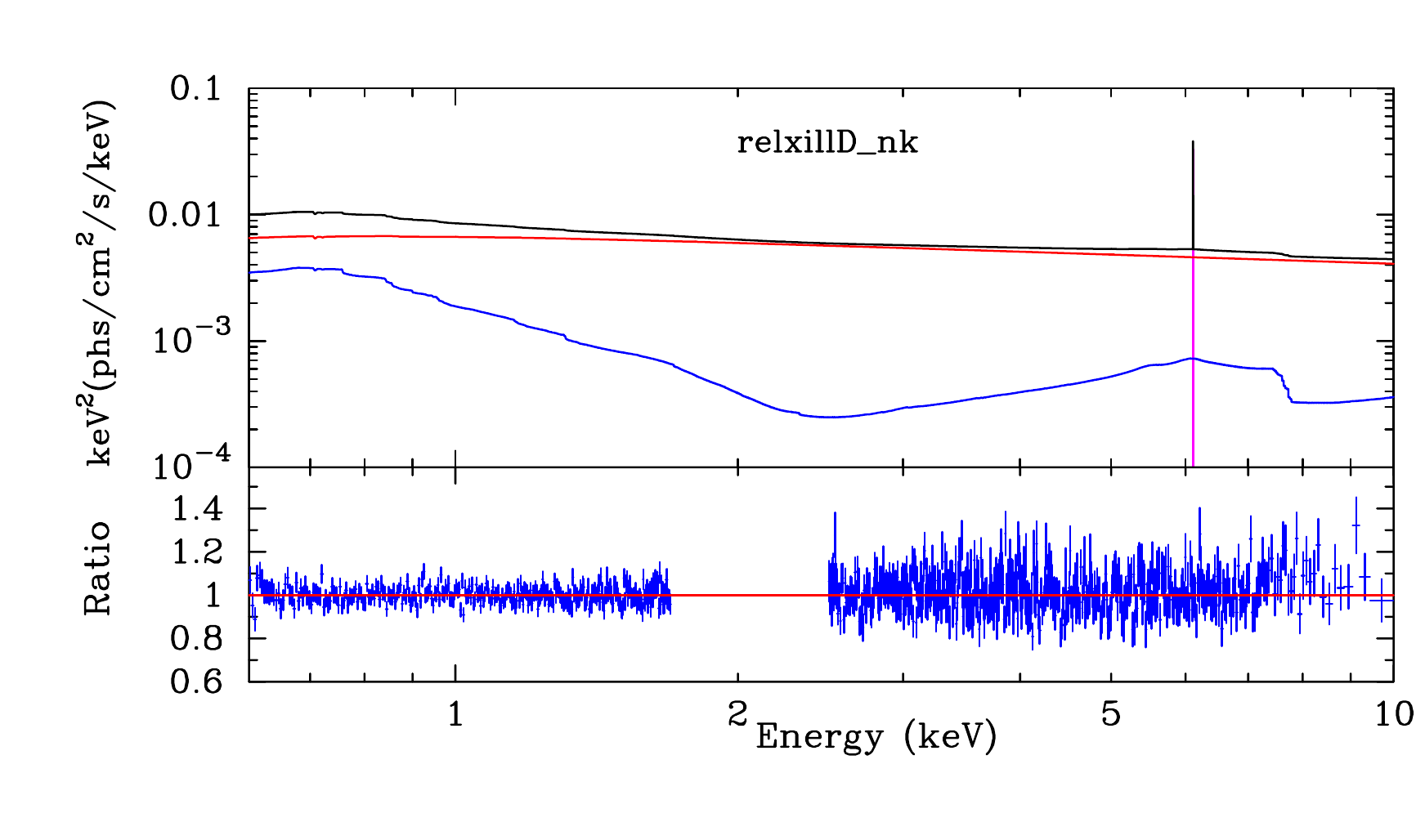}
\includegraphics[width=8.5cm,trim={1.0cm 0 3cm 17.5cm},clip]{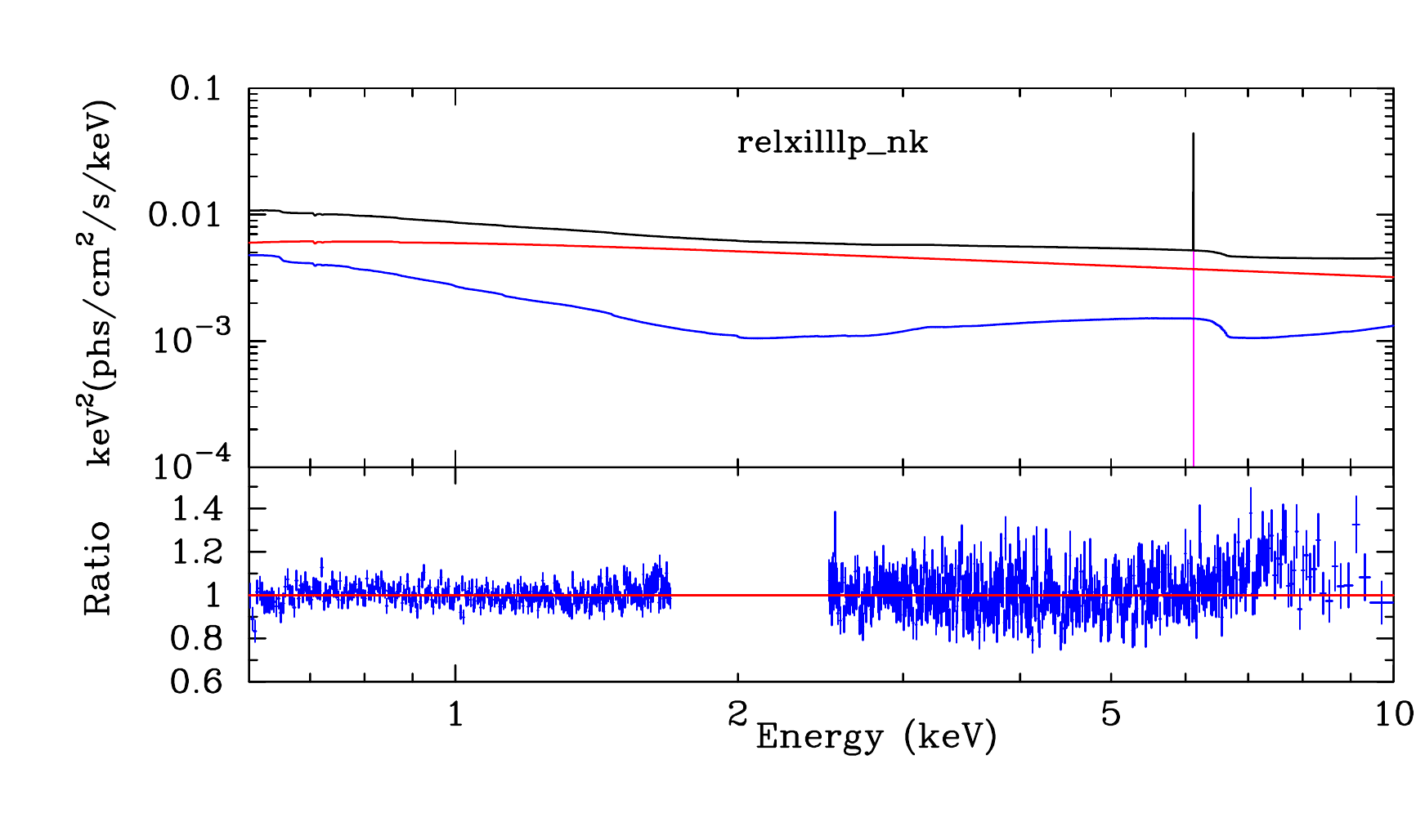}
\end{center}
\vspace{-0.7cm}
\caption{Spectra of the best fit models with the corresponding components (upper panels) and data to best-fit model ratios (lower panels) for PKS~0558--504 when the reflection component is modeled by {\sc relxill\_nk} (top left panel), {\sc relxillCp\_nk} (top right panel), {\sc relxillD\_nk} (bottom left panel), and {\sc relxilllp\_nk} (bottom right panel). The total spectra are in black, power law components from the coronas are in red, the relativistic reflection components from the disk are in blue, and the gaussian line is in magenta. \label{r-pks}}
\end{figure*}

\begin{figure*}[t]
\begin{center}
\includegraphics[width=8.5cm,trim={0cm 2cm 0cm 1cm},clip]{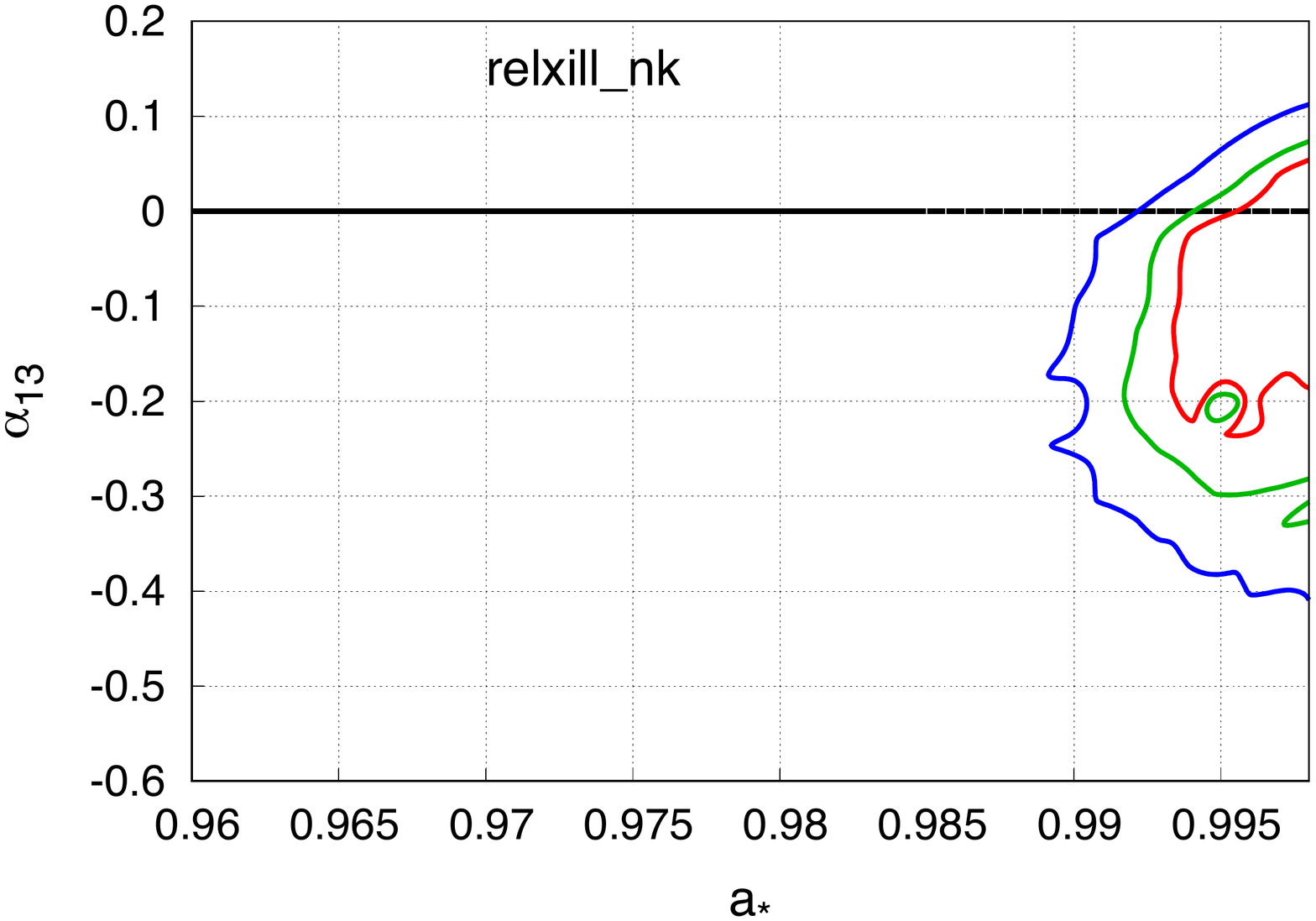}
\includegraphics[width=8.5cm,trim={0cm 2cm 0cm 1cm},clip]{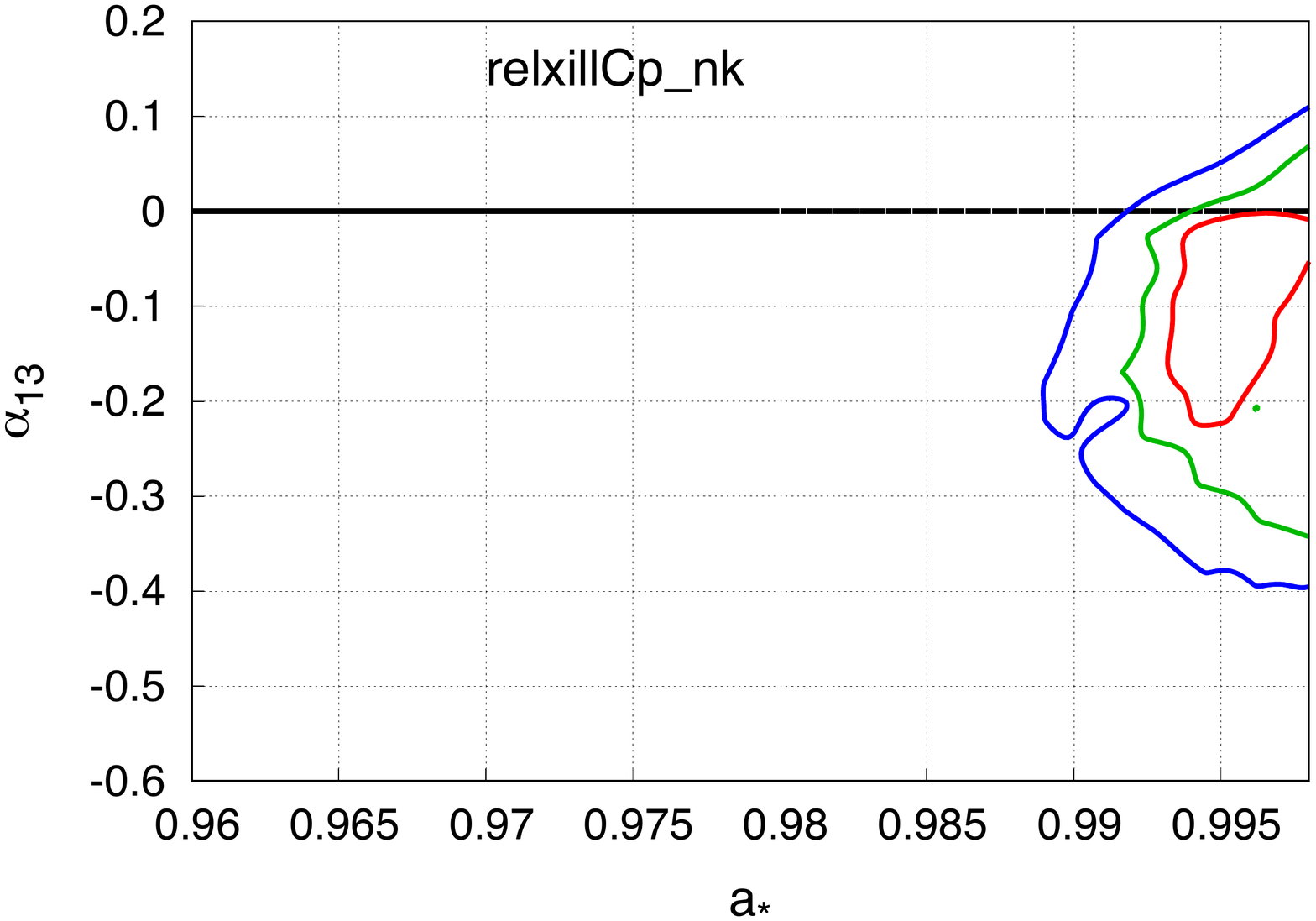} \\
\includegraphics[width=8.5cm,trim={0cm 2cm 0cm 1cm},clip]{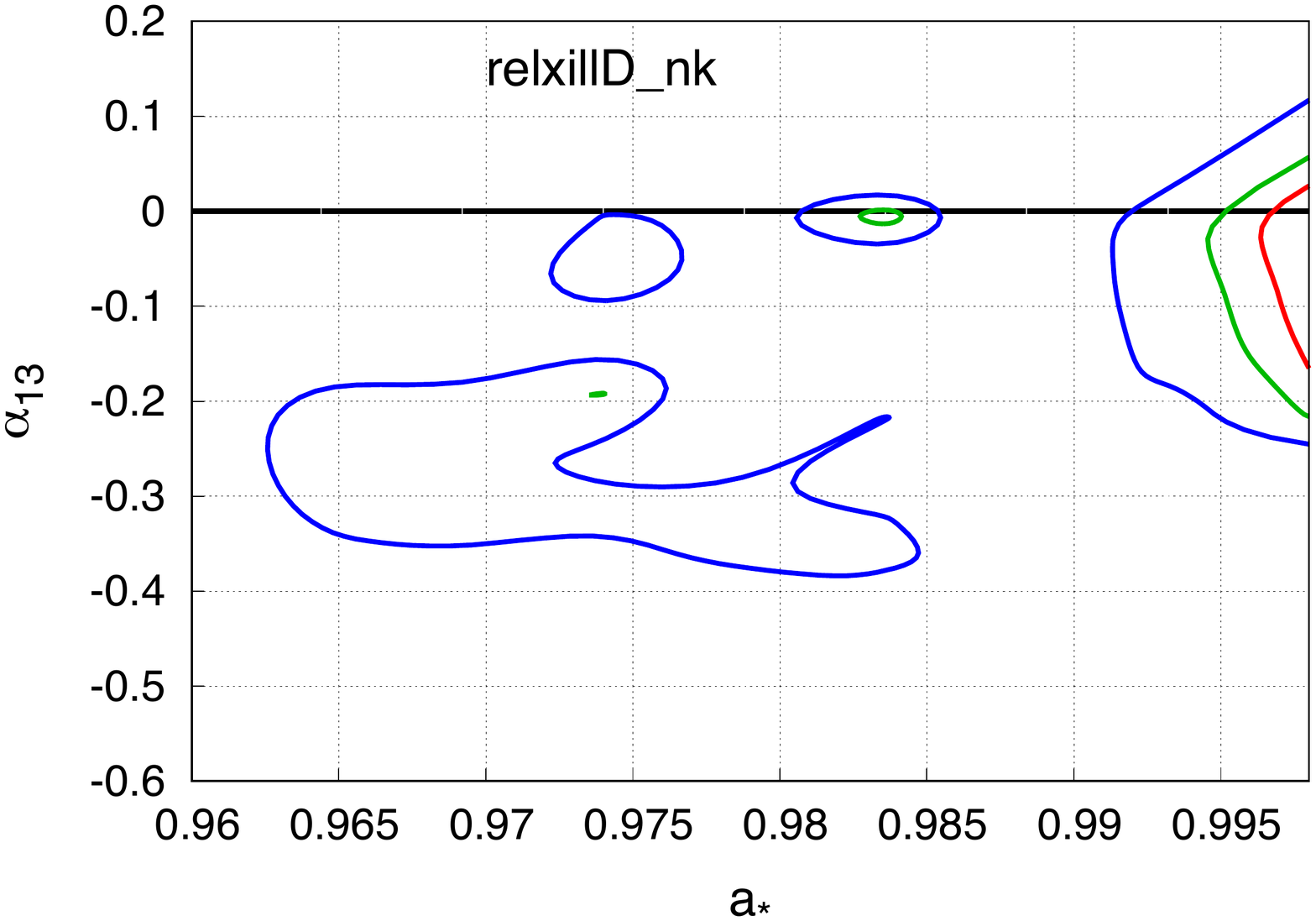}
\includegraphics[width=8.5cm,trim={0cm 2cm 0cm 1cm},clip]{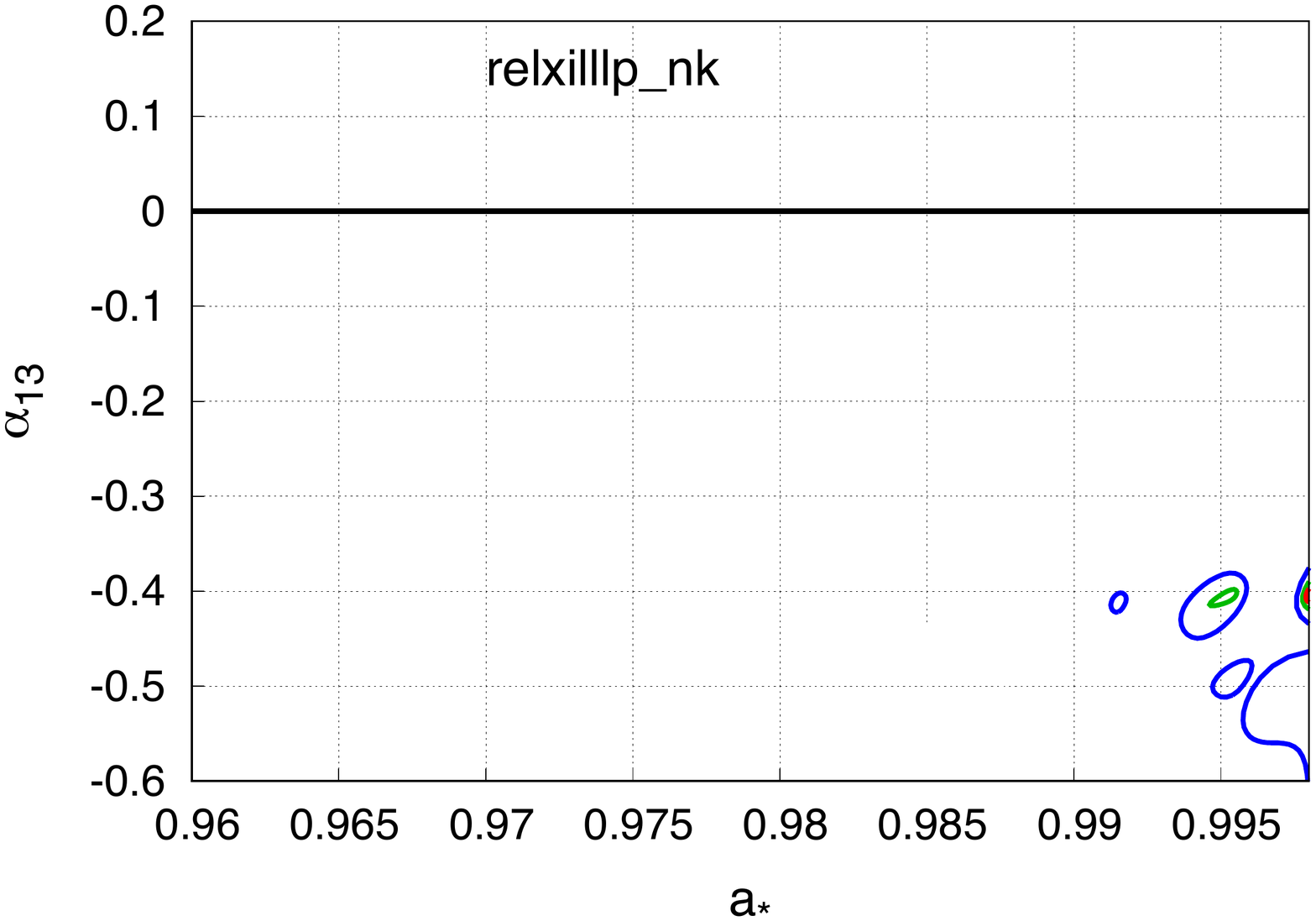}
\end{center}
\vspace{-0.7cm}
\caption{Constraints on the spin parameter $a_*$ and the Johannsen deformation parameters $\alpha_{13}$ for PKS~0558--504 when the reflection component is modeled by {\sc relxill\_nk} (top left panel), {\sc relxillCp\_nk} (top right panel), {\sc relxillD\_nk} (bottom left panel), and {\sc relxilllp\_nk} (bottom right panel). The red, green, and blue curves are, respectively, the 68\%, 90\%, and 99\% confidence level boundaries for two relevant parameters. \label{c-pks}}
\end{figure*}

\section{Impact of different {\sc relxill} flavors \label{s-flav}}

As the {\sc relxill} package for the Kerr spacetime, even in {\sc relxill\_nk} there are several ``flavors'', namely some variants in the model. The main flavors are listed below and more details can be found in the original paper~\cite{noi-nk2}. 
\begin{enumerate}
\item {\sc relxill\_nk}: this is the default model, where the intensity profile of the reflection component is described by a broken power-law (three parameters: inner emissivity index $q_{\rm in}$, outer emissivity index $q_{\rm out}$, and breaking radius $R_{\rm br}$), the spectrum of the corona is described by a power-law with an exponential cut-off (two parameters: photon index $\Gamma$ and cut-off energy $E_{\rm cut}$), and the electron density of the disk is fixed to $n_{\rm e} = 10^{15}$~cm$^{-3}$.
\item {\sc relxillCp\_nk}: the spectrum of the corona is now described by a thermally Comptonized continuum (two parameters: photon index $\Gamma$, coronal temperature $T_{\rm e}$). This is thought to be a better approximation than a power-law with an exponential cut-off.
\item {\sc relxillD\_nk}: the electron density is still constant over the whole disk, but its value can vary in the range $n_{\rm e} = 10^{15} - 10^{19}$~cm$^{-3}$. In this model, it is assumed $E_{\rm cut} = 300$~keV, so the number of the model parameters is the same as in {\sc relxill\_nk} and {\sc relxillCp\_nk}.
\item {\sc relxilllp\_nk}: the intensity profile of the disk is calculated assuming that the corona is a point-like source along the spin axis of the black hole (lamppost corona). The profile mainly depends on the height of the corona, $h$, while the exact spacetime metric plays a marginal role (but still it is properly taken into account). The normalization of the reflection component can either be free or tied to the normalization of the corona spectrum for a lamppost corona at that height.
\end{enumerate}

In the next subsections, we present the results of the analyses of \textsl{Suzaku} data of five ``bare'' active galactic nuclei (AGNs), where bare here means that they are sources with no intrinsic absorption, with the four {\sc relxill} flavors listed above. This is a very preliminary study to figure out the impact of different choices in the model to test the Kerr hypothesis. The sources are Ton~S180, Ark~120, 1H0419--577, Swift~J0501.9--3239, and PKS~0558--504. These sources were already studied with the default model {\sc relxill\_nk} in Ref.~\cite{noi-bare}. Here we repeat the analysis even with {\sc relxill\_nk} because in Ref.~\cite{noi-bare} we employed a previous version of the model. To simplify the discussion, we only consider the possibility of a non-vanishing $\alpha_{13}$, while the other deformation parameters are set to zero. Details on the observations and the data reduction can be found in Ref.~\cite{noi-bare}. Since in several cases the spectrum is dominated by the reflection component, as done in Ref.~\cite{noi-bare} we prefer to model the component from the corona with  {\sc zpowerlw} ({\sc nthcomp} when we use {\sc relxillCp\_nk}) and set the reflection fraction to $-1$ in our reflection model\footnote{In the {\sc relxill} and {\sc relxill\_nk} packages, if the reflection fraction is frozen to $-1$ the model returns only the reflection component. If the reflection fraction is positive, the model returns both the reflection component from the disk and the power-law component (thermally Comptonized continuum in {\sc relxillCp\_nk}). See Refs.~\cite{noi-nk2,noi-bare} for more details.}

\subsection{Ton~S180}

For Ton~S180, we analyze the \textsl{Suzaku} observation of 2006 shown in Tab.~\ref{t-obs}. The XSPEC model is
\be
\text{\sc tbabs*(zpowerlw + relxill\_nk)} \, . \nonumber
\ee
{\sc tbabs} describes Galactic absorption~\cite{tbabs}. Here and in the next subsections, {\sc relxill\_nk} generically indicates one of the four {\sc relxill} flavors (i.e. {\sc relxill\_nk}, {\sc relxillCp\_nk}, {\sc relxillD\_nk}, and {\sc relxilllp\_nk}) and {\sc zpowerlw} indicates the spectrum from the corona (so the XSPEC model employed is {\sc nthcomp} when we use {\sc relxillCp\_nk}). Tab.~\ref{t-ton} shows the best-fits for the four flavors. Fig.~\ref{r-ton} shows the spectra and the specific components of the best-fit models (top panels) and the data to best-fit model ratios (bottom panels). The constraints on the spin parameter $a_*$ and the deformation parameter $\alpha_{13}$ are reported in Fig.~\ref{c-ton}, where the red, green, and blue curves are, respectively, the 68\%, 90\%, and 99\% confidence level boundaries for two relevant parameters.

We note that {\sc relxill\_nk}, {\sc relxillCp\_nk}, and {\sc relxillD\_nk} provide quite similar fits and constraints, which means that the quality of these \textsl{Suzaku} data is not enough to distinguish the three models. The fit with {\sc relxilllp\_nk} is surely worse, even if one may argue that it makes more sense physically because here the reflection component is subdominant with respect to the power-law component from the corona, while with the other three flavors we find that the spectrum is completely dominated by the reflection component, which is difficult to explain (see Fig.~\ref{r-ton}). Despite this substantial difference, it seems that the constraints on the spin parameter $a_*$ and the deformation parameter $\alpha_{13}$ obtained with {\sc relxilllp\_nk} are not too different from the other models.

\subsection{Ark~120}

The 2007 \textsl{Suzaku} observation of Ark~120 (see Tab.~\ref{t-obs}) is fitted with the following XSPEC model 
\be
\text{\sc tbabs*(zpowerlw + relxill\_nk + xillver} \nonumber\\
\text{\sc  + zgauss + zgauss)} \, . \nonumber 
\ee
In addition to the component from the corona and the relativistic reflection component from the disk, described respectively by {\sc zpowerlw} and {\sc relxill\_nk}, we have a non-relativistic reflection component from some cold material at larger distance ({\sc xillver}; {\sc xillverCp} when we use {\sc relxillCp\_nk}), a narrow absorption line ({\sc zgauss}), and a narrow emission line consistent with Fe~XXVI ({\sc zgauss}). Best-fits, spectra and data to best-fit model ratios, and constraints on the spin parameter $a_*$ and the deformation parameter $\alpha_{13}$ are shown, respectively, in Tab.~\ref{t-ark}, Fig.~\ref{r-ark}, and Fig.~\ref{c-ark}.

{\sc relxill\_nk} and {\sc relxillCp\_nk} provide essentially the same result, which can be expected because our analysis is based on data up to 10~keV and the difference between {\sc zpowerlw} and {\sc nthcomp} should be more evident at higher energies. With {\sc relxillD\_nk}, the uncertainties on $a_*$ and $\alpha_{13}$ are substantially larger, but still consistent with the Kerr hypothesis. The fit with {\sc relxilllp\_nk} is worse, but as in the case of Ton~S180 one may argue it makes more sense because the reflection component is subdominant with respect to the power-law component. However, the inclination angle found with {\sc relxilllp\_nk} is definitively too high for an unobscured AGN, indicating some problem with the lamppost coronal geometry as well. The Kerr metric is now recovered at a higher confidence level.

\subsection{1H0419--577}

For 1H0419--577, we have a 179~ks observation of \textsl{Suzaku} in 2007. In the spectrum we see a component from the corona, a relativistic reflection component from the accretion disk, and a non-relativistic reflection component from some cold material far from the source. The XSPEC model is
\be
\text{\sc tbabs*(zpowerlw + relxill\_nk + xillver)} \, . \nonumber 
\ee
Best-fits, spectra and data to best-fit model ratios, and constraints on the spin parameter $a_*$ and the deformation parameter $\alpha_{13}$ are shown, respectively, in Tab.~\ref{t-1h}, Fig.~\ref{r-1h}, and Fig.~\ref{c-1h}.

The quality of the fits with the four {\sc relxill} flavors is quite similar. With {\sc relxilllp\_nk}, we do not recover the Kerr solution at 90\% confidence level, confirming that the choice of the correct intensity profile is one of the most important issues towards precision tests of the Kerr metric using this technique.

\subsection{Swift~J0501.9--3239}

For the \textsl{Suzaku} observation of Swift~J0501.9--3239 in 2008, we employ the following XSPEC model:
\be
\text{\sc tbabs*(zpowerlw + relxill\_nk + xillver)} \, . \nonumber 
\ee
Best-fits, spectra and data to best-fit model ratios, and constraints on the spin parameter $a_*$ and the deformation parameter $\alpha_{13}$ are shown, respectively, in Tab.~\ref{t-swift}, Fig.~\ref{r-swift}, and Fig.~\ref{c-swift}. Fig.~\ref{c-swift2} is an enlargement of Fig.~\ref{c-swift} to better visualize the constraints from {\sc relxill\_nk}, {\sc relxillCp\_nk} and {\sc relxillD\_nk}.

The fits with {\sc relxill\_nk}, {\sc relxillCp\_nk} and {\sc relxillD\_nk} are all very similar, as well as the final constraints on the deformation parameter $\alpha_{13}$. Once again, {\sc relxilllp\_nk} provides a substantially worse fit, but we find the more natural result that the reflection component is subdominant with respect to the power-law from the corona, while in the fits with {\sc relxill\_nk}, {\sc relxillCp\_nk} and {\sc relxillD\_nk} we have that the total spectrum is dominated by the reflection component, which is difficult to explain. Even in the lamppost model we perfectly recover the Kerr metric with $\alpha_{13} = 0$, but the constraint on the deformation parameter is weaker. This suggests that for this source, like for Ton~S180 and Ark~120, the relativistic features produced by the spacetime metric are stronger and not very correlated with the choice of the intensity profile.

\subsection{PKS~0558--504}

In the case of PKS~0558--504, we have five short observations of \textsl{Suzaku} in 2007 that can be combined together. As in Ref.~\cite{noi-bare}, eventually we fit the data with the XSPEC model  
\be
\text{\sc tbabs*(zpowerlw + relxill\_nk + zgauss)} \, . \nonumber 
\ee
The summary of the best-fit values of this source for the four {\sc relxill} flavors is reported in Tab.~\ref{t-pks}. Fig.~\ref{r-pks} shows the spectra and the specific components for every best-fit as well as the data to best-fit model ratios. The constraints on the spin and the deformation parameter $\alpha_{13}$ are reported in Fig.~\ref{c-pks}.

Like for the previous sources, {\sc relxill\_nk} and {\sc relxillCp\_nk} provide very similar fits and it is not possible to distinguish the two models with the available data. {\sc relxillD\_nk} provides a better fit, but the constraints on the spin parameter and the deformation parameter $\alpha_{13}$ are very similar to those from {\sc relxill\_nk} and {\sc relxillCp\_nk}. {\sc relxilllp\_nk} provides a substantially worse fit and this is also the only case in which we do not recover the Kerr solution at $\alpha_{13} = 0$ at a high confidence level. For this source, the choice of the intensity profile seems thus to be very important.

\section{Concluding remarks \label{s-con}}

Thanks to a new generation of observational facilities, it is today possible to start testing Einstein's Theory of General Relativity in the strong field regime and black hole tests with electromagnetic and gravitational wave techniques are becoming a hot topic among both the astrophysics and the theoretical physics communities. The results of our group using X-ray reflection spectroscopy are currently the only constraints from an electromagnetic method on the spacetime metric near black holes, and in this review paper we have summarized the state of the art of the measurements of supermassive black holes. Our current work is devoted to improve these results by looking for more suitable sources/data and by upgrading our theoretical model to reduce systematic uncertainties.

The choice of the right source and the right data is extremely important in our studies because we want to analyze small features in the spectrum of the source and reduce the systematic uncertainties. The desired properties of sources/data seem to be:
\begin{enumerate}
\item Supermassive black holes seem to be more suitable than the stellar-mass ones, because their spectrum is easier to model and they naturally have a very high spin parameter (point 2 below).
\item Sources with very high spins ($a_* > 0.9$) in order to have the inner edge of the disk very close to the compact object and maximize the relativistic effects. For sources with low or moderate value of the spin parameter, relativistic effects are too weak and it is impossible to break the parameter degeneracy. 
\item Objects without intrinsic absorption (e.g. bare AGNs), so we do not have to worry about systematic uncertainties related to the absorption model.
\item We need a good energy resolution at the iron line (which is the most informative feature of the spectrum about the spacetime metric) and a broad energy band (which helps to break parameter degeneracy). Considering current X-ray facilities, we would need, for instance, simultaneous observations with \textsl{XMM-Newton} and \textsl{NuSTAR}.
\item Sources with a prominent broad iron line, because the latter is the most informative feature about the spacetime metric. Note that strong constraints on the deformation parameters may also be obtained from sources with weak iron lines and a strong soft excess. However, the nature of the soft excess is still controversial and the soft X-ray band is more significantly affected by Galactic absorption. 
\item The source should have a luminosity between 5\% to about 30\% of its Eddington limit. This is indeed the standard condition to have the accretion disk be well described by the Novikov-Thorne model with the inner edge at the radius of the innermost stable circular orbit.
\item It would be desirable to know the coronal geometry and that the corona is compact and very close to the black hole. Indeed, only if the coronal geometry is known can we properly take all relativistic effects into account. If the corona is compact and close to the black hole, most of the reflection radiation comes from the very inner part of the accretion disk, so the reflection spectrum is more significantly affected by the strong gravity region of the source. 
\end{enumerate}

The possibility of performing precision tests of General Relativity using X-ray reflection spectroscopy will eventually depend on our capability of improving current reflection models to reduce, and have under control, all systematic uncertainties. In Section~\ref{s-sys}, we have listed the main simplifications in our theoretical model and work is underway to upgrade our calculations. In Section~\ref{s-flav}, we have presented a very preliminary study of the impact of different choices of the model by fitting the X-ray spectrum of five bare AGNs with the four main flavors of our package, namely {\sc relxill\_nk}, {\sc relxillCp\_nk}, {\sc relxillD\_nk}, {\sc relxilllp\_nk}. The \textsl{Suzaku} data that we have analyzed are in the range $0.5-10$~keV, which does not permit us to fit the Compton hump and measure the coronal temperature. For this reason {\sc relxill\_nk} and {\sc relxillCp\_nk} provide the same results. When we employ the model {\sc relxillD\_nk} to allow a variable disk electron density, usually we get the same (or very similar) measurement of the deformation parameter $\alpha_{13}$ as {\sc relxill\_nk} and {\sc relxillCp\_nk}, with the exception of two sources (Ark~120 and PKS~0558--504) in which {\sc relxillD\_nk} provides a slightly better fit with a larger uncertainty on $\alpha_{13}$ for Ark~120 and essentially the same measurement for PKS~0558--504. With {\sc relxilllp\_nk}, we typically get worse fits, which suggests that our sources are not illuminated by a lamppost corona. However, in some cases we obtain a more realistic picture of the astrophysical system, in which the reflection component is subdominant with respect to the spectrum of the corona. The constraints on $\alpha_{13}$ found with {\sc relxilllp\_nk} are not dramatically different from the constraints from the other flavors, which means that still the choice of the intensity profile may have an impact on the reflection spectrum weaker than a deformation of the Kerr background. An exception is the case of PKS~0558--504, in which the analysis with {\sc relxilllp\_nk} would point out to deviations from Kerr at high confidence level. Work is underway to perform a more accurate study of the impact of these flavors with \textsl{NuSTAR} data.

%%%%%%%%%%%%%%%%%%%%%%%%%%%%%%%

{\bf Acknowledgments --}
This work was supported by the Innovation Program of the Shanghai Municipal Education Commission, Grant No.~2019-01-07-00-07-E00035, and Fudan University, Grant No.~IDH1512060. A.B.A. also acknowledges support from the Shanghai Government Scholarship (SGS). 
%K.C. also acknowledges support from the China Scholarship Council (CSC), Grant No.~2015GXYD34. 
A.T. also acknowledges support from the China Scholarship Council (CSC), Grant No.~2016GXZR89. S.N. acknowledges support from the Alexander von Humboldt Foundation and the Excellence Initiative at Eberhard-Karls Universit\"at T\"ubingen.

%%%%%%%%%%%%%%%%%%%%%%%%%%%%%%%

\end{document}